\documentclass[aps,twocolumn,showpacs,pra,superscriptaddress,floatfix,10pt]{revtex4-1}

\usepackage{graphicx,amsmath, amssymb, amsthm, amsfonts, mathtools}
\usepackage[usenames,dvipsnames]{xcolor}
\usepackage{float}

\usepackage{hyperref}			   
\usepackage{braket}

\def\beq{\begin{equation}}
\def\eeq{\end{equation}}
\def\beqa{\begin{eqnarray}}
\def\eeqa{\end{eqnarray}}
\def\bsub{\begin{subequations} \begin{align}}
\def\esub{\end{subequations} \end{align}}

\begin{document}

\title{Kink dynamics and quantum simulation of supersymmetric lattice Hamiltonians}

\author{Ji\v{r}\'{i} Min\'{a}\v{r}} 
\affiliation{Institute for Theoretical Physics, University of Amsterdam, Science Park 904, 1098 XH Amsterdam, the Netherlands}
\affiliation{QuSoft, Science Park 123, 1098 XG Amsterdam, the Netherlands}

\author{Bart van Voorden} 
\affiliation{Institute for Theoretical Physics, University of Amsterdam, Science Park 904, 1098 XH Amsterdam, the Netherlands}

\author{Kareljan Schoutens} 
\affiliation{Institute for Theoretical Physics, University of Amsterdam, Science Park 904, 1098 XH Amsterdam, the Netherlands}
\affiliation{QuSoft, Science Park 123, 1098 XG Amsterdam, the Netherlands}

\date{\today}

\begin{abstract}
We propose a quantum simulation of a supersymmetric lattice model using atoms trapped in a 1D configuration and interacting through a Rydberg dressed potential.	
The elementary excitations in the model are kinks or (in a sector with one extra particle) their superpartners - the skinks. The two are connected by supersymmetry and display identical quantum dynamics. We provide an analytical description of the kink/skink quench dynamics and propose a protocol to prepare and detect these excitations in the quantum simulator. We make a detailed analysis,  based on numerical simulation, of the Rydberg atom simulator and show that it accurately tracks the dynamics of the supersymmetric model.
\end{abstract}

\maketitle

\emph{Introduction. }
Models of strongly interacting fermions are key to our understanding of condensed matter systems. 
At the same time, they are notoriously hard to track, even with sophisticated tools ranging from numerical approaches such as quantum Monte-Carlo \cite{Suzuki_1993_Book, Gubernatis_2016_Book,Becca_2017_Book} and tensor networks \cite{Orus_2014_AnnPhys, Montangero_2018_Book} to application of gauge-gravity duality \cite{Zaanen_2015_Book}. 
One strategy to make progress is to consider models with special symmetries. A non-standard but intriguing choice is to consider \emph{supersymmetry} as an explicit symmetry on the lattice
\cite{Nicolai_1976_JPA, Fendley_2003_PRL, Fendley_2003_JPhysA,Fu_2017_PhysRevD,Sannomiya_2017_PhysRevD,OBrien_2017_PhysRevLett}
or as an emergent symmetry 
\cite{Grover_2014_Science,Huijse_2014_PRL,Yu_2010_PRL}.

${\cal N}=2$ supersymmetry in a lattice model or a quantum field theory comes with a number of tools, such as the Witten index
\cite{Witten_1982_NuclPhysB}, 
which facilitate the analysis. Exploiting these tools unveils remarkable features such as extensive ground state degeneracies, a phenomenon dubbed superfrustration \cite{vEerten_2005_JMathPhys,Fendley_2005_PRL} which can lead to multicriticality \cite{Chepiga_2021_arXiv}.

Despite the supersymmetry, many hard questions remain, such as the nature of the quantum phases in higher spatial dimensions. Here a quantum simulator might provide ingenious insights to these questions.
We make a step in this direction and propose such a simulator using arrays of neutral atoms trapped in optical potentials and dressed to their Rydberg state. This is motivated by the high versatility of these platforms~\cite{Browaeys_2020_NatPhys,Barredo_2016_Science,Endres_2016_Science,Barredo_2018_Nature,Wang_2019,Schauss_2015_Science, Labuhn_2016_Nature,Bernien_2017, deLeseleuc_2019_Science, Helmrich_2020_Nature} and by the fact that an off-resonant dressing \cite{Balewski_2014_NJP, Jau_2016_NatPhys, Zeiher_2016_NatPhys, Arias_2019_PRL} naturally implements the constrained dynamics inherent to the supersymmetric lattice model. Specifically, we consider a so-called M$_1$ model for spin-less fermions on a 1D chain \cite{Fendley_2003_PRL}. 
As a function of a parameter $\lambda$, this model interpolates between a trivial ($\lambda=0$) and a quantum critical ($\lambda=1$) phase, the latter connecting to superconformal field theory \cite{Fendley_2003_PRL, Huijse_2011_JStatMech}. 
The value of the Witten index $W=2$ indicates the existence of two supersymmetric vacua and points at kinks connecting these two vacua as elementary excitations. Furthermore, in a sector with one particle added, the excitations correspond to the \emph{superpartners} of the kinks, which we call the skinks. We propose a protocol for the (s)kink preparation and detection and solve for their dynamics following a quench (we note a recent study of kink-antikink pair dynamics in a spin chain \cite{Milsted_2020}). 
We show that it is identical in both cases and accurately reproduced by the quantum simulator which is a direct consequence and a clear-cut sign of the underlying supersymmetry.

\begin{figure}[t!]
	\centering
	\includegraphics[width=0.5\textwidth]{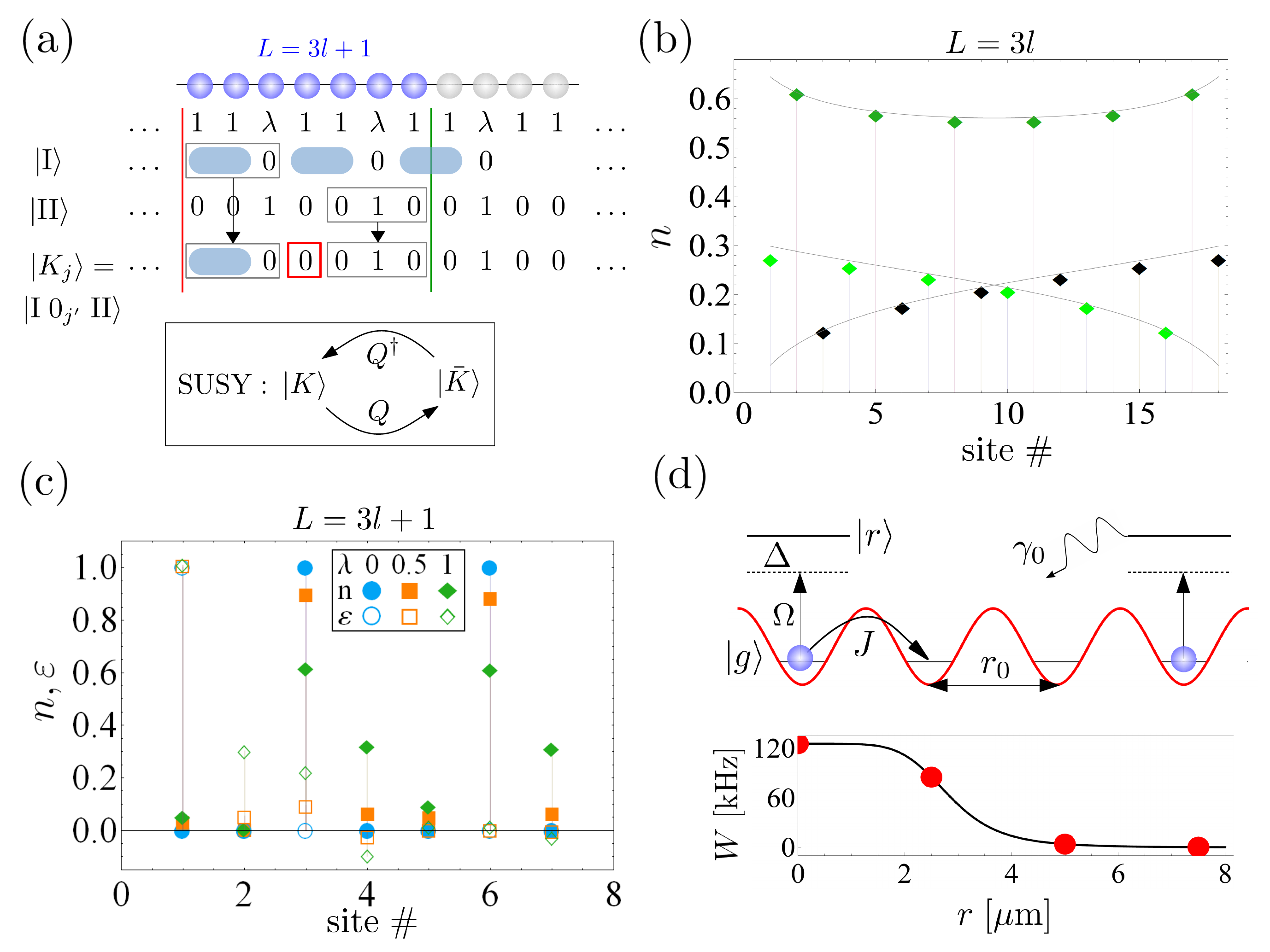}
	\caption{
	{\bf (a)} An infinite chain with staggering $11\lambda$ can accommodate two ground states $\ket{{\rm I}},\ket{{\rm II}}$. The lowest energy states for an open chain of length $L=3l+1$ and $l$ particles are kinks $\ket{K_j}$. A supercharge $Q$ acting on a kink creates a particle at the kink location. The blue oval represents the triplet.
	{\bf (b)} Particle densities in the ground state of an open chain, $L=3l$, $\lambda=1$ with an apparent ${\mathbb Z}_3$ pattern highlighted by the light green, dark green and black datapoints.
	{\bf (c)} Particle (filled symbols) and energy (empty symbols) densities of $\ket{K_1}$ for $\lambda=0,0.5,1$ (blue circles, orange squares, green diamonds).	
	{\bf (d)} Scheme of the proposed realization. Atoms in their electronic ground state $\ket{g}$ tunnel in an optical lattice with spacing $r_0$ at rate $J$ subject to dressing to a Rydberg state $\ket{r}$. Lower graph shows the dressed potential $W(r)$ for the $\ket{84{\rm}{\rm S}}$ state of ${}^6{\rm Li}$ with $\Omega = 2\pi \times 10\,{\rm MHz}$, $\Omega/\Delta=1/10$, $C_6=645\,{\rm GHz}\cdot \mu{\rm m}^6$.}
	\label{fig:scheme}
\end{figure}

\emph{The M$_1$ model.} 
An ${\cal N}=2$ supersymmetric lattice Hamiltonian for spinless fermions can be defined as 
\beq
\label{eq:HQ}
H_Q =\{Q,Q^\dag\},
\eeq
where $Q$ is the nilpotent supercharge, $Q^2=0$, and the brackets denote the anti-commutator. The M$_1$ model \cite{Fendley_2003_PRL} (on a bipartite graph) arises when $Q=\sum_i Q_i$ with $Q_i =(-1)^i\lambda_i c^\dag_i P_{\braket{i}}$, where $c_i$ are fermionic annihilation operators, $\{c_i,c_j\} = \{c_i^\dag,c_j^\dag\}=0$, $\{c_i,c^\dag_j\}=\delta_{ij}$, and $\lambda_i \in {\mathbb C}$. The M$_1$ model constraint, stipulating that fermions are not allowed to occupy nearest neighbour sites $\braket{ij}$, is implemented via the projector $P_{\braket{i}}=\prod_{j \in \braket{ij}} P_j$, with $P_j = 1 - n_j$, $n_j=c_j^\dag c_j$. The Hamiltonian $H_Q$ describes nearest neighbour hoppings and local interactions; it preserves the number of particles, $[H_Q,\sum_i n_i]=0$. 

We now specialize to 1D and specify real $\vec{\lambda}=\{\lambda_1, \lambda_2,\ldots,\lambda_L\}$, where $L$ is the length of the chain, $\lambda \geq 0$ and $\lambda_i$ repeats every 3 sites in a pattern $1\, 1\, \lambda$. 
For this choice of staggering, the M$_1$ model is known to be integrable \cite{Fendley_2011_JStatMech}. We refer to $\lambda=0$ as {\em extreme staggering}.

\emph{Supersymmetric groundstates. }
Let us first consider periodic boundary conditions, $L=3l$, $l \in {\mathbb N}$ and  $\vec{\lambda}=(1,1,\lambda,\ldots,1,1,\lambda)$. In this case, there are two supersymmetric ground states with $E=0$, each at 1/3 filling. 
At extreme staggering, they are $\ket{{\rm I}} \equiv \ket{t_{1,2} 0_3 \ldots t_{L-2,L-1} 0_L}$, $\ket{{\rm II}} \equiv \ket{0_1 0_2 1_3 \ldots 0_{L-2} 0_{L-1} 1_L}$, where $t_{j,j+1}=1/\sqrt{2}(c_j^\dag + c_{j+1}^\dag)\ket{{\rm vac}}$ is the triplet state and $\ket{\rm vac}$ the fermionic vacuum, see Fig. \ref{fig:scheme}a. 
For an \emph{open} chain of length $L=3l$, the degeneracy is lifted and we have a single $E=0$ ground state. 
Ref. \cite{Fendley_2011_JStatMech} analysed the particle densities $\braket{n_i}$ in this groundstate, perturbatively in $1/\lambda$. The same particle densities have been studied at the critical point $\lambda=1$ by invoking conformal field theory which provides closed form expressions for the associated scaling functions \cite{Huijse_2011_JStatMech, Suppl}. 
The corresponding particle densities constitute a direct experimental probe of the M$_1$ model as they follow a characteristic ${\mathbb Z}_3$ pattern indicated by the grey lines in Fig. \ref{fig:scheme}b together with the data points (diamonds) for $l=6$ \cite{Suppl}.

\emph{Kinks at extreme staggering. }
For an open chain of $L=3l+1$ there are no supersymmetric groundstates. Instead, at extreme staggering the lowest energy states with $l$ particles interpolate between the ground state configurations $\ket{{\rm I}}$ and $\ket{{\rm II}}$, with an empty site at position $i=3j-2$, with $j=1,\ldots,l+1$. We write these {\em bare kink} states as $\ket{K_j} = \ket{{\rm I}_{[1,i-1]} 0_{i} {\rm II}_{[i+1,L]}}$, where ${\rm I}_{[a,b]}$, ${\rm II}_{[a,b]}$ denote the part of the ground state configuration located between sites $a$ and $b$. They all have energy $E=1$. The labels $j=1 \; (j=l+1)$ correspond to the leftmost (rightmost) kink, see Fig. \ref{fig:scheme}a. 
Acting with the supercharge on the kink increases the number of particles by one creating the kink's superpartner, the \emph{skink},
$\ket{\bar{K}_j} \equiv Q \ket{K_j} = \ket{{\rm I}_{[1,i-1]} 1_{i} {\rm II}_{[i+1,L]}}$. Consequently, $Q^\dag \ket{\bar{K}_j}=\ket{K_j}$ such that $\ket{K_j}$ and $\ket{\bar{K}_j}$ form doublets under supersymmetry, see Fig. \ref{fig:scheme}a \cite{Fokkema_2017_SciPost}.
To characterize the kinks, we introduce a local energy density $h_i = \frac{1}{2}\left( \{Q,Q^\dag_i\} + \{Q^\dag,Q_i\} \right)$ such that $H_Q=\sum_{i=1}^L h_i$. 

Fig. \ref{fig:scheme}c shows the particle density $n=\braket{ n_i }$ and energy density $\varepsilon= \braket{h_i}$ for the leftmost kink $\ket{K_1}$ for $\lambda=0$ (blue data). The kink is clearly located at the left end of the chain with a corresponding peak in the energy density.

\begin{figure}[t!]
	\centering
	\includegraphics[width=0.37\textwidth]{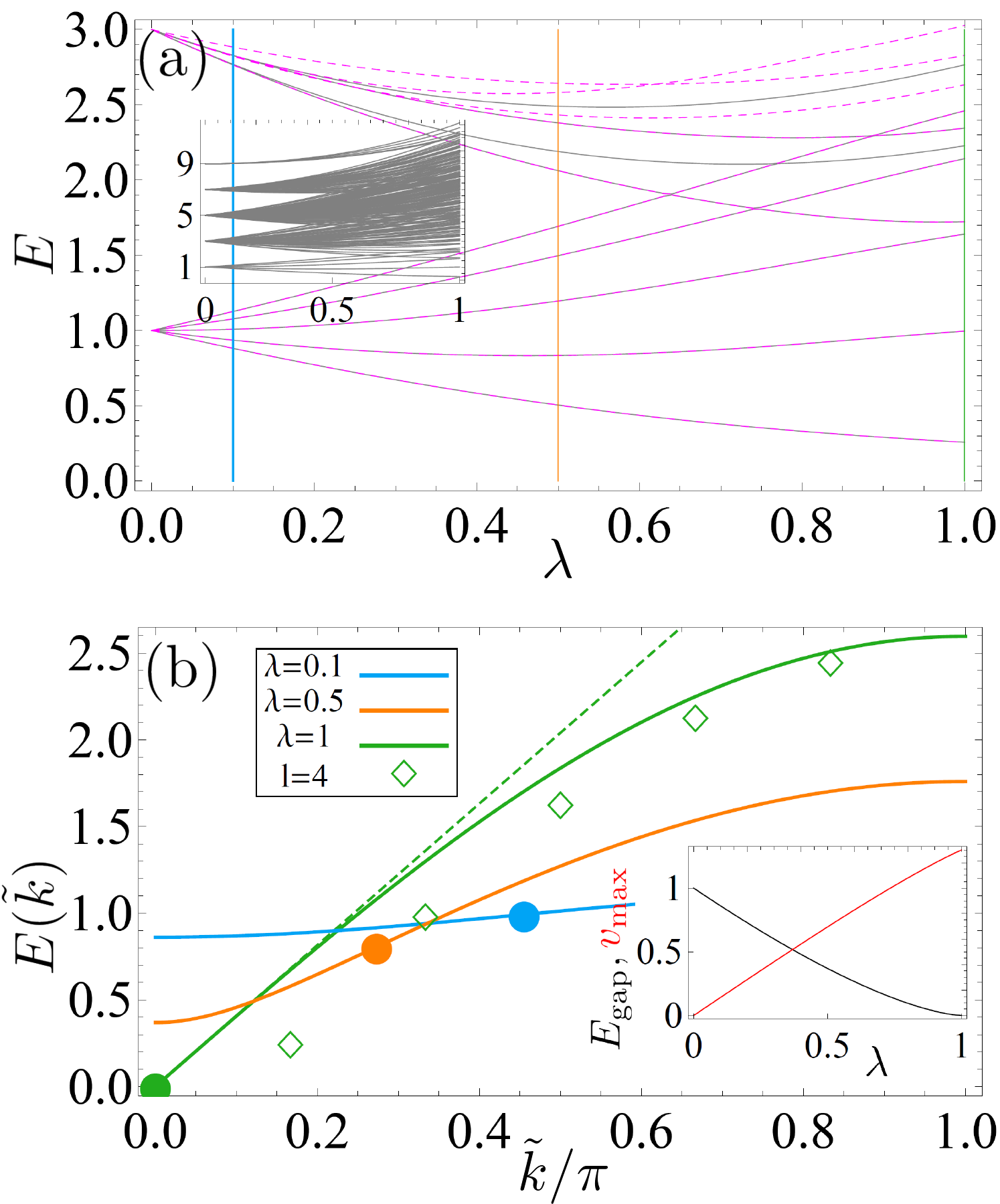}
	\caption{	
	{\bf (a)} Spectrum (inset) and nine lowest eigenenergies for $l=4$ in $l$ (grey) and $l+1$ (dashed magenta) particle-number sector of $H_Q$.
	{\bf (b)} The dispersion Eq. (\ref{eq:Dispersion}) for $\lambda=0.1,0.5,1$ (blue, orange, green). The filled circles correspond to the fastest mode $\tilde{k}$ with $v_{\rm max}$. The green diamonds denote the exact eigenenergies for $l=4$ and $\lambda=1$. The green dashed line is an eye-guide depicting the linear dispersion at the origin. The inset shows the gap, i.e. the lowest energy, (black) and $v_{\rm max}$ (red) vs. $\lambda$.	
	}	
	\label{fig:2}
\end{figure}

\emph{Kinks at general $\lambda$. }
We claim that the notion of 1-kink (and multi-kink) states is well defined also away from extreme staggering, when $0<\lambda \leq 1$. To illustrate this, we present in the inset of Fig. \ref{fig:2}a the spectrum of the system for $l=4$. The energies become degenerate for $\lambda=0$ taking odd positive values corresponding to the 1-kink, 3-kink, etc. states. The unavoided level crossings, characteristic for integrability, allow us to unambiguously characterize states as multi-kink states for all $\lambda$. 

Fig. \ref{fig:2}a shows the low-lying part of the spectrum, which includes a band of $l+1$ 1-kink eigenstates $\ket{v_k}$ of energy $E_k$. We define a localized kink as \cite{InPrep}
\beq
	\ket{K_j} = \sqrt{\frac{2}{l+2}} \sum_{k=1}^{l+1} \sin( \tilde{k} j) \ket{v_k},
	\label{eq:Kinks}
\eeq
where $\tilde{k}=\pi k/(l+2)$. 

In Fig. \ref{fig:scheme}c the orange and green data points show the particle and energy densities in the state $\ket{K_1}$ obtained numerically using Eq.~(\ref{eq:Kinks}) for $\lambda=0.5,1$. We see that, even for $\lambda=1$, the kink is well defined with most of its energy localized at the kink position. 

\begin{figure}[t!]
	\centering
	\includegraphics[width=0.5\textwidth]{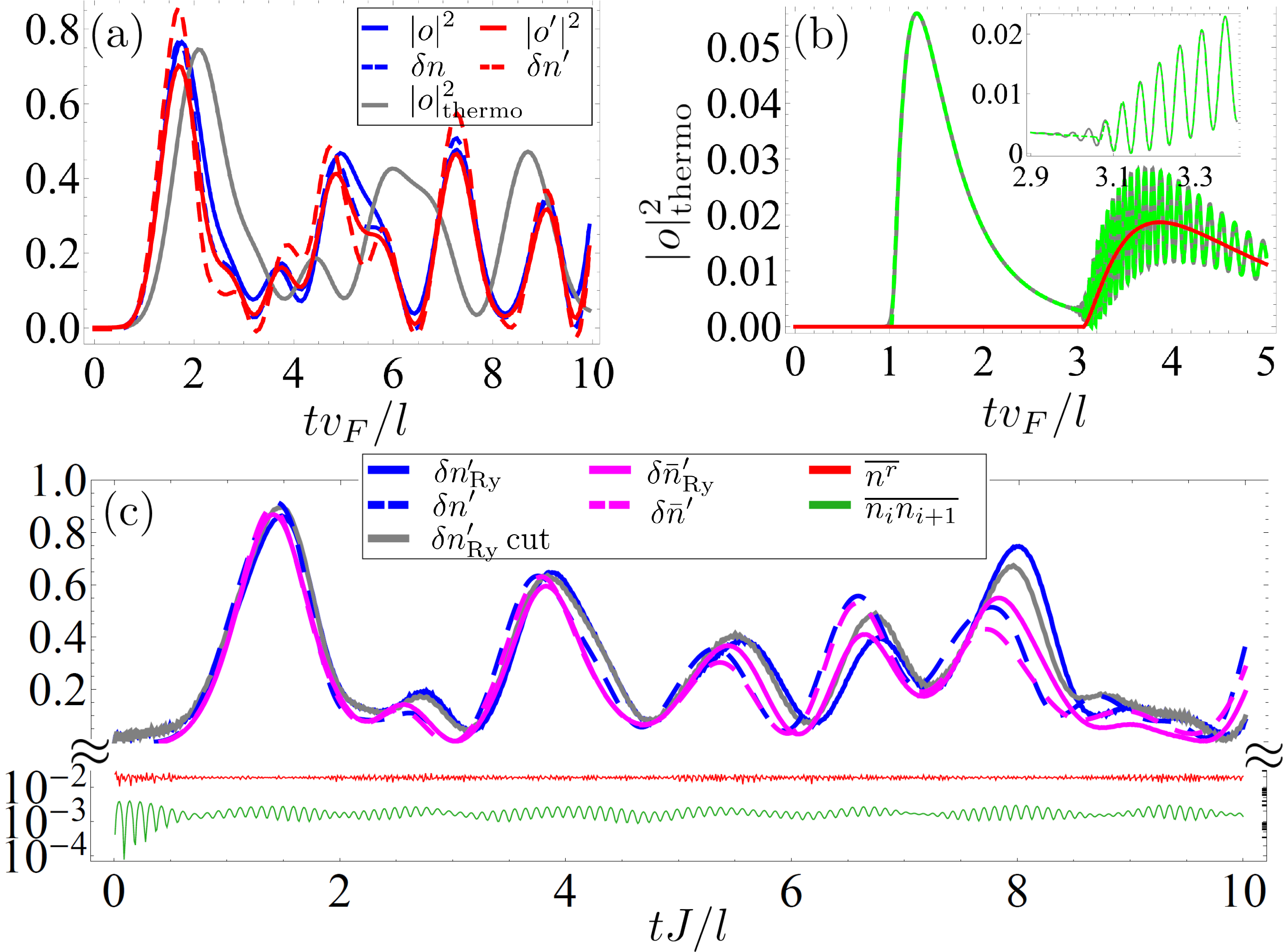}
	\caption{
	{\bf (a)} Time evolution of $|o(t)|^2$, Eq. (\ref{eq:Overlap}), (solid lines) and $\delta n$ (dashed lines) for a quench from the exact (blue) and reduced fidelity (red) kink state $\ket{K_1}$ for $l=4$. The grey line corresponds to $|o(t)|^2$ evaluated with Eq. (\ref{eq:Dispersion}) for the eigenenergies.
	{\bf (b)} Numerical evaluation (grey, Eq. (\ref{eq:Overlap})) and saddle point approximation (green dashed, red, Eq. (\ref{eq:Overlap_saddle})) of the overlap $|o(t)|^2$ for $l=101$. The green (red) lines correspond to considering the first two (only the second) saddle points. The inset shows the onset of oscillations around $t v_F/l = 3$.
	{\bf (c)} Quench dynamics for 10-site chain ($l=3$), with initial states $\ket{K_1'}$ (blue) and $\ket{\bar{K}_1'}$ (magenta). Solid lines show dynamics under the full Hamiltonian $H_{\rm Ry}$, while the grey curve is for a truncation of $H_{\rm Ry}$, neglecting interactions beyond next-nearest neighbours. Dashed lines show dynamics under $H_Q$. The red line shows the average population in the Rydberg state $\overline{n^r}=1/l\sum_i n^r_i$, while the green line tracks the nearest-neighbour occupation of ground state atoms, $\overline{n_i n_{i+1}}=1/l\sum_{i'} n_{i'}n_{i'+1}$. Parameters as in Fig. \ref{fig:scheme}d.
	}	
	\label{fig:3}
\end{figure}

\emph{Kink dynamics. }
We now proceed with the evaluation of the kink dynamics. We start from the leftmost kink $\ket{K_1}$ and consider overlap at time $t$ with the rightmost kink, $o(t) \equiv \braket{K_{l+1}|K_1(t)}$, where $\ket{K_1(t)}={\rm e}^{-i H_Q t} \ket{K_1}$.
It follows from Eq.~(\ref{eq:Kinks}) that
\beq
	o(t) = \frac{2}{l+2} \sum_{k=1}^{l+1} \sin \left( \tilde{k} \right) \sin \left( \tilde{k} (l+1) \right) {\rm e}^{-i E_k t}.
	\label{eq:Overlap}
\eeq
 
For simplicity, from now on we focus on the critical case $\lambda=1$.
In Fig. \ref{fig:3}a we show $|o(t)|^2$ for $l=4$ (solid blue line). 
At criticality, the fastest mode propagates with the Fermi velocity $v_F$, see the discussion after Eq.~(\ref{eq:Dispersion}). This results in the onset of the overlap at $t v_F/l \approx 1$, with the maximum achieved for a later time, $t v_{F}/l \approx 1.75$.

\emph{Kink detection.}
To make a connection with experimentally observable quantities, we construct an observable $\delta n$ which detects the presence of a kink at the right end of the system, by requiring that
$\braket{K_i | \delta n | K_j} \approx \delta_{i,l+1} \delta_{j,l+1}$. Taking $\delta n = \alpha(\lambda,l) \left[ 1 - \beta(\lambda,l) \left(n_{L-1} + n_{L+1} \right) \right]$ we find $\alpha(0,l)=1$, $\beta(0,l)=1$ for any $l$ and $\alpha(1,l)\approx 1.08$, $\beta(1,l) \approx 1.09$ for $l=3,4$ \cite{Suppl}. The numerically obtained result for $\delta n(t)$ is shown as a blue dashed line in Fig. \ref{fig:3}a and corresponds to a good accuracy to $|o|^2$.

\emph{Kink preparation.}
An important question is how the spatially localized kink $\ket{K_1}$ can be prepared in practice. To this end we note that the kink site and its nearest neighbours remain approximately empty for all $\lambda$, cf. Fig. \ref{fig:scheme}c. 
We thus consider an adiabatic preparation of a ground state $\ket{K_1'}$ of the final Hamiltonian $H_f=H_Q + \mu (n_1 + n_2)$, where $\mu \rightarrow \infty$ ensures the kink condition on the first two sites.
The initial Hamiltonian is chosen such that its ground state is a kink at extreme staggering $\lambda=0$ (and similarly for skinks below), cf. \cite{Suppl}.
For $l=4$ we find the fidelities $F \in [0.95,1]$, where $F=|\braket{K_1' | K_1}|$, with the highest (lowest) value at extreme staggering (criticality).
In Fig. \ref{fig:3}a we show the numerically evaluated overlap $|o'|^2=|\braket{K_{l+1}|K_1'(t)}|^2$ and the corresponding observable $\delta n'$ as solid (dashed) red lines. We find that despite the limited fidelity of the initial state,  $|o'|^2$ and $\delta n'$ agree well with $|o|^2$ and $\delta n$. 

\emph{Skinks.}
Supersymmetry guarantees that the 1-skink energies (the lower dashed magenta lines in Fig. \ref{fig:2}a) in the sector with $l+1$ particles are identical to the 1-kink energies $E_k$. As a consequence, the quench dynamics for the skinks is again given by Eq.~(\ref{eq:Overlap}). For the detection of $\ket{\bar{K}_{l+1}}$ we propose $\delta \bar{n} = -\bar{\alpha}(\lambda,l) \left[ 1 - \bar{\beta}(\lambda,l) \left(n_{L-2} + n_{L-1} + n_{L+1} \right) \right]$ with $\bar{\alpha}(0,l)=2$, $\bar{\beta}(0,l)=1$ and $\bar{\alpha}(1,l)\approx 1.46$, $\bar{\beta}(1,l)\approx 0.98$ for $l=3,4$. For the preparation we find that the ground state $\ket{\bar{K}_1'}$ of $H_f=H_Q + 3(-n_1 + n_2-0.5 n_3)$ corresponds well to $\ket{\bar{K}_1}$ \cite{Suppl}. The $l=4$ fidelities are $\bar{F} \in [0.93,1]$ with $\bar{F}=|\braket{\bar{K}_1' | \bar{K}_1}|$.

\emph{Kink/skink dynamics at large $l$.}
Surprisingly, the kink arrival amplitude Eq.~(\ref{eq:Overlap}) is analytically tractable, for general $\lambda$, in the large-$l$ limit. A key element for this is the continuum limit $E(\tilde{k})$ of the kink dispersion relation. Exploiting a relation between the M$_1$ model and the XYZ spin-1/2 chain \cite{Fendley_2010_JPhysA}, we have found \cite{InPrep}
\beq
	E(\tilde{k}) = \frac{\left(3 \lambda +s\right)^{3/2} \sqrt{1-\left(1-\frac{\left(-3 \lambda +s\right)^3 \left(\lambda +s\right)}{\left(-\lambda +s\right) \left(3 \lambda +s\right)^3}\right) \cos ^2\left(\frac{\tilde{k} }{2}\right)}}{2 \sqrt{2}
   \sqrt{\lambda +s}},
	\label{eq:Dispersion}
\eeq
where $s=\sqrt{8+\lambda ^2}$. In Fig. \ref{fig:2}b we show the dispersion for $\lambda=0.1, 0.5, 1$. We denote by $v_{\rm max}(\lambda)$ the maximum value of the group velocity $v(\tilde{k})=\partial_{\tilde{k}}E(\tilde{k})$.  At criticality, $v_{\rm max}(\lambda=1)=v_F=3\sqrt{3}/4$, with $v_F$ the Fermi velocity. This gives real space velocity (since kinks hop three sites at a time) $3 v_F=9\sqrt{3}/4$, in agreement with \cite{Huijse_2011_JStatMech}. 

In Fig.~ \ref{fig:3}a the grey line shows the overlap Eq.~(\ref{eq:Overlap}) evaluated with the energies $E(\tilde{k})$ instead of $E_k$ (blue line). 
The difference is a consequence of finite $l$, cf. the green diamonds vs. green solid line in Fig.~\ref{fig:2}b.

Using the dispersion $E(\tilde{k})$ we can evaluate the large-$l$ limit of Eq.~(\ref{eq:Overlap}) in a saddle point approximation \cite{Suppl}, giving
\beqa
	o(t) &\approx & \frac{2}{l+2}\sum_{s=1}^\infty \theta\left( \frac{v_{\rm max}t}{l+2} - (2s-1)\right) \sin \left( \tilde{k}_s \right)^2 \nonumber \\
	&& {\rm e}^{i \left[(2s-1)\pi k_s + E(\tilde{k}_s)t\right] + i \frac{5\pi}{4}} \sqrt{\frac{2\pi}{-E''(\tilde{k}_s)t}},
	\label{eq:Overlap_saddle}
\eeqa
where $\theta$ is the Heaviside step function, $k_s = \left.E'\right.^{-1} \left((2s-1)\pi/t \right)$, $E'=\partial_{k} E(\tilde{k})$, $E''=\partial^2_{k}E(\tilde{k})$  and $s$ labels the saddle point corresponding to the arrival times $t = (2s-1)(l+2)/v_{\rm max} \approx (2s-1)l/v_{\rm max}$, $s \in \mathbb{N}$, of the kink front (maximum velocity mode). At criticality, where $E(\tilde{k})=2v_F \sin(\tilde{k}/2)$, the saddle point expression takes a simple closed form \cite{Suppl}.

In Fig. \ref{fig:3}b we show an example of the dynamics for $l=101$ evaluated using Eq. (\ref{eq:Overlap}) (grey line) together with the prediction of Eq. (\ref{eq:Overlap_saddle}) (green dashed line). We see a close to perfect agreement, with the inset showing the details around $t v_F/l=3$, where the second saddle point, $s=2$, starts to generate the characteristic modulation of the overlap due to the interference of the kink front propagating at $v_F$ incident on the right edge (after it has undergone one round trip) and the kink tail. We note the frequency chirp of the modulation due to the non-trivial time dependence of $\tilde{k}_s$. Here we do not show the observable $\delta n(t)$ as for large $l$ the Hamiltonian cannot be diagonalized exactly. 

\emph{Experimental implementation. }
We now discuss how $H_Q$ can be engineered using Rydberg dressed atoms \cite{Henkel_2010_PRL,Pupillo_2010_PRL}. 
We consider effectively two-level atoms with the ground and Rydberg states $\ket{g}$, $\ket{r}$, where the ground state atoms experience an optical lattice potential and the atoms in a Rydberg state a repulsive Van der Waals interaction described by
\beqa
	H_{\rm Ry} &=& -J \sum_{i=1}^{L-1}\left( c^\dag_{i+1} c_i + c^\dag_i c_{i+1} \right) + \sum_{i=1}^L \mu_i n_i \nonumber \\
		& & + \sum_{i=1}^L \Omega \sigma^x_i + \Delta n^r_i +  \sum_{i>j=1}^{L-1} V_{ij} n^r_i n^r_j.
	\label{eq:HRy}
\eeqa
Here, $J>0$ is the hopping amplitude, $\sigma^x=\ket{r}\bra{g}+\ket{g}\bra{r}$, $n^r=\ket{r}\bra{r}$ and $V_{ij}=C_6/(r_0|i-j|)^6$ with $C_6$ the Van der Waals coefficient and $r_0$ the lattice spacing.
We consider a regime of large detuning $\Omega/\Delta \ll 1$, where the ground state atoms interact, up to order $\Omega^4$, through an effective flat-top potential $W(r=r_0|i-j|) =  2\Omega^4 V_{ij}/[\Delta^3(V_{ij}+2 \Delta)]$, cf. Fig. \ref{fig:scheme}d. To obtain the supersymmetric $H_Q$, the interaction and chemical potentials $W$, $\mu$ and the hopping $J$ need to be tuned as follows.

For simplicity, we refer the discussion of general $\lambda$ to \cite{Suppl} and focus on $\lambda=1$. In this case
the chemical potential terms in $H_Q$ become site-independent up to the boundary terms originating from $P_1 P_3$ and $P_{L-2} P_L$, which can be accounted for by setting $\mu_1 = \mu_L = J$.

Next, the M$_1$ model Hamiltonian forbids nearest neighbour occupation while the potential terms are of the form $P_{i-1}P_{i+1}$, with no interactions beyond lattice distance 2. For this to be captured by the flat-top potential we need $W(r_0)/W(2 r_0) \gg 1$ and $W(2 r_0)/W(3 r_0) \gg 1$ with the maximum achieved in the limit $r_0 \rightarrow \infty$. However, to counteract experimental imperfections \cite{Suppl}, one should reduce the duration of the simulation by maximizing the relevant energy scale, here $W(2 r_0)$, which happens for $r_0 \rightarrow 0$
and one has to set $J=W(2 r_0)$. 
This corresponds to the optimal approximation of $H_Q$ using single dressing. 
Importantly, we show in \cite{Suppl} that $H_Q$ can be reached in principle with arbitrary number of dressings with already a tenfold increase in $W(r_0)/W(2 r_0)$ and $W(3 r_0)/W(2 r_0)$ for a double dressing with realistic parameters.


As a specific example, we consider the fermionic ${}^6{\rm Li}$ dressed with the $\ket{84{\rm S}}$ state with $C_6=645\,{\rm GHz}\cdot \mu{\rm m}^6$ \cite{Sibalic_CompPhysComm_2017,ARC_package} and lattice spacing $r_0=2.5\,\mu{\rm m}$. The resulting dressed potential is shown in Fig. \ref{fig:scheme}d. We get $W(2 r_0) = J \approx 4\,{\rm kHz}$, which for the optical lattice laser wavelength $\lambda=2 r_0 = 5 \,\mu{\rm m}$ corresponds to lattice depth $\approx 5.5 E_r$, $E_r$ being the recoil energy \cite{Arzamasovs_2017} 
\footnote{In order to be well in the deep lattice limit where the tight-binding approximation is applicable, one might further reduce $\Omega/\Delta$. This would in turn reduce $J$ and the achievable $L_{\rm max}$. To overcome this limitation, one could use a Raman-assisted hopping as we discuss in \cite{Suppl}.}.

Fig. \ref{fig:3}c shows the \emph{quantum simulation} of $H_Q$, where we compare the dynamics generated by the Rydberg Hamiltonian (\ref{eq:HRy}) with that of $H_Q$ quenching from $\ket{K_1'}$ and $\ket{\bar{K}_1'}$, see caption for details. We draw two main conclusions. First, the quantum simulator accurately tracks the dynamics set by the model hamiltonian $H_Q$ and, second, the dynamics in the $l$-particle sector (blue lines) is highly similar to that in the $l+1$ particle sector (magenta lines). The latter observation is direct evidence of the supersymmetry of $H_Q$.


\emph{Outlook. }%
We have proposed a realization of a supersymmetric lattice Hamiltonian $H_Q$ based on atoms interacting through a Rydberg dressed potential \cite{Weimer_APS_2012,Weimer_DPG_2012}. Our results constitute a stepping stone to quantum simulations of supersymmetric lattice models in higher dimensions 
\cite{Fendley_2005_PRL, Huijse_2008_PRL, Huijse_2012_NJP, Galanakis_2012_PRB,Surace_2020}, 
which can require $n$-body, rather than 2-body, interactions. In this context, it would be interesting to consider a scheme relying on coupling the Rydberg atoms with phonons \cite{Gambetta_PRL_2020} or to use cold molecules with permanent or electric-field induced dipole moments, avoiding the need for off-resonant dressing \cite{Buchler_2007,Carr_2009_NJP,Baranov_2012_ChemRev,Balakrishnan_2016_JChemPhys}.
Another interesting avenue is to exploit the mapping of the supersymmetric lattice Hamiltonians to spins \cite{Fendley_2003_JPhysA,
Fendley_2010_JPhysA,Chepiga_2021_arXiv, InPrep}
which would allow for simulations with platforms such as superconducting devices with $n$-body interactions \cite{Pedersen_2019_PRR,Roy_2020_PRApp}.

\emph{Acknowledgments. }
We are very grateful to P. Fendley, I. Lesanovsky, Y. Miao, E. Ilievski, N. Chepiga, F. Schreck, K. van Druten, R. Spreeuw, R. Gerritsma, T. Lahaye, B. Pasquiou, V. Barb\'{e} and A. Urech for stimulating discussions. 
This work is part of the Delta ITP consortium, a program of the Netherlands Organisation for Scientific Research (NWO) that is funded by the Dutch Ministry of Education, Culture and Science (OCW).






%


\clearpage

\onecolumngrid

\setcounter{equation}{0}
\setcounter{figure}{0}
\renewcommand{\theequation}{S\arabic{equation}}
\renewcommand{\thefigure}{S\arabic{figure}}

\begin{center}
{\Large{SUPPLEMENTAL MATERIAL}}
\end{center}

\section{Ground state particle densities}
\label{app:GS}

In this section we recall the expressions for single particle densities for the M$_1$ model at criticality, on an open chain of length $L=3l$. We used these expressions to produce the grey lines in Fig. 1b. Exploiting the tools of conformal field theory, the model can be mapped to a free boson and the particle densities can be expressed in terms of the correlators of the bosonic vertex operators with a characteristic ${\mathbb Z}_3$ pattern \cite{Huijse_2011_JStatMech}. Specifically, they read
\begin{subequations}
	\label{eq:n_GS}
	\begin{align}
		n_{3j-2} &= \frac{1}{3}-\frac{2A}{3} \frac{s(x)}{{\mathfrak s}(x)} \\
		n_{3j-1} &= \frac{1}{3}+\frac{2A}{3} \frac{c(x-L'/2)}{{\mathfrak s}(x)} \\
		n_{3j} &= \frac{1}{3}+\frac{2A}{3} \frac{s(x-L')}{{\mathfrak s}(x)},
	\end{align}
\end{subequations}
where $x \in [2,L]$, $L'=L+3$ and
\begin{subequations}
	\begin{align}
		s(x)&=\left( \frac{\pi}{2 L'} \right)^{\frac{1}{3}} \sin\left( \frac{\pi x}{3L'} \right) \\
		c(x)&=\left( \frac{\pi}{2 L'} \right)^{\frac{1}{3}} \cos\left( \frac{\pi x}{3L'} \right)  \\
		{\mathfrak s}(x)&=\sin\left( \frac{\pi x}{L'} \right)^{\frac{1}{3}}.
	\end{align}
\end{subequations}
Here, the parameter $A$ has been determined numerically as $A=0.77$ \cite{Huijse_2011_JStatMech}. We note that analogous results hold for $L=3l-1$ \cite{Huijse_2011_JStatMech}.

\section{(S)kink profiles and design of the observables}
\label{app:Observables}

\hspace*{-1cm}
\begin{figure}[h!]
\centering
\includegraphics[scale=0.6]{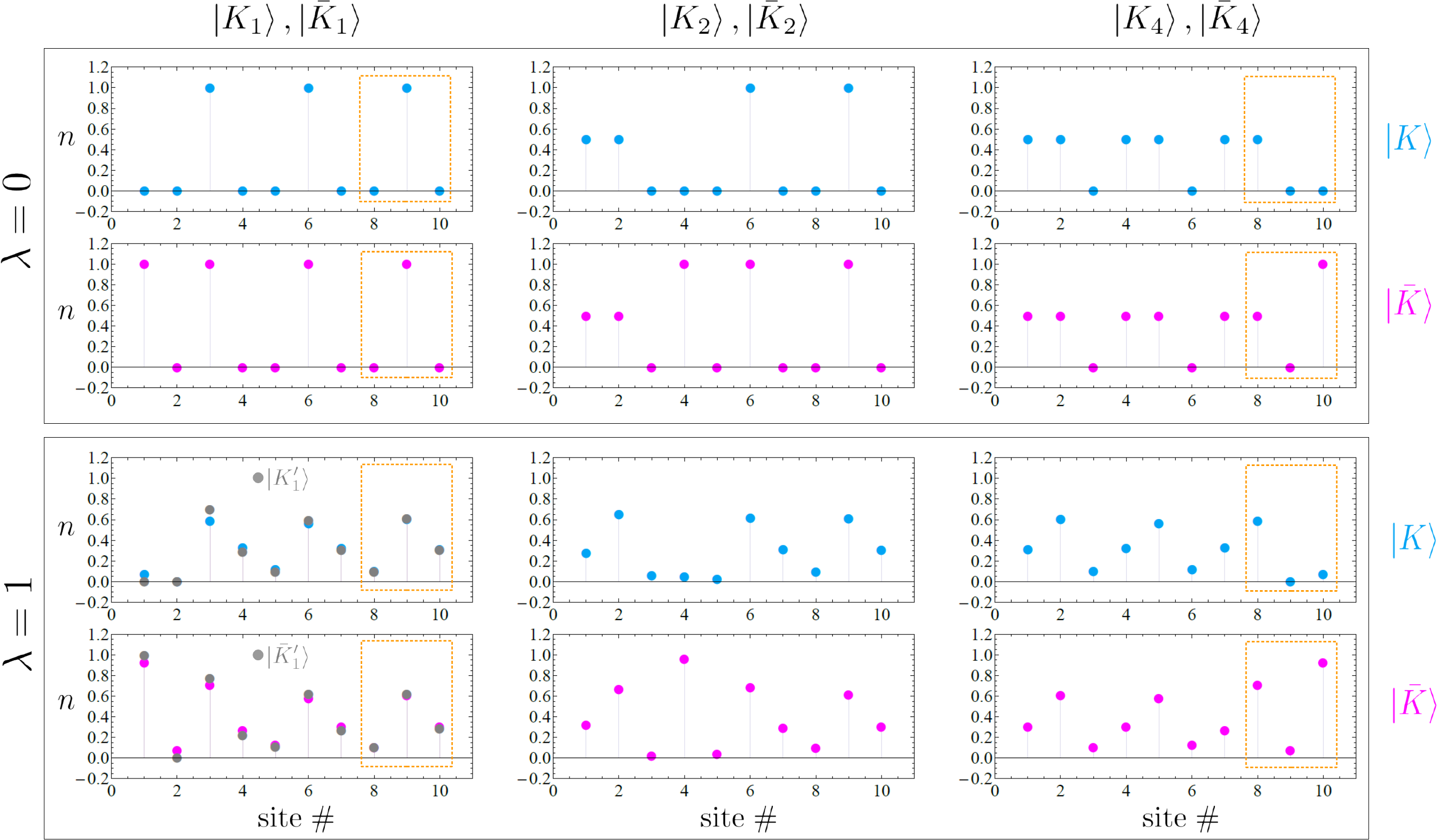}
\caption{
Single site particle densities of kinks (blue data points) and skinks (magenta data points) for $l=3$ at extreme staggering $\lambda=0$ (first two rows) and at criticality $\lambda=1$ (third and fourth row) for the leftmost (s)inks $\ket{K_1}, \ket{\bar{K}_1}$ (first column), $\ket{K_2}, \ket{\bar{K}_2}$ (second column) and the rightmost (s)kinks $\ket{K_4}, \ket{\bar{K}_4}$ (third column). The gray data points for $\lambda=1$ indicate the densities evaluated with $\ket{K_1'}$ and $\ket{\bar{K}_1'}$ for kinks and skinks respectively, see text for details.
}
\label{fig:profiles}
\end{figure}

To motivate the choice of observables, we will first examine the kink and skink profiles. Fig.~\ref{fig:profiles} shows the single site particle densities of kinks (blue data points) and skinks (magenta data points) for $l=3$ at extreme staggering, $\lambda=0$, (first two rows) and at criticality, $\lambda=1$, (third and fourth row) for the leftmost (s)inks $\ket{K_1}, \ket{\bar{K}_1}$ (first column), $\ket{K_2}, \ket{\bar{K}_2}$ (second column) and the rightmost (s)kinks $\ket{K_4}, \ket{\bar{K}_4}$ (third column). We wish to design observables which capture the properties of the overlaps $o(t)=\braket{K_{l+1}|K_1(t)}=\braket{\bar{K}_{l+1}|\bar{K}_1(t)}=\bar{o}(t)$ signalling the arrival of the leftmost kink to the right edge. Motivated by the (s)kink profiles, we in particular examine the situation in the last unit cell corresponding to the last three sites of the chain and to staggering $1 \lambda 1$ indicated by the orange boxes in Fig.~\ref{fig:profiles}. For the observable we consider a function of the particle densities on these three sites, $\delta n = \delta n(n_{L-2}, n_{L-1}, n_L)$, and require that $\braket{K_i | \delta n | K_j} \approx \delta_{i,l+1} \delta_{j,l+1}$. We find that these conditions are satisfied by imposing
\begin{subequations}
	\label{eq:dn_def}
	\begin{align}
		\braket{K_1 | \delta n | K_1} &= 0 \\
		\braket{K_{l+1} | \delta n | K_{l+1}} &= 1
	\end{align}
\end{subequations}
and similarly for the skinks. 

\begin{figure}[h!]
\centering
\includegraphics[scale=0.6]{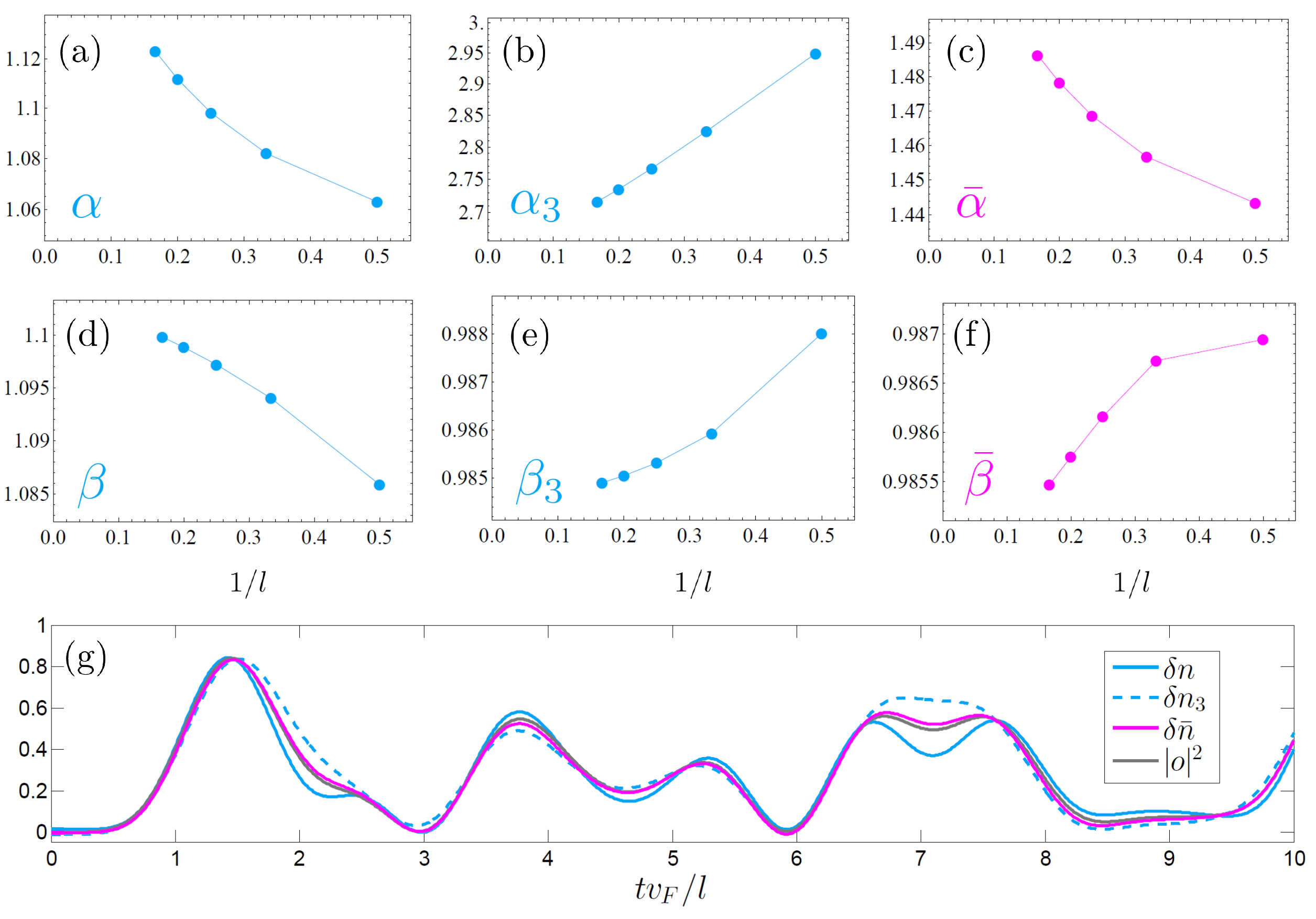}
\caption{
	Scaling of the coefficients {\bf(a,d)} $\alpha,\beta$, {\bf(b,e)} $\alpha_3, \beta_3$ and {\bf(c,f)} $\bar{\alpha}, \bar{\beta}$ parametrizing the observables (\ref{eq:dn_par}) with $l$ at criticality, $\lambda=1$. 
	{\bf(g)} Time evolution of the observables (\ref{eq:dn_par}) for $l=3$ following a quench from $\ket{K_1}$ (blue lines) and $\ket{\bar{K}_1}$ (magenta line) together with the square of the overlap $|o|^2 = |\braket{K_{l+1}|K_1(t)}|^2 = |\braket{\bar{K}_{l+1}|\bar{K}_1(t)}|^2$ (gray line).
}
\label{fig:ab_scaling}
\end{figure}

At extreme staggering, the occupation of the last unit cell is 1 and 1/2 for $\ket{K_1}$ and $\ket{K_{l+1}}$ (1 and 3/2 for $\ket{\bar{K}_1}$ and $\ket{\bar{K}_{l+1}}$). We also note, that for kinks the total particle number at the last \emph{two} sites is different for $\ket{K_1}$ and $\ket{K_{l+1}}$, but the same for the skinks. In contrast to the extreme staggering, at criticality (and generally for $\lambda > 0$) the total particle number in the last unit cell varies and is not equal to integer or half-integer values. This motivates us to introduce the following observables 
\begin{subequations}
	\label{eq:dn_par}
	\begin{align}
		\delta n &= \alpha(\lambda,l) \left[ 1 - \beta(\lambda,l) \left(n_{L-1} + n_{L+1} \right) \right] \\
		\delta n_3 &= \alpha_3(\lambda,l) \left[ 1 - \beta_3(\lambda,l) \left(n_{L-2} + n_{L-1} + n_{L+1} \right) \right] \\
		\delta \bar{n} &= -\bar{\alpha}(\lambda,l) \left[ 1 - \bar{\beta}(\lambda,l) \left(n_{L-2} + n_{L-1} + n_{L+1} \right) \right],
	\end{align}
\end{subequations}
where the coefficients $\alpha,\beta$ in general depend on $\lambda$ and the chain length $L=3l+1$, where the latter is to be expected, similarly to the presence of such scaling in the particle densities (\ref{eq:n_GS}) of the ground state of chain of length $L=3l$. At extreme staggering, the value of the coefficients can be read-off directly from Fig.~\ref{fig:profiles} and we obtain
\begin{subequations}
	\begin{align}
		\lambda=0: & \nonumber \\
		(\alpha, \beta) &=(1,1) \\
		(\alpha_3, \beta_3) &=(2,1) \\
		(\bar{\alpha}, \bar{\beta}) &=(2,1)
	\end{align}
\end{subequations}
which are independent of $l$. For $\lambda=1$, the values are obtained from the defining property (\ref{eq:dn_def}) and their scaling with $l$ is shown in Fig.~\ref{fig:ab_scaling}a-f (up to $l=6$). We use these values in the analysis of the critical dynamics.

To examine the observables, in Fig.~\ref{fig:ab_scaling}g we show the (square of the) overlap (gray) together with $\delta n, \delta n_3$ (solid and dashed blue) and $\delta \bar{n}$ (magenta) for a quench from $\ket{K_1}$ and $\ket{\bar{K}_1}$ respectively with the dynamics generated by the supersymmetric Hamiltonian $H_Q$, Eq.~(\ref{eq:HQ}). We see that the proposed observables nicely capture the overlap function $|o|^2$.

\hspace*{-1cm}

~\\
~\\
\section{State preparation}
\label{app:Preparation}

In this section we study state preparation by means of adiabatic following for the two scenarios considered in the main text, an open chain of length $L=3l$ and staggering $1 \lambda 1 ... 1 \lambda 1$ \emph{(scenario 1)} and an open chain of length $L=3l+1$ and staggering $11\lambda \ldots 1\lambda 1$ \emph{(scenario 2)}. Motivated by the relative simplicity of the experimental implementation, we focus on the critical case $\lambda=1$. We comment on the (un)suitability of the adiabatic protocol for critical systems at the end of the section.

We consider a preparation of a ground state of a final Hamiltonian $H_f$ by adiabatic following from $H_i$, such that the time-dependent Hamiltonian can be expressed as
\beq
	H(t) = (1-\mathcal{F}(t))H_i + \mathcal{F}(t) H_f.
	\label{eq:Ht}
\eeq
We consider three distinct cases for $H_f$. \emph{(i)} $H_f=H_Q$ in \emph{(scenario 1)}, \emph{(ii)} $H_f=H_Q +\mu(n_1+n_2)$ with $\mu \rightarrow \infty$ for kink and \emph{(iii)} $H_f=H_Q +\bar{\mu}(-n_1+n_2-\bar{\nu} n_3)$ with $\bar{\mu},\bar{\nu} \approx 3,1/2$ for skink preparation in \emph{(scenario 2)}. $H_f$ in \emph{(ii)} is chosen to ensure no occupation of the first two sites in preparation of the leftmost kink $\ket{K_1}$, a condition well satisfied even at criticality. This is apparent from Fig.~\ref{fig:profiles} ($\lambda=1, \ket{K_1}$ pane, blue data points) where we plot the densities corresponding to the ground state $\ket{K'_1}$ of $H_f$ as gray data points. For skinks, enforcing a particle at first site and zero particle on the second yields a density on the third site which is too small as compared with the ideal skink $\ket{\bar{K}_1}$. To rectify that, we introduce the $H_f$ described in \emph{(iii)} and optimize the skink fidelity by tuning $\bar{\mu}, \bar{\nu}$ with optimal values indicated above. The corresponding densities are shown as gray data points in Fig.~\ref{fig:profiles} ($\lambda=1$, $\ket{\bar{K}_1}$ pane).

The function ${\cal F}(t)$ in Eq.~(\ref{eq:Ht}) has to be chosen such that it satisfies the adiabatic theorem \cite{Messiah_1999, Teufel_2003_Book, Albash_2018_RMP}. In particular, ${\mathcal F}(0)=0$, ${\mathcal F}(T)=1$, and it is at least twice differentiable. Here $T$ is the duration of the adiabatic sweep. In order to provide a specific example, we consider
\beq
	{\mathcal F}(t) = \frac{1}{2}\left[1 - \cos\left(\pi \frac{t}{T}\right) \right].
	\label{eq:Ft}
\eeq
While we focus on a critical system, as we are concerned with experiments with a finite number of atoms, the system will remain gapped (we exploit the rigorous results for the finite size scaling of the gap in Sec. \ref{app:Coherent}). To this end we consider an approximate but qualitatively sufficient picture that the time duration of the adiabatic sweep $T$ should be much larger than the inverse of the spectral gap $1/\Delta_{sg}$ (see e.g. \cite{Messiah_1999, Teufel_2003_Book, Albash_2018_RMP, Avron_1998_PRA, Rigolin_2012_PRA} for discussion of adiabaticity conditions). We define the spectral gap as the energy difference between the lowest and the first excited state of the Hamiltonian
\footnote{This is in contrast to the gap $E_{\rm gap}$ used in the main text in scenario \emph{(ii)}, which was the energy of the lowest excited state, i.e. its distance from the zero energy corresponding to the true ground state of the supersymmetric Hamiltonian on a chain with periodic boundaries. In the thermodynamic limit the lowest excited states become however equidistant as a consequence of the conformal symmetry and $E_{\rm gap}$ coincides with $\Delta_{sg}$.}.

\begin{figure}[h!]
\centering
\includegraphics[scale=0.6]{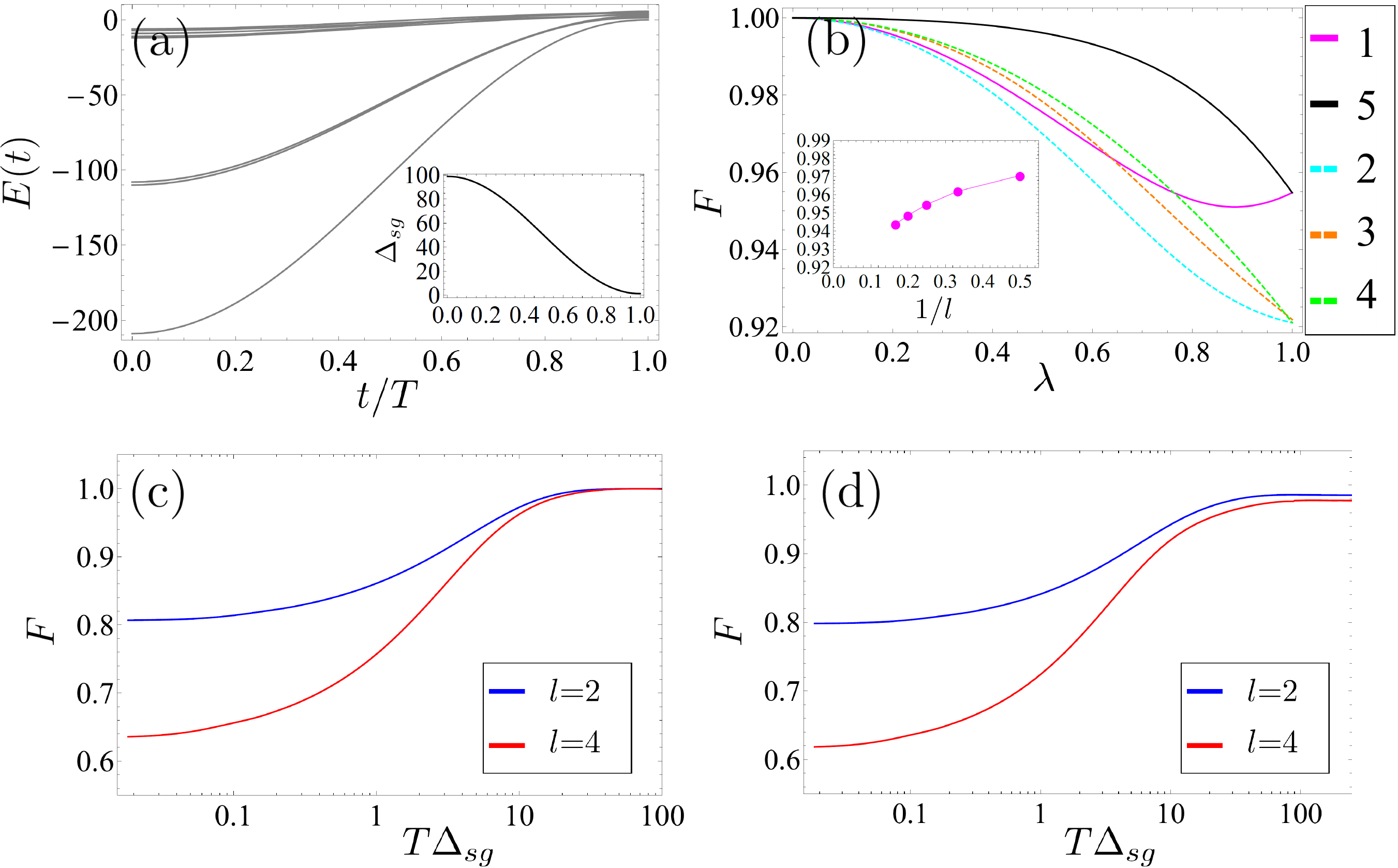}
\caption{Adiabatic state preparation.
{\bf (a)} Example of the spectrum of the Hamiltonian (\ref{eq:Ht}) for case \emph{(i)}, i.e. $L=3l$ with $l=2$, as a function of time. The inset shows the spectral gap.
{\bf (b)} Fidelity $|\braket{K_j'|K_j}|$ of the kink states as a function of $\lambda$ for case \emph{(ii)}, $L=3l+1$ with $l=4$. The solid (dashed) lines are for the boundary, i.e. rightmost and leftmost, (bulk) kinks respectively. Here, $\ket{K_j'}$ are the lowest energy states of the Hamiltonian $H+\mu \sum_{i \in j-{\rm th\;kink}}n_i$, where the $\mu$-term enforces no occupation (for $\mu \rightarrow \infty$) on the sites of $j$-th kink, $i=(3(j-1),3(j-1)+1,3(j-1)+2)$ for $1<j<l+1$ and $i=1,2$ ($i=L-1,L$) for the leftmost (rightmost) kink $\ket{K_1}$ ($\ket{K_{l+1}}$), see also the main text. The inset shows the finite size scaling of the fidelity of the leftmost kink at criticality for $l=2,3,4,5,6$.
{\bf (c)} Fidelity of the preparation of the ground state as a function of the adiabatic sweep length $T$ for $l=2,4$ (blue, red), i.e. case \emph{(i)}, $L=3l$, $\lambda=1$.
{\bf (d)} Fidelity of the preparation of the leftmost kink $\ket{K_1}$ as a function of the adiabatic sweep length $T$ for $l=2,4$ (blue, red), i.e. case \emph{(ii)}, $L=3l+1$, $\lambda=1$. In (c) and (d) the horizontal axes are in terms of $T \Delta_{sg}$, where $\Delta_{sg}=\Delta_{sg}(T)$.
}
\label{fig:state_prep}
\end{figure}

~\\ \noindent
{\emph{Note on the initial Hamiltonian.}} In order to ensure that no level-crossing occurs as the Hamiltonian is swept from $H_i$ to $H_f$, some care has to be taken in the choice of $H_i$. One possibility is to choose $H_i$ such that its ground state is the lowest energy state of $H_Q$ at extreme staggering. For the specific lengths and staggerings considered, the $\lambda=0$ states are particularly simple as they are given by product states of the form (see Table \ref{tab:GS_config} for a summary of low energy spectral properties of the M$_1$ model at extreme staggering)
\begin{subequations}
	\label{eq:psi_lowest}
	\begin{align}
		(i) & \; L=3l, \; 1\lambda 1 \ldots \; \rightarrow \; \ket{\psi_i} = \ket{010010...} \\
		(ii) & \; L=3l+1, \; 11\lambda \ldots \; \rightarrow \; \ket{\psi_i} = \ket{001001...} \\
		(iii) & \; L=3l+1, \; 11\lambda \ldots \; \rightarrow \; \ket{\psi_i} = \ket{101001...}
	\end{align}	
\end{subequations}
We thus take the initial Hamiltonian, which enforces the $l$ particles to be trapped at sites $s=\{2,5,...\}$, $s=\{3,6,...\}$ for cases \emph{(i)}, \emph{(ii)} and $l+1$ particles at sites $s=\{1,3,6,...\}$ for \emph{(iii)}, to be of the form
\beq
	H_i = - \sum_i (\alpha i + \mu_0) n_i - \mu \sum_{i \in s} n_i.
	\label{eq:Hi}
\eeq 
The first summand is meant to lift remaining degeneracies and in principle more complicated functions of the position $i$ can be considered. In practice, as we are mainly concerned with preserving the gap between the lowest energy and the first excited state, the details of this function are not essential as far as $\mu \gg \mu_0, \alpha i$ for all $i$, which is used in the following. The state is prepared as
\beq
	\ket{\psi(t)} = {\cal T} {\rm e}^{-i \int_{0}^{t} {\rm d}t' \, H(t')} \ket{\psi_i},
	\label{eq:time_ev_prep}
\eeq
where $H(t)$ is given by (\ref{eq:Ht}), $\cal{T}$ is the usual time ordering operator and $\ket{\psi_i}$ are given by (\ref{eq:psi_lowest})
\footnote{
We numerically implement the time evolution using Crank-Nicholson discrete time-step evolution given by $\ket{\psi(t+\Delta t)} = (1+ i H(t) \Delta t/2)^{-1} (1 - i H(t) \Delta t/2) \ket{\psi(t)}$ \cite{Press_2007}.
}. \phantom{\cite{Press_2007}} \hspace*{-0.5cm}
We further denote the prepared state at the end of the adiabatic evolution as $\ket{\psi_{\rm prep}} \equiv \ket{\psi(T)}$. 
We then quantify the fidelity of the prepared state as \emph{(i)} $F=|\braket{\psi_{\rm prep}|\psi_0}|$, where $\ket{\psi_0}$ is the ground state of $H_Q$ and \emph{(ii)} $F_j =|\braket{\psi_{\rm prep}|K_j}|$, \emph{(iii)} $\bar{F}_j =|\braket{\psi_{\rm prep}|\bar{K}_j}|$. The results for \emph{(i)} and \emph{(ii)} are shown in Fig. \ref{fig:state_prep}, see the caption for details (the results of \emph{(iii)} not shown are similar to \emph{(ii)}). In summary, the fidelity of preparation $F>0.92$ ($F>0.95$ for the boundary kinks $\ket{K_1},\ket{K_{l+1}}$ and $\bar{F}>0.91$ for the boundary skinks $\ket{\bar{K}_1},\ket{\bar{K}_{l+1}}$) for system sizes we analyzed numerically, i.e. $l \leq 6$.

~\\ \noindent
\emph{Optimization.} Few remarks are in order with respect to the adiabatic procedure considered. A first technical one is that one can optimize the sweep function (\ref{eq:Ft}) which adapts the rate of change to the instantaneous gap $\Delta_{sg}(t)$ evolving at a slower rate for smaller gaps,
which has the potential to significantly reduce the time required to achieve a desired fidelity for a given system size \cite{Albash_2018_RMP}. This is desirable as the final (and minimal) gap $\Delta_{sg}(t=T)$ is decreasing with system size, so that using the same ${\mathcal F}(t)$ for different system sizes will result in larger times in order to achieve the same target fidelity as is illustrated in Fig. \ref{fig:state_prep}. The other remark is qualitative and is about the inadequacy of adiabatic protocols for preparing \emph{critical} states with a vanishing gap in the thermodynamic limit, requiring an infinitely slow sweep akin to the Kibble-Zurek mechanism. While we consider the adiabatic preparation protocol even in this case, we note that other schemes using spatiotemporal quenches have been proposed recently \cite{Agarwal_2018_PRL}. Whether such a scheme can be implemented in our setup goes beyond the scope of the present work and we leave it for future investigations. 

\hspace*{-1cm}

\begin{table}	
	\centering
	\begin{tabular}{|c||c|c|c|}	
	\hline
			 & $1 1 \lambda$ \ldots & $1 \lambda 1 \ldots$ & $\lambda 1 1 \ldots$ \\
	\hline \hline
		$3l-1$ & 1,0,2 & 1,0,1 & 1,0,1 \\
	\hline
		$3l$   & 1,0,1 & 1,0,2 & 1,0,1 \\
	\hline
		$3l+1$ & $l+1$,1,2 & $l+1$, 1,2 & 1,0,1 \\
	\hline
	\end{tabular}
	\caption{Summary of the low energy properties of the M$_1$ model at extreme staggering $\lambda=0$. The three possible staggerings and system sizes are listed in the top row/left column respectively. The three numbers are $({\rm deg}_{\rm lowest},E_{\rm lowest}, \Delta_{sg})=$(degeneracy of the lowest energy manifold, energy of the lowest energy states, spectral gap between the lowest and first excited state).}
	\label{tab:GS_config}
\end{table}

\section{Saddle point approximation}
\label{app:Saddle_point}

\begin{figure}[h!]
\centering
\includegraphics[scale=0.4]{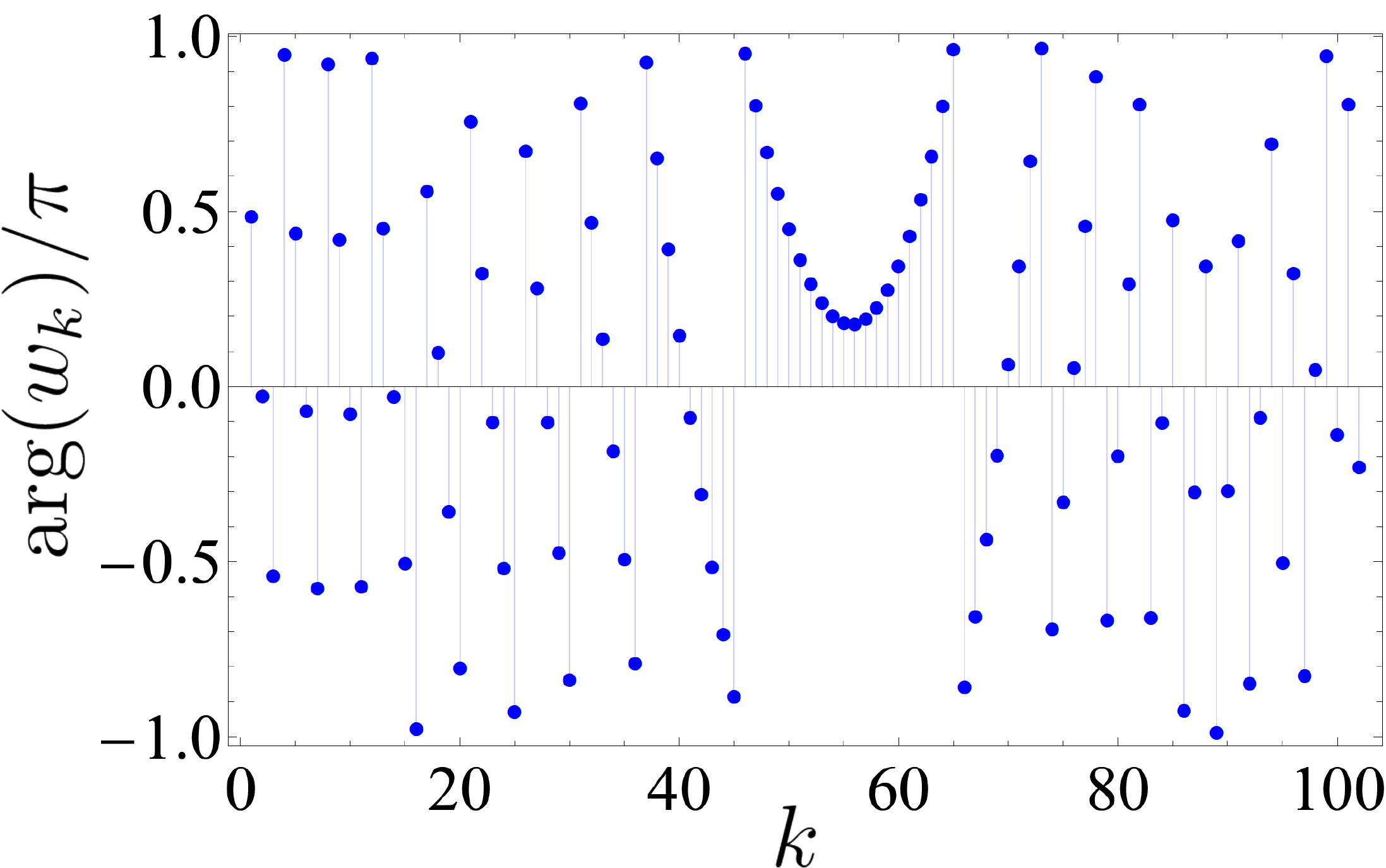}
\caption{Argument of the summands $w_k$ in (\ref{eq:OverlapS}) for $l=101, t=120$ and $s=1$.}
\label{fig:saddle}
\end{figure}

When analyzing the kink dynamics, we have introduced the overlap between the time evolved lefttmost kink and the rightmost kink, which, in the thermodynamic limit of large $l$, where $E_k \rightarrow E(\tilde{k})$, becomes
\beq
	o(t) = \braket{K_{l+1}|K_1(t)} \approx \frac{2}{l+2} \sum_{k=1}^{l+1} \sin \left( \tilde{k} \right) \sin \left( \tilde{k} (l+1) \right) {\rm e}^{-i E(\tilde{k}) t} =: \sum_{k=1}^{l+1} w_k
	\label{eq:OverlapS}
\eeq
with $\tilde{k} = \pi k/(l+2)$. We note that the form of the overlap is reminiscent of the saddle point approximation for a purely imaginary exponent
\beq
	o(t) \sim \int {\rm d}z\,f(z) {\rm e}^{i g(z)} \approx {\rm e}^{i g(z_0) \pm i\frac{\pi}{4}} f(z_0) \sqrt{\frac{\pm 2 \pi}{g''(z_0)}},
	\label{eq:saddle_def}
\eeq
where $g(z) \in {\mathbb R}$, the saddle point $z_0$ satisfies $g'(z_0)=\left. \partial_z g(z)\right|_{z=z_0}=0$ and the signs are chosen such that the term under the square root is positive. Expanding the second sine term in Eq. (\ref{eq:OverlapS}) we obtain
\beq
	\sin \left( \tilde{k} \right) \sin \left( \tilde{k} (l+1) \right) = \cos \left( \pi k \right) \sin \left( \tilde{k} \right)^2 = e^{i(2s-1)\pi k}\sin \left( \tilde{k} \right)^2, \; s \in {\mathbb Z}.
\eeq
Absorbing the phase factor in the exponential term in (\ref{eq:OverlapS}), by comparison with (\ref{eq:saddle_def}) we define
\beq
	g(k)=(2s-1)\pi k - E(k) t.
\eeq
Differentiating with respect to $k$ and imposing the saddle point condition, we can formally write the $s$-dependent solution for the saddle point as
\beq
	k_s(t) = (E')^{-1}\left( \frac{(2s-1) \pi}{t} \right), \; s=1,2,\ldots,
	\label{eq:ks}
\eeq
where the positivity of $s$ follows from the non-negativity of $E'(k)$ in $g'(k)=0$. We thus get the saddle point which needs to be evaluated at each time $t$. The situation for a specific time is depicted in Fig. \ref{fig:saddle}, where we show the argument of the summands in (\ref{eq:OverlapS}) for $l=101$ and $t=120$. We see a clear rapid oscillatory behaviour resulting in the cancellation of terms from most $k$ but the ones in the vicinity of saddle point, which (here for $s=1$) is evaluated from (\ref{eq:ks}) to $k_s = 56$.

Next, we write the condition $g'(k)=0$ for the saddle point as
\beq
	\frac{v_{\rm max} t}{l+2} \tilde{v}(k) - (2s-1)=0,
	\label{eq:Heaviside_cond}
\eeq
where we have used $\partial_k \tilde{k} = \pi/(l+2)$ and the definition of the group velocity $v(\tilde{k})=\partial_{\tilde{k}}E(\tilde{k})$, which we have written as $v(\tilde{k}) = v_{\rm max} \tilde{v}(k)$, so that $0 \leq \tilde{v}(k) \leq 1$. 
It follows from (\ref{eq:Heaviside_cond}), that with increasing $t$, an increasing number of saddle-points $s=1,2,\ldots$ contribute, with the limiting values of $t$ corresponding to the fastest mode, $\tilde{v}(k)=1$. This leads us to the final result, namely Eq. (\ref{eq:Overlap_saddle}) in the main text. 

We note that the expressions can be simplified in particular cases, such as at criticality, $\lambda=1$, where the dispersion takes a simple form $E(\tilde{k})=2 v_F \sin(\tilde{k}/2)$. For the first saddle point ($s=1$), for $x = (l+2) / (v_F t) <1$, the overlap Eq.~(\ref{eq:Overlap_saddle}) evaluates to
\beq
|o(t)| = \frac{2}{\pi} \sin^2 (\tilde{k}_1) \sqrt{4\pi \over v_F t \sin(\tilde{k}_1/2)} = \frac{16}{\sqrt{\pi(l+2)}} (1-x^2)^{3 \over 4} x^{5 \over 2}.
\eeq
It follows that the probability $|o(t)|^2$ (green-dashed curve in Fig.~\ref{fig:3}b) peaks at arrival time
\beq
x_{\rm max}=\sqrt{5 \over 8} \qquad \Rightarrow \quad t_{\rm max}=\sqrt{8 \over 5}{l+2 \over v_F}.
\eeq
Similarly, we get the contribution from the second saddle point ($s=2$, red curve in Fig.~\ref{fig:3}b) by substituting $x=3 (l+2)/(v_F t)$ in the above formula.

\section{Experimental implementation}
\label{app:Implementation}

In order to realize the blockade mechanism, we assume the ground state atoms to be excited off-resonantly with detuning $\Delta$ to the Rydberg state $\ket{r}$ as described by the Hamiltonian (\ref{eq:HRy}). We assume that the sites can be addressed individually such that a ground state $\ket{g}$ of an atom at site $i$ is coupled to $\ket{r}$ with Rabi frequency $\Omega_i$, while the detuning $\Delta$ is kept constant for all atoms. When $\Delta \gg \Omega_i$ for all $i$, one can adiabatically eliminate the many-body Rydberg states by means of Brillouin-Wigner perturbation theory carried out to fourth order in $\Omega/\Delta$ \cite{Henkel_2010_PRL,Pupillo_2010_PRL}, which leads to a so-called flat-top potential. Here we present a useful shortcut derivation for two atoms, which, to the order considered, coincides with the results of the systematic adiabatic elimination and which has been also invoked in the analysis of Rydberg atoms in optical lattices \cite{Macri_2014_PRA}.

\subsection{Shortcut derivation of the dressed atomic potential}
\label{app:Elimination}

The Rydberg Hamiltonian of a system of two atoms located at positions ${\bf r}_i$ and ${\bf r}_j$ with $r=|{\bf r}_j-{\bf r}_i|$ and $i<j$ reads
\beq
	H = \Omega_i \sigma^x_i + \Omega_j \sigma^x_j + \Delta \left( n^r_i + n^r_j \right) + V n^r_i n^r_j.
\eeq
In the basis $\{ \ket{gg}, \ket{gr}, \ket{rg}, \ket{rr} \}$ it can be written as
\beq
	H=
	\begin{pmatrix}
		0 & \Omega_j & \Omega_i & 0 \\
		\Omega_j & \Delta & 0 & \Omega_i \\
		\Omega_i & 0 & \Delta & \Omega_j \\
		0 & \Omega_i & \Omega_j & 2 \Delta + V,
	\end{pmatrix}
\eeq
where $V=C_6/r^6$. The Schr\"{o}dinger equation for the coefficients of the wavefunction is
\beq
	i \dot{{\bf c}} = H {\bf c},
	\label{eq:Schrodinger_two_atom}
\eeq 
where
\beq
	{\bf c} = (c_{gg},c_{gr},c_{rg},c_{rr})^T.
\eeq
Since $\Omega_{i,j} \ll \Delta$, we eliminate the rapidly oscillating components by setting $\dot{c}_{gr}=\dot{c}_{rg}=\dot{c}_{rr} = 0$ in (\ref{eq:Schrodinger_two_atom}). Solving for these three components and substituting in the remaining equation for $c_{gg}$, which is of the form $i \dot{c}_{gg}=W'(r) c_{gg}$, yields the effective potential
\beq
	W'(r) = -\frac{2 \Omega_i \Omega_j (2 \Delta +V(r))}{\Delta (2 \Delta +V(r))-2 \Omega_i \Omega_j}.
	\label{eq:Wprime}
\eeq
Expanding in $\Omega$ and subtracting a global offset $-2\Omega_i \Omega_j/\Delta$ we obtain the effective potential between the two ground state atoms
\beq
	W(r) = \frac{2 (\Omega_i \Omega_j)^2 V(r)}{\Delta^3 (2\Delta +V(r) )} + O(\Omega^6),
	\label{eq:W}
\eeq
with amplitude $W(0)-W(\infty)=2(\Omega_i \Omega_j)^4/\Delta^3$, which sets the maximum available energy for realizing the blockade. In practice, one needs to accommodate the lattice spacing $r_0$ so that the blockade energy is $W(r_0)$ and the energy scale of the Hamiltonian is $W(2 r_0)$. We note that the chemical potentials $\mu_i$ of the ground state atoms are not affected by the dressing.
For future convenience let us parametrize the dressed interaction between atoms at site $i$ and $i+n$ by writing the the Eq.~(\ref{eq:W}) as
\beq
	W_i(n r_0) = \frac{2 (\Omega_i \Omega_{i+n})^2 V(n r_0)}{\Delta^3 (2\Delta +V(n r_0) )},
	\label{eq:Wi}
\eeq
where we have ommited the higher order contributions $O(\Omega^6)$.

\subsection{Implementing the M$_1$ model off criticality}
\label{app:OffCritical}

\begin{figure}[h!]
\centering
\includegraphics[scale=0.5]{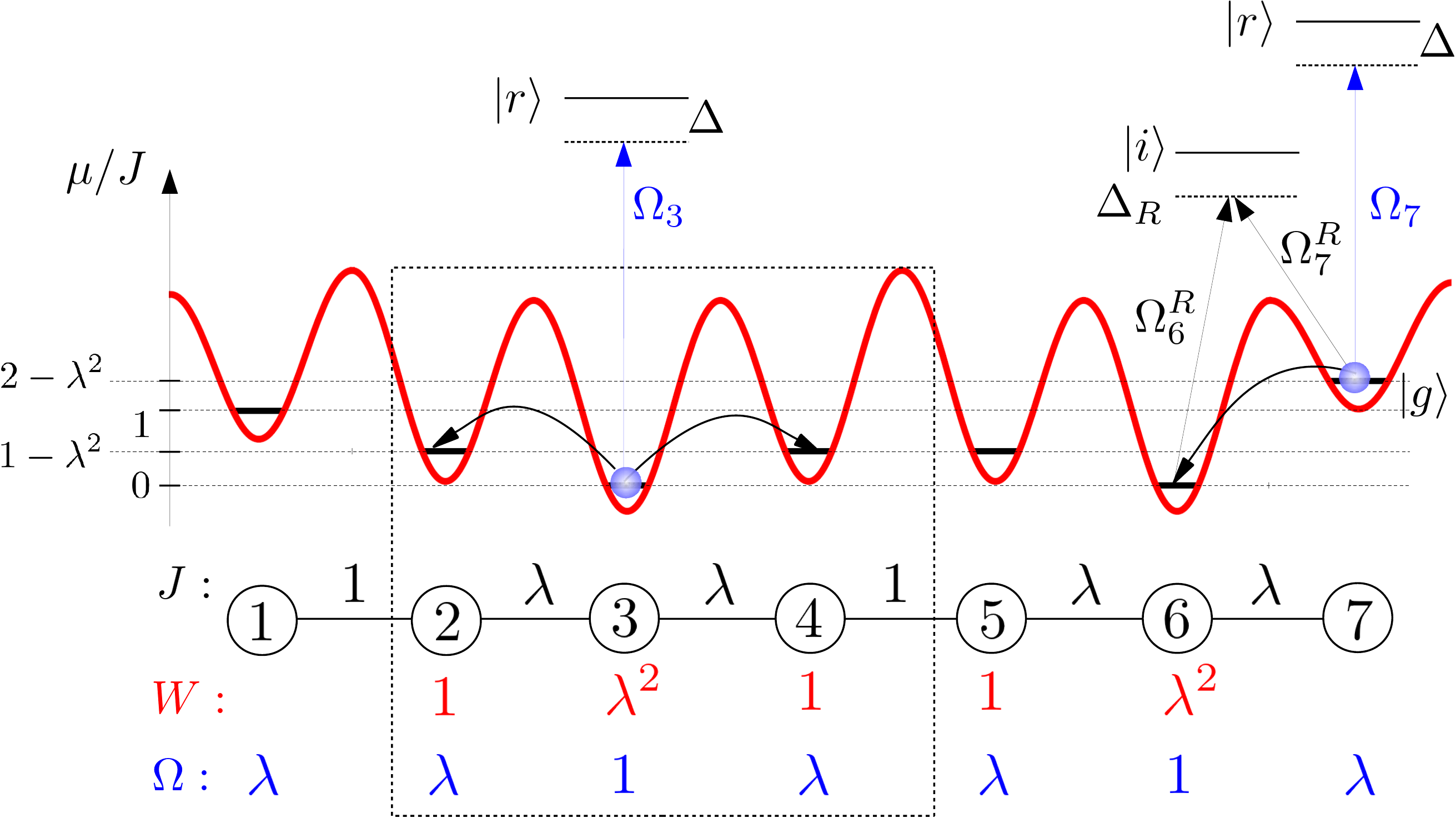}
\caption{Scheme of the possible experimental implementation of the M$_1$ model for a generic $\lambda$. The pattern of tunneling amplitudes $J$, next-nearest neighbour interactions $W$ and the corresponding on-site Rydberg laser Rabi frequencies $\Omega$ is given at the bottom in black, red, blue respectively. The necessary periodicity can be achieved with a bi-chromatic optical lattice with ratio 1/3 between the lattice wavelengths. In order to realize the required tunneling amplitudes, one might use Raman assisted hoppings with Raman Rabi frequencies $\Omega^R_i$ and detunings $\Delta_R$ via an intermediate state $\ket{i}$. The dashed rectangle denotes the basic building block which, when repeated, constitutes the whole chain (with the appropriate boundary chemical potentials as indicated).}
\label{fig:scheme_off}
\end{figure}

Here we discuss how to implement the M$_1$ model off criticality, $0 \leq \lambda <1$. It is instructive to write $H_Q$ explicitly. For $l=2$, $L=3l+1$ and staggering pattern $11\lambda \ldots$ it reads 
\footnote{
Here it becomes apparent why we have introduced the site-dependent factor $(-1)^i$ in the definition of the supercharge $Q = \sum_i (-1)^i \lambda_i c^\dag_i P_{\braket{i}}$ - this is required, for $\lambda_i$ real, for the kinetic term to be \emph{negative}. Another consequence of this choice is the appearance of triplets in the ground state $\ket{\rm I}$. Conversely, the omission of the $(-1)^i$ factor leads to positive kinetic term and singlets instead of triplets in $\ket{{\rm I}}$}
\beqa
	H_Q &=& -\sum_{i=1}^{L-2} {\tilde{J}}_{i,i+1} P_{i-1} \left(c_i^\dag c_{i+1} + {\rm H.c.}\right)P_{i+2} + P_2 + \sum_{i=1}^{L-2} {\tilde{W}}_{i+1} P_{i}P_{i+2} + P_{L-1},\nonumber \\	
	J H_Q &=& -\sum_{i=1}^{L-2} J_{i,i+1} P_{i-1} \left(c_i^\dag c_{i+1} + {\rm H.c.}\right)P_{i+2} + \sum_{i=1}^L \mu_i n_i + \sum_{i=1}^{L-2} W_{i}(2 r_0) n_{i}n_{i+2}.
	\label{eq:HQ_dimful}
\eeqa
In the first line, $H_Q$ is expressed in dimensionless units with $\vec{\tilde{J}}=(1,\lambda,\lambda,1,\lambda,\lambda)$, $\vec{\tilde{W}}=(1,\lambda^2,1,1,\lambda^2)$. In the last line, we have expanded the projectors $P_i=(1-n_i)$ (except in the kinetic term) and restored the dimensions to make connection with the physical Rydberg Hamiltonian, $\vec{J}=J \vec{\tilde{J}}$, $\vec{W}=J \vec{\tilde{W}}$, where $J=W(2 r_0)$. Omitting a constant factor, we have for the chemical potentials $\vec{\mu}=-J(1,1+\lambda^2,2,1+\lambda^2,1+\lambda^2,2,\lambda^2)$. 
The situation is summarized in Fig. \ref{fig:scheme_off}, see the caption for details. We thus have a repeated pattern of period three (starting at the second site for $L=3l+1$) of tunneling amplitudes, next-nearest neighbour interaction potentials and chemical potentials, denoted by the black dashed box in Fig. \ref{fig:scheme_off}. We now comment on the details of implementation related to each of these three types of Hamiltonian contributions.
~\\
~\\
\noindent
\emph{Chemical potentials.} The chemical potentials can be realized by a bichromatic optical lattice with the two lattice wave vectors having ratio of 1/3 as depicted in Fig. \ref{fig:scheme_off}. We note that, due to the boundary conditions, the chemical potentials $\mu_1, \mu_L$ on the first and last site get an extra offset which can be realized by for instance additional optical fields.
~\\
~\\
\noindent
\emph{Interaction potential.} It is straightforward to show from Eq.~(\ref{eq:W}), that the off-critical potential pattern $\vec{W}$ can be realized by the pattern of on-site Rabi frequencies $\vec{\Omega} \propto (\lambda, \lambda,1,\lambda,\lambda,1,\lambda)$. In principle, the atoms in the ground and the Rydberg state will experience different polarizability leading to a different AC Stark shift originating both in the driving Rabi field $\Omega$ and the optical lattice potential. Since $\Omega \gg J$, the leading contribution to the AC Stark shift will be from the Rabi frequencies and is proportional to $\Omega^2/\Delta \ll \Delta$ so that it has been neglected in the derivation of Eq. (\ref{eq:W}) assuming identical detunings $\Delta$ for all lattice sites.
~\\
~\\
\noindent
\emph{Tunneling amplitudes.} Similarly to the chemical and interaction potentials, one needs to tune the tunneling amplitudes, which can be achieved in principle by means of Raman assisted hoppings \cite{Jaksch_2003_NJP,Aidelsburger_2011_PRL,Miyake_2013_PRL}, with $J_{i,i+1} \propto \Omega^R_i \Omega^R_{i+1}/\Delta^R$, where $\Omega^R_i$ and $\Delta^R$ label the on-site (single-photon) Raman Rabi frequencies and detunings, respectively, see Fig. \ref{fig:scheme_off}. These additional laser beams will also contribute to the ground and Rydberg state polarizabilities, however, as $J \ll \Delta$, they will contribute only a subleading correction to the dressed potential as discussed in the previous paragraph.

\subsubsection{Effect of the interaction tails off-criticality}
\label{app:Tails}

\begin{figure}[h!]
\centering
\includegraphics[scale=0.7]{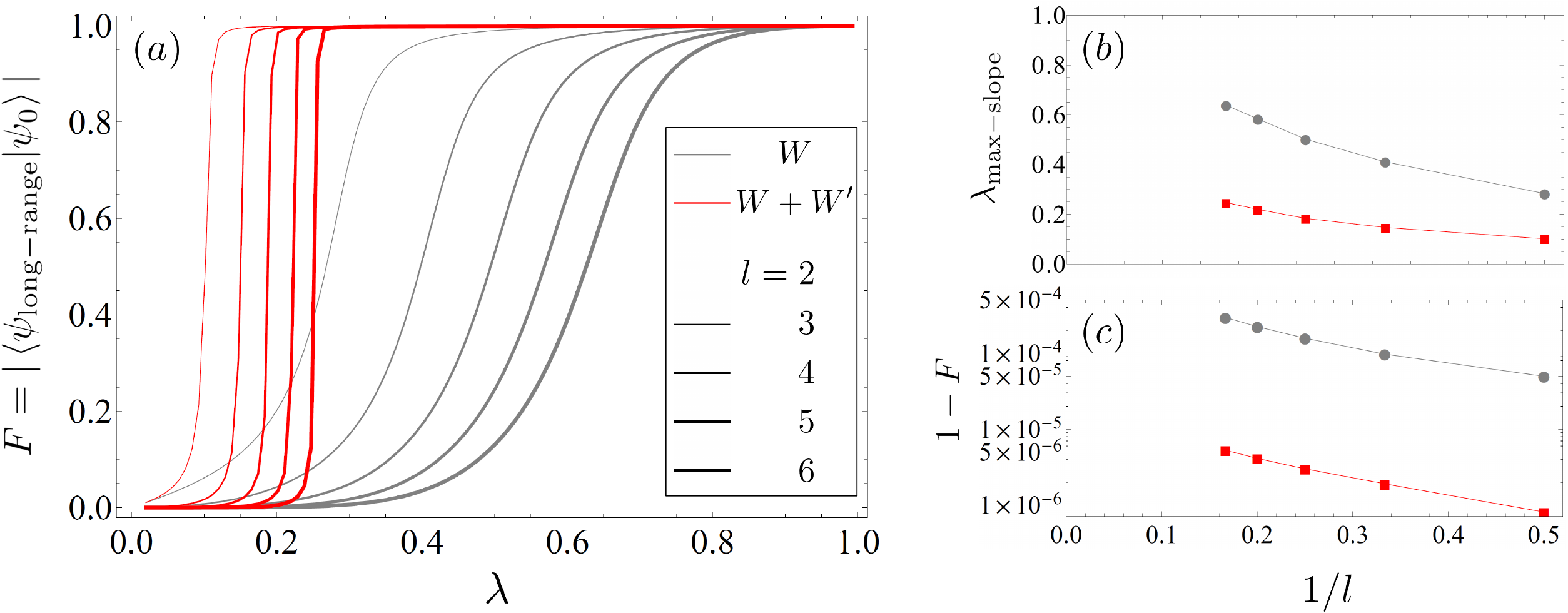}
\caption{
{\bf (a)} Fidelity (\ref{eq:F_longrange}) vs. $\lambda$ for the single and double dressing schemes (gray and red lines respectively) corresponding to the dressed potentials $W$ and $W_{\rm tot} = W + W'$, see Sec.~\ref{app:Double_dressing}. The line thickness indicates varying system size from $l=2$ (thin lines) to $l=6$ (thick lines).
{\bf (b)} Finite-size scaling of $\lambda_{\rm max-slope}$, Eq.~(\ref{eq:lambda_maxslope}), corresponding to the largest gradient of the fidelity in (a) for the single (gray) and double (red) dressing scheme.
{\bf (c)} Finite-size scaling of the error $1-F$ at criticality ($\lambda=1$) for the single (gray) and double (red) dressing scheme.
}
\label{fig:F_off}
\end{figure}

It is instructive to consider the effect of the dressed interaction beyond next-to-nearest neighbours. Using the pattern of the on-site Rabi frequencies $\Omega_i$ shown in Fig.~\ref{fig:scheme_off} and the Eq.~(\ref{eq:Wi}) we get for the patterns of the next-to and the next-to-next-to nearest neighbours
\beqa
	\vec{\Omega} &=& \Omega (\lambda, \lambda, 1, \lambda, \lambda, 1, \ldots) \nonumber \\
	& \Rightarrow & \nonumber \\
	\vec{W}(2 r_0) &=& (W_1(2 r_0), W_2(2 r_0), W_3(2 r_0), W_4(2 r_0), \ldots ) \propto \Omega^4 (\lambda^2, \lambda^4, \lambda^2, \lambda^2, \ldots) \nonumber \\
	\vec{W}(3 r_0) &=& (W_1(3 r_0), W_2(3 r_0), W_3(3 r_0), \ldots ) \propto \Omega^4 (\lambda^4, \lambda^4, 1, \ldots)
\eeqa
It is thus apparent that for sufficiently small $\lambda$, the neglected interactions beyond next-to-nearest neighbours $W(3 r_0)$ will become dominant (the ``1'' terms in $\vec{W}(3 r_0)$) and cannot be neglected anymore. We would like to emphasize that this observation does not affect the discussion of the kink dynamics at the critical point $\lambda=1$ in the main text, but it clearly limits the exploration of the off-critical regime. To quantify the effect of the long-range interactions we consider system sizes $L=3 l$ featuring a unique ground state, cf. the Table~\ref{tab:GS_config}, and the dressed Hamiltonian (\ref{eq:HQ_dimful}) which now includes all the terms beyond $W(2 r_0)$
\beq
	J H_Q^{\rm long-range} = J H_Q + \sum_{n=3}^{L-1} \sum_{i=1}^{L-n} W_{i}(n r_0) n_{i}n_{i+n}
	\label{eq:HQ_long_range}
\eeq
Let us denote by $\ket{\psi_{\rm long-range}}$, $\ket{\psi_0}$ the ground states of (\ref{eq:HQ_long_range}) and (\ref{eq:HQ_dimful}), $\ket{\psi_0}$ corresponding to the notation used in Sec.~\ref{app:Preparation}. In Fig. \ref{fig:F_off}a we show the fidelity 
\beq
	F = |\braket{\psi_{\rm long-range} | \psi_0}|
	\label{eq:F_longrange}
\eeq
vs. $\lambda$ for various system sizes (gray lines). It is apparent that the inclusion of the long-range tails limits the exploration of the off-critical regime to $\lambda$ asymptotically approaching 1 in the limit of infinite system sizes. Panes (b,c) of Fig.~\ref{fig:F_off} then show the finite-size scaling of the value of $\lambda$ corresponding to maximum slope (largest gradient) of the fidelity
\beq
	\lambda_{\rm max-slope} = \max\limits_{\lambda} \left(  \partial_\lambda F(\lambda) \right),
	\label{eq:lambda_maxslope}	
\eeq
pane (b), and the error $1-F$ in the ground state fidelity at criticality, pane (c) (gray lines). 

In the following section we discuss how to significantly suppress the unwanted effect of the tails of the interaction using doubly-dressed Rydberg potential. This not only allows to extend the region of high fidelity ground states to smaller $\lambda$ and larger system sizes but also achieves tenfold improvement in the $W(2 r_0)/W(3 r_0)$ ratio over the single dressing scheme.

\subsection{Improving the scheme using double dressing}
\label{app:Double_dressing}

The idea behind the improvement is to suppress the long-range tail $W_i(n r_0)$, $n>2$, of the dressed potential (\ref{eq:Wi}) by a second potential with asymptotically the same behaviour in the long separation limit, but with opposite sign. Specifically, from (\ref{eq:Wi}) we have
\beq
	\left. W_i(n r_0) \right|_{n \rightarrow \infty} = \frac{(\Omega_i \Omega_{i+n})^2}{\Delta^4} V(n r_0).
\eeq
With a slight abuse of notation, let us denote the quantities corresponding to the second potential with a prime (not to be confused with (\ref{eq:Wprime})). We thus require
\beq
	\left. W_i(n r_0) \right|_{n \rightarrow \infty} = - \left. W'_i(n r_0) \right|_{n \rightarrow \infty} \;\; \Rightarrow \;\; \Omega' = \Omega \left| \frac{\Delta'}{\Delta} \right| \left| \frac{C_6}{C_6'} \right|^\frac{1}{4},
	\label{eq:Omega0_cond}
\eeq
where $\Omega^{(')}$ is the amplitude of the pattern of the on-site Rabi frequencies, $\vec{\Omega}^{(')} = \Omega^{(')} (\lambda, \lambda, 1, \ldots)$, and we recall that the bare Rydberg interaction reads $V^{(')}(n r_0) = C^{(')}_6/(n r_0)^6$ and $C_6^{'} < 0$. The Eq.~(\ref{eq:Omega0_cond}) thus represents a condition for the amplitude of the second dressing Rabi frequency with other parameters given. We also note that in order to avoid the resonance $\Delta' + V'(r)=0$, which leads to the vanishing denominator in (\ref{eq:Wi}) and can be exploited for instance to realize a spin model featuring both attractive and repulsive interaction \cite{Lan_2015_PRL}, we take $\Delta' < 0$.

\begin{figure}[h!]
\centering
\includegraphics[scale=0.5]{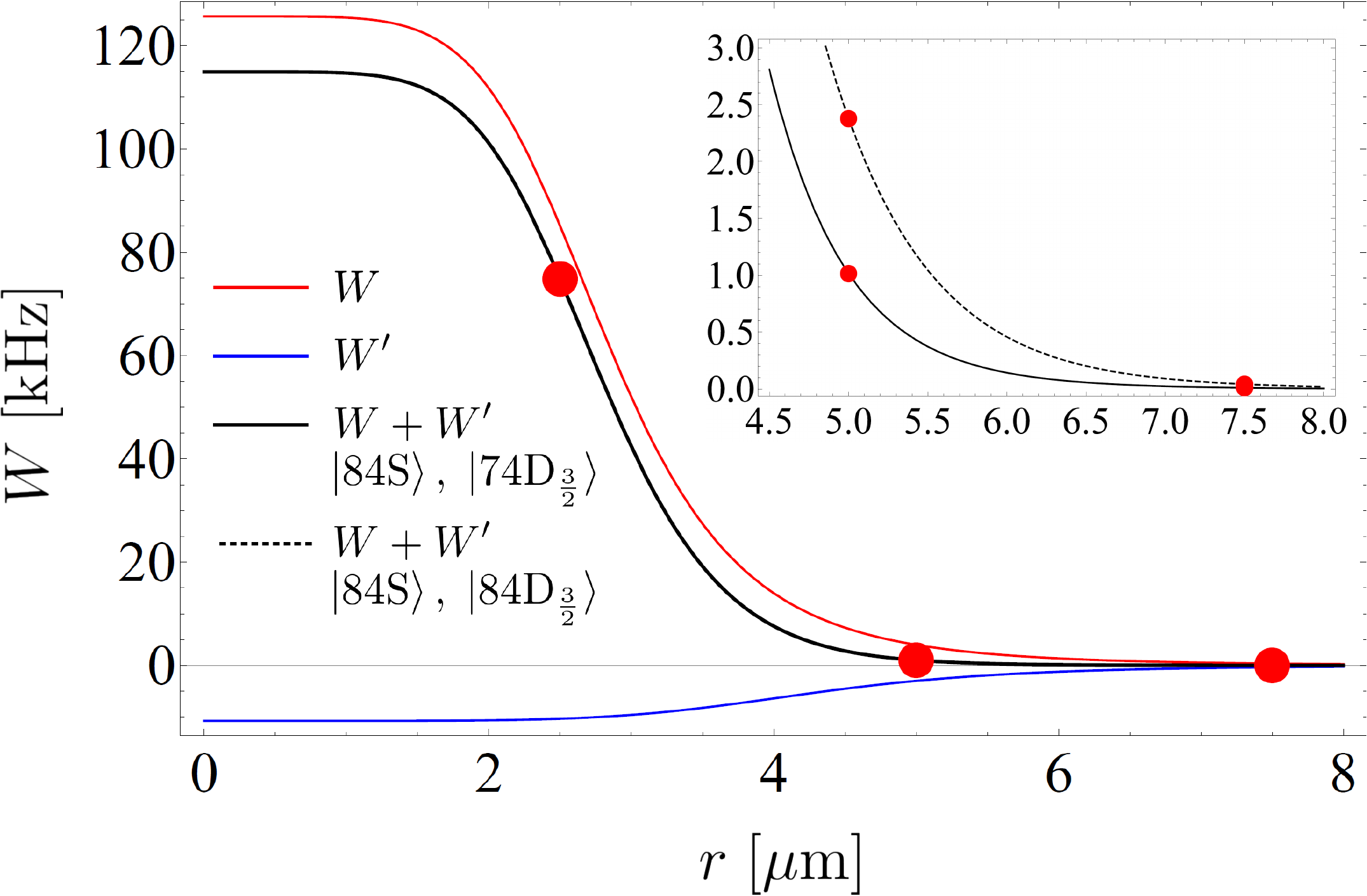}
\caption{
The doubly dressed potential $W_{\rm tot}=W + W'$ (solid black line) composed of the repulsive $W$ (solid red line) and attractive $W'$ (solid blue line) as a function of the distance $r$. The red circles indicate the potential at the positions of the atoms spaced by $r_0=2.5 \, \mu{\rm m}$. 
Parameters used for $W$ are that of Fig.~\ref{fig:scheme}, i.e. $\Omega = 2\pi \times 10\,{\rm MHz}$, $\Omega/\Delta=1/10$, $C_6=645\,{\rm GHz}\cdot \mu{\rm m}^6$ corresponding to the state $\ket{84{\rm}{\rm S}}$ of ${}^6{\rm Li}$. 
Parameters used for $W'$ are $\Delta' = -500 \,{\rm MHz}$, $C_6'=-6\,005\,{\rm GHz}\cdot \mu{\rm m}^6$ for $\ket{74{\rm D}_\frac{3}{2}}$ with $\Omega' = 2\pi \times 4.5 \,{\rm MHz}$ given by the Eq.~(\ref{eq:Omega0_cond}).
The inset shows, in addition to $W_{\rm tot}$ of the main plot (solid black line), $W_{\rm tot}$ corresponding to taking $\ket{84{\rm D}_\frac{3}{2}}$ instead of $\ket{74{\rm D}_\frac{3}{2}}$ for the second Rydberg state with parameters $C_6'=-24\,200\,{\rm GHz}\cdot \mu{\rm m}^6$ yielding $\Omega' = 2\pi \times 3.2 \,{\rm MHz}$ (dashed black line).
}
\label{fig:FigS_potential}
\end{figure}

The considered construction is allowed by an appropriate choice of the additional Rydberg state which features \emph{attractive} rather than repulsive interaction. Specifically, in addition to the $\ket{{\rm nS}}$ state of ${}^6{\rm Li}$ which has $C_6 >0$ we choose the $\ket{{\rm n D}_{\frac{3}{2}}}$  state with $C_6'<0$. An example of $W$, $W'$ and $W+W'$ is shown in Fig.~\ref{fig:FigS_potential} as solid red, blue and black lines respectively. Here we have assumed $\lambda=1$, i.e. $\Omega_i = \Omega \; \forall i$ for the sake of simple illustration and considered the Rydberg level $\ket{74 {\rm D}_\frac{3}{2}}$ with $C_6' = -6\,005 \, {\rm GHz} \cdot \mu{\rm m}^6$ \cite{Sibalic_CompPhysComm_2017,ARC_package} in addition to the $\ket{84 {\rm S}}$ one. Clearly, the details of the resulting potential depend on the precise choice of the parameters in the now extended parameter space spanned by $\Omega, \Delta, C_6, \Delta', C_6'$ (with $\Omega'$ given by (\ref{eq:Omega0_cond})). The detailed exploration of this parameter space goes beyond the scope of the present work, however we note that the resulting choice is typically a compromise between maximizing the energy scale $W_{\rm tot}(2 r_0)$ and maximizing the ratios $W_{\rm tot}(r_0)/W_{\rm tot}(2 r_0)$ and $W_{\rm tot}(2 r_0)/W_{\rm tot}(3 r_0)$ which characterize the blockade and the suppression of the long-range tails respectively. Here $W_{\rm tot} = W + W'$ is the total dressed potential. To illustrate this, the inset of Fig.~\ref{fig:FigS_potential} shows the detail of the potential for the next-to and next-to-next-to nearest neighbours for $\ket{84 {\rm D}_\frac{3}{2}}$ with $C_6' = -24\,200 \, {\rm GHz} \cdot \mu{\rm m}^6$ \cite{Sibalic_CompPhysComm_2017,ARC_package} (dashed line) in addition to the potential stemming from the $\ket{74 {\rm D}_\frac{3}{2}}$ state (solid line). We get for these two situations
\beqa
	\ket{84 {\rm S}},\ket{74 {\rm D}_\frac{3}{2}}: \; (W_{\rm tot}(2 r_0) = J, W_{\rm tot}(r_0)/W_{\rm tot}(2 r_0), W_{\rm tot}(2 r_0)/W_{\rm tot}(3 r_0)) &\approx& (1 \, {\rm kHz},74,94) \nonumber \\
	\ket{84 {\rm S}},\ket{84 {\rm D}_\frac{3}{2}}: \; (W_{\rm tot}(2 r_0) = J, W_{\rm tot}(r_0)/W_{\rm tot}(2 r_0), W_{\rm tot}(2 r_0)/W_{\rm tot}(3 r_0)) &\approx& (2.4 \, {\rm kHz},35,56).
\eeqa
which illustrate the point discussed, i.e. the increase of the effective Hamiltonian energy scale $W_{\rm tot}(2 r_0)$ at the expense of reducing the quality of the blockade and the suppression of the interaction tails. In particular the values $(W_{\rm tot}(r_0)/W_{\rm tot}(2 r_0), W_{\rm tot}(2 r_0)/W_{\rm tot}(3 r_0))$ should be contrasted with the values 21 and 11 respectively of the single dressing scheme. We thus see that both of these ratios can be enhanced by a factor 5-10 with the double dressing scheme.

Two minor comments are in order. Firstly, as $\Omega'$ is constrained by the Eq.~(\ref{eq:Omega0_cond}) one has to check whether it still satisfies $\Omega' \ll |\Delta'|$ required for the perturbative description to hold. Since we have chosen $|\Delta'| \approx |\Delta|$ and we have $|C_6'| > |C_6|$, it follows from (\ref{eq:Omega0_cond}) that it is indeed the case (specifically, for the chosen value $\Delta' = -500 \, {\rm MHz}$ we get $|\Omega'/\Delta'| \approx 1/17.5$ with $\Omega' \approx 2\pi \times 4.5 \, {\rm MHz}$ for $\ket{74 {\rm D}_\frac{3}{2}}$ and $|\Omega'/\Delta'| \approx 1/24.5$ with $\Omega' \approx 2\pi \times 3.2 \, {\rm MHz}$ for $\ket{84 {\rm D}_\frac{3}{2}}$). Secondly, one should also verify whether the separation between adjacent Rydberg levels is larger than the considered detunings so that one still selectively addresses the target Rydberg state. For the range $n \sim 70-80$ of the principal quantum numbers, the typical separation between adjacent $\ket{{\rm nS}}-\ket{{\rm (n+1)S}}$ and $\ket{{\rm nD}}-\ket{{\rm (n+1)D}}$ states is of the order of 10 GHz which is well above the values of the considered detunings $|\Delta|, |\Delta'| \sim 0.5 \, {\rm GHz}$.

Finally, let us go back to the discussion of Sec.~\ref{app:Tails} about the effect of the long-range tails of the interactions as quantified by the ground state fidelity of the supersymmetric Hamiltonian (\ref{eq:HQ_dimful}). The fidelity, $\lambda_{\rm max-slope}$ and the ground state error $1-F$ at criticality corresponding to the doubly-dressed potential are shown in red in Fig.~\ref{fig:F_off}. We see that the double dressing scheme significantly improves over the single dressing one reducing $\lambda_{\rm max-slope}$ by a factor of $\gtrsim 3$ and the error at criticality by two orders of magnitude for the same system size.

\subsection{Arbitrary number of dressing potentials}
\label{app:Arbitrary_dressing}

In this section we generalize the above considerations and address the following question: Provided an arbitrary number of dressing potentials $W$ of the form (\ref{eq:W}) is available, is it possible to generate a total potential $W_{\rm tot} = \sum_{\{W\}} W$ which matches the target potential implementing the supersymmetric model, i.e. satisfying (we again focus on the critical case $\lambda=1$ for simplicity)
\begin{subequations}
	\label{eq:pot_conditions}
	\begin{align}
		\frac{W_{\rm tot}(r_0)}{W_{\rm tot}(2 r_0)} & \rightarrow \infty \;\;\; \text{(perfect blockade)}\\
		\frac{W_{\rm tot}(2 r_0)}{W_{\rm tot}(n r_0)} & \rightarrow \infty \;\; \text{for} \; n > 2 \;\;\; \text{(no interaction tails)} \label{eq:pot_conditions_b}
	\end{align}
\end{subequations}
with $W(2 r_0)$ the relevant energy scale? To this end we first write the $j$-th potential (\ref{eq:W}) in a customary form as
\begin{subequations}
	\begin{align}
		W(r,\rho_j) &= A_j \frac{1}{(\rho_j r)^6 + 1} \\
		A_j &= \frac{2 \left. \Omega^{(j)} \right.^4}{\left. \Delta^{(j)} \right.^3} \\
		\rho_j &= \left( \frac{2 \Delta^{(j)}}{C^{(j)}_6} \right)^\frac{1}{6},
	\end{align}
\end{subequations}
where $A_j$ is the potential amplitude and $\rho_j$ the characteristic inverse radius. The functions $W(r,\rho_j)$ constitute a set on which we wish to decompose our target potential $W_{\rm target}$. If we allow for an infinite such set with smoothly varying $A$ and $\rho$, we can formulate this requirement as
\beq
	W_{\rm target}(r) = \int_0^\infty {\rm d}\rho \, W(r,\rho) A(\rho).
	\label{eq:Fredholm}
\eeq
This is nothing but the inhomogeneous Fredholm equation of the first kind with kernel $W(r,\rho)$ and an unknown function $A(\rho)$. In order to proceed, we need to specify the target potential $W_{\rm target}$, which has to be chosen to satisfy the conditions (\ref{eq:pot_conditions}). Motivated by the fact that at long distances the dressed potentials decay as $r^{-6}$, we consider a function
\beq
	\frac{W_{\rm target}(n r_0)}{W_{\rm target}(2 r_0)} = 
	\begin{cases}
		s \;\;\; \text{for} \;\;\; n=1 \\
		1 \;\;\; \text{for} \;\;\; n=2 \\
		\frac{1}{s} \;\;\; \text{for} \;\;\; n=3 \\
		\frac{1}{s} \left( \frac{3}{n} \right)^6 \;\;\; \text{for} \;\;\; n>3,
	\end{cases}
	\label{eq:Wtarget}
\eeq
where $s$ is the suppression factor such that $W_{\rm target}(r_0)/W_{\rm target}(2 r_0) = W_{\rm target}(2 r_0)/W_{\rm target}(3 r_0) = s$ and\\ $W_{\rm target}(n r_0)/W_{\rm target}((n+1) r_0) = (n+1)^6/n^6$. One obtains the ideal supersymmetric potential in the limit $s \rightarrow \infty$.

Next, we need to determine the potential amplitudes $A(\rho)$. To proceed, we consider a discrete set of separations $\{ r_i \}$ and inverse characteristic radii $\{ \rho_j \}$ such that we get from the Fredholm equation (\ref{eq:Fredholm})
\beq
	W_{\rm target}(r_i) = \sum_j W(r_i, \rho_j) A(\rho_j).
	\label{eq:Fredholm_dicrete}
\eeq
The equation (\ref{eq:Fredholm_dicrete}) is thus a matrix equation for the amplitudes $A$ which can be inverted using \emph{pseudoinverse} $W^+$ as
\beq
	A = W^+ W_{\rm target}.
\eeq
Here we consider the pseudoinverse since we are dealing in general with a rectangular rather than a square matrix $W(r_i,\rho_j)$. In fact, even when dealing with a square matrix, it might not be invertible (and in general is not). Since we use a finite set $\{ r_i \}$ in (\ref{eq:Fredholm_dicrete}), the resulting total potential
\beq
	W_{\rm tot}(r) = \sum_j W(r,\rho_j) A(\rho_j)
	\label{eq:Wtot_Fredholm}
\eeq
will differ from $W_{\rm target}$ for a general $r \not\in \{r_i\}$.

\begin{figure}[h!]
\centering
\includegraphics[scale=0.6]{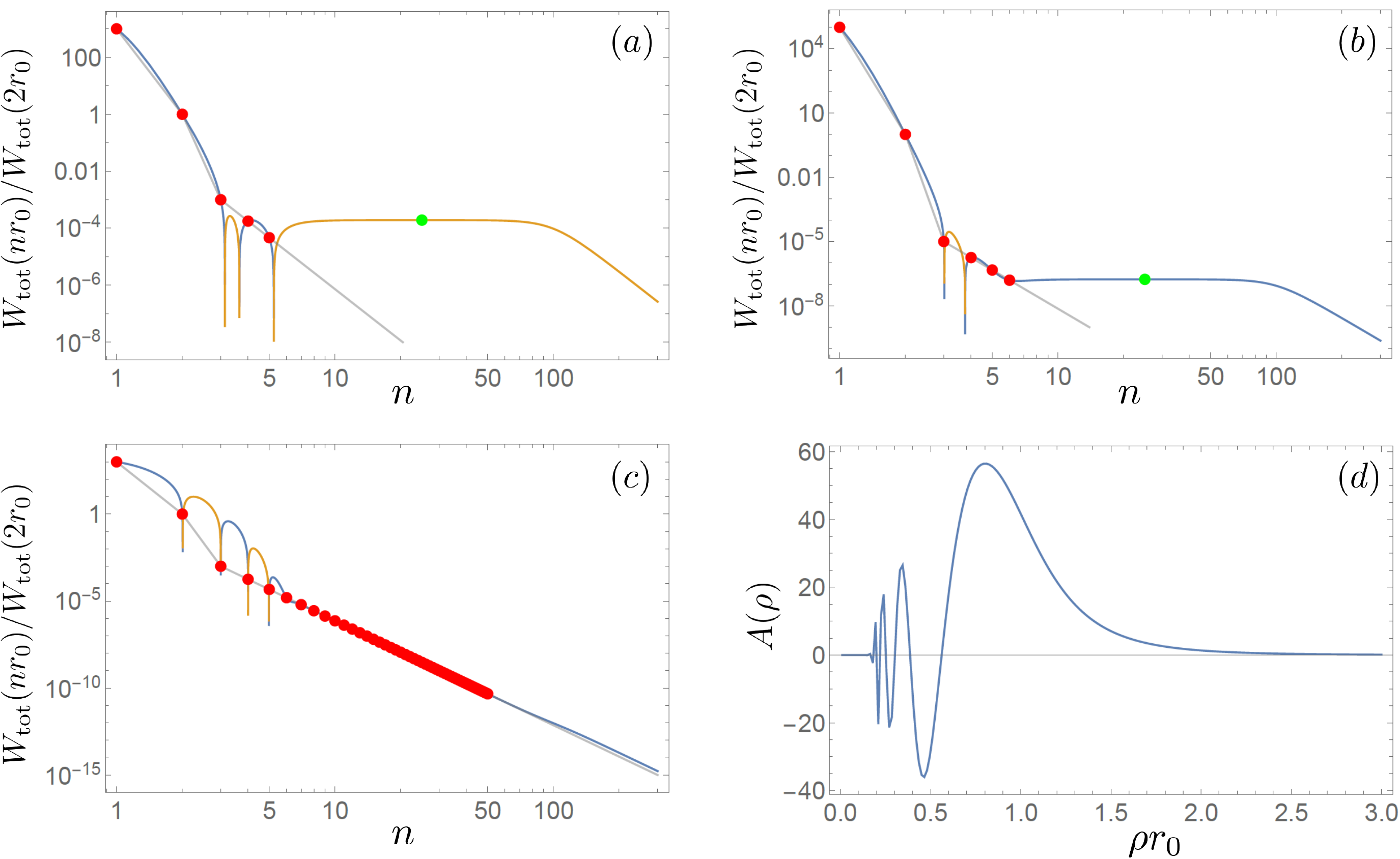}
\caption{
The total potential $W_{\rm tot}$ (blue and yellow solid lines in (a-c)), Eq.~(\ref{eq:Wtot_Fredholm}), obtained from the discrete version of the Fredholm equation (\ref{eq:Fredholm_dicrete}) for suppression factors $s=10^3$ (a,c) and $s=10^5$ (b). The red circles indicate the set $\{ r_i \}$ of atomic distances used in (\ref{eq:Fredholm_dicrete}). In (a,b) [(c)] we use 8 [200] equally spaced inverse radii $\rho_j$ from the interval $\rho_j r_0 \in [0.01,2]$ [$\rho_j r_0 \in [0.01,3]$]. Since we use the logarithmic plot, the blue (yellow) colour indicate the positive (negative) part of $W_{\rm tot}$. One can see from (a-c) that the total potential matches the specified points $\{ r_i \}$ but deviates from $W_{\rm target}$, Eq.~(\ref{eq:Wtarget}) (solid gray line), for $r \not\in \{ r_i \}$. In particular, one has to verify that the condition (\ref{eq:pot_conditions_b}) is satisfied for large $r$. In (a), (b) we get $W_{\rm tot}(2 r_0)/W_{\rm tot}(25 r_0) \approx -5200$ and $6 \cdot 10^6$ respectively, where the reference point $n=25$ is indicated by the green circle. Furthermore, in (d) we show the form of the amplitude function $A(\rho_j)$ for the parameters used in (c).
}
\label{fig:arbitrary_dressing}
\end{figure}

It would be desirable to investigate the mathematical properties of (\ref{eq:Fredholm}),(\ref{eq:Fredholm_dicrete}) in detail. For the purpose of demonstrating that such approach can yield arbitrary suppression $s$, yielding the ideal potential satisfying Eqs.~(\ref{eq:pot_conditions}), we consider specific examples illustrated in Fig.~\ref{fig:arbitrary_dressing}. We refer the reader to the caption for all relevant details.

The results presented in the Fig.~\ref{fig:arbitrary_dressing} indicate that arbitrary suppression $s$, and thus the ideal supersymmetric potential corresponding to the next-to-nearest neighbour interaction (\ref{eq:pot_conditions}) can be achieved \emph{in principle}, provided a sufficient number of suitable dressing potentials is available. To assess an experimental feasibility of such approach would however require a detailed study of available van der Waals couplings provided by the Rydberg levels and realistic Rabi frequencies and detunings. More involved dressing schemes would also increase the decoherence rates as we discuss in the next section. While such a detailed study of experimental feasibility goes beyond the scope of the present work, the possibility of engineering step-like potentials through multiple dressings remains an interesting opening which it might be interesting to address in the future.




\subsection{Coherent evolution}
\label{app:Coherent}
\begin{figure}[h!]
\centering
\includegraphics[scale=0.6]{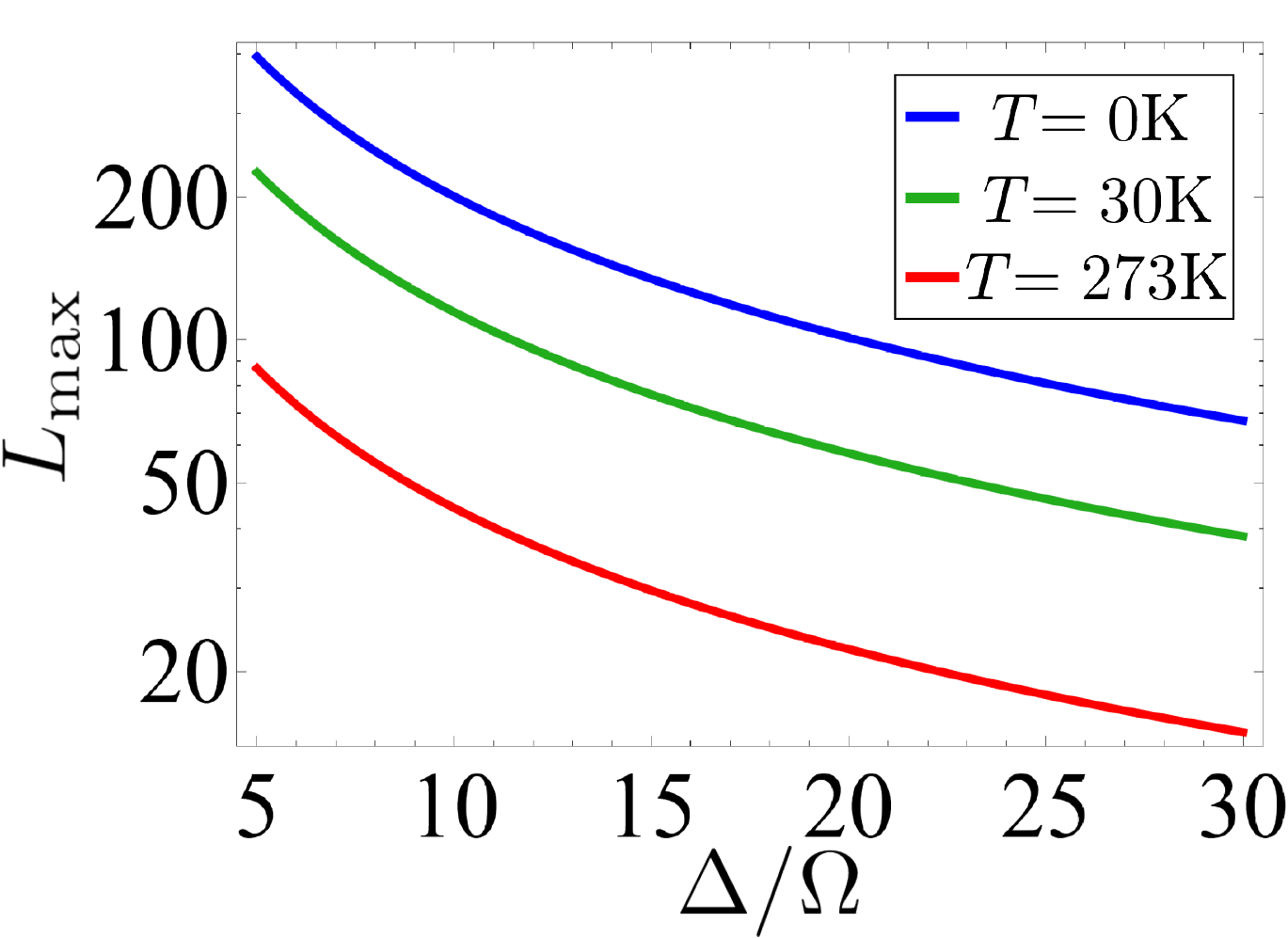}
\caption{
Maximum chain length $L_{\rm max}$ for a coherent evolution using Rydberg state $\ket{84{\rm S}}$ of ${}^6{\rm Li}$ for temperatures $T=(0,30,273)\,{\rm K}$ (blue, green, red) corresponding to $\tau_0=(8.6,2.8,0.42)\,{\rm ms}$ \cite{Sibalic_CompPhysComm_2017,ARC_package}.
}
\label{fig:Lmax}
\end{figure}

In practice, any experiment is prone to detrimental effects, such as decoherence due to off-resonant scattering from the Rydberg state. On the one hand, one can reduce the scattering, the rate of which $\propto (\Omega/\Delta)^2$, by reducing $\Omega/\Delta$. On the other hand, to limit the influence of such scattering, it is desirable to reduce the time of the experiment, by increasing the energy scale of $H_{\rm Ry}$. Since $J = W(2r_0) \propto \Omega^4/\Delta^3$, it can be achieved by increasing $\Omega$ and reducing $\Delta$ while still in the perturbative regime $\Omega/\Delta \ll 1$. 
For this reason, to weigh between these two effects, we choose as a figure of merit the achievable system size, for which $\ket{K_1}$ traverse the chain coherently in time $t=L T_c$, where $T_c = 1/(3 v_F W(2r_0))$ is the characteristic propagation time between lattice sites. To obtain $L_{\rm max}$, $t$ should be equated with the effective decay time $\tau = 1/\gamma$, where $\gamma \cong L/3 \gamma_0 (\Omega/\Delta)^2$ is the sum of $\approx L/3$ individual atomic far-detuned decay rates \cite{Wuster_2011_NJP}. Combining these expressions we get
\beq
	L_{\rm max} = 3 \sqrt{\frac{\tau_0 v_F W(2 r_0)}{\left(\frac{\Omega}{\Delta}\right)^2}} = 3 \Omega \sqrt{\frac{2 \tau_0 v_F}{\Delta\left(1+\frac{2\Delta}{V_{2}}\right)}}.
	\label{eq:L}
\eeq
Fig.~\ref{fig:Lmax} shows $L_{\rm max}$ vs. $\Delta/\Omega$ for the state $\ket{84{\rm S}}$ of ${}^6$Li and parameters used in the main text, $C_6=645\,{\rm GHz}\cdot \mu{\rm m}^6$, $r_0=2.5 \,\mu{\rm m}$.
It is apparent from the Figure that system sizes of the order of 100 sites might be achievable for realistic parameters. While this is encouraging, it is known that the dressing schemes are sensitive to detrimental effects, such as line broadening \cite{Goldschmidt_2016_PRL} leading to avalanche dephasing \cite{Boulier_2017_PRA,Young_2018_PRA}. It has been suggested, however, that these effects can be mitigated by cooling to reduce the black-body radiation or quenching the contaminant states \cite{Boulier_2017_PRA}.

~\\
Next, in analogy to the derivation of Eq. (\ref{eq:L}) we can derive the scaling of the system size taking into account both adiabatic preparation, cf. Sec. \ref{app:Preparation}, and the subsequent time evolution. We note that the estimation Eq. (\ref{eq:L}) holds for any $\lambda$, in which case $v_F$ should be replaced by $v_{\rm max}(\lambda)$. 

To proceed, we quantify the preparation time as a multiple $\kappa$ of the inverse spectral gap,
\beq
	T = \frac{\kappa}{\Delta_{sg}}.
	\label{eq:tprep}
\eeq
Working at criticality, we use the formula for finite size scaling of the gap \cite{Huijse_2011_JStatMech}
\beq
	\Delta_{sg} \approx \frac{2 \pi E_{\rm SCFT} 3 v_F W(2r_0)}{L},
	\label{eq:gap_critical}
\eeq
where $E_{\rm SCFT}=1/3 \; (2/3)$ for $L=3l+1$ ($L=3l$) respectively. We write the dispersive decay rates of $l \approx L/3$ atoms as
\beq
\Gamma = \frac{L}{3}\frac{1}{\tau_0} \left( \frac{\Omega}{\Delta}\right)^2 \bar{p} , \qquad 
\gamma = \frac{L}{3} \frac{1}{\tau_0} \left( \frac{\Omega}{\Delta}\right)^2,
\label{eq:gammas}
\eeq
where $\bar{p}$ stands for the average probability of being in the Rydberg state during the adiabatic sweep and depends on the form of the sweep. For the function (\ref{eq:Ft}) it is taken to be
\beq
	\left( \frac{\Omega}{\Delta} \right)^2 \bar{p} = \frac{1}{T} \int_0^T {\rm d}t\, \left( \frac{\Omega(t)}{\Delta} \right)^2 =  \left( \frac{\Omega}{\Delta} \right)^2 \frac{1}{T} \int_0^T {\rm d}t\, \sqrt{{\mathcal F}(t)} = \left( \frac{\Omega}{\Delta} \right)^2 \frac{2}{\pi} \; \rightarrow \; \bar{p} = \frac{2}{\pi},
\eeq
where the ramping of the Hamiltonian is implemented by tuning the Rabi frequency so that ${\mathcal F}(t) \propto \Omega(t)^4$ and correspondingly for the tunnelings $J(t)$. Setting
\beq
	\Gamma T + \gamma t = 1 
	\label{eq:L_cond}
\eeq
and recalling that $t = L T_c = L/[3 v_F W(2 r_0)]$, we can express $L$ after substituting (\ref{eq:tprep}-\ref{eq:gammas}) to (\ref{eq:L_cond}) as
\beq
	L_{\rm max} = 3 \Omega \sqrt{\frac{\tau_0 2 v_F}{\Delta\left(1+\frac{2\Delta}{V_{2}}\right)} \frac{1}{\left( \frac{1}{2\pi}E_{\rm SCFT} \kappa \bar{p} + 1 \right)}}.
	\label{eq:Lmax_comb}
\eeq
We note that setting $\kappa=0$ corresponds to ignoring the preparation stage and we recover the formula (\ref{eq:L}). Similarly, ignoring the time evolution, as in the case of the preparation of the ground states for chains with $L=3l$, amounts to neglecting the ``+1'' term in the second denominator.

\subsection{Energy scales and quality of the approximations}
\label{app:Scales}

\noindent
\emph{Estimates of energy scales.}
It is apparent from the previous sections that engineering the supersymmetric Hamiltonian is a tradeoff between a number of requirements. On one hand one wishes to maximize the energy scales, namely the dressing Rabi frequency $\Omega$, which determines the interaction Eq.~(\ref{eq:W}) to overcome the decoherence, cf. Eq.~(\ref{eq:Lmax_comb}). Tuning the Hamiltonian to the supersymmetric point $J=W(2 r_0)$, it also maximizes the tunneling rate, cf. Eq.~(\ref{eq:HQ_dimful}). Same effects result from reducing $\Delta$.

On the other hand, one wishes to minimize $\Omega/\Delta$ to be well in the adiabatic approximation regime, where the flat-top interaction potential (\ref{eq:W}) remains valid, the population in the Rydberg state is negligible and where the Rydberg Hamiltonian, Eq.~(\ref{eq:HRy}) of the main text, approaches the supersymmetric one.

While precise quantification of this trade-off goes beyond the scope of the present article, for illustration purposes it is instructive to consider specific example. Focusing on fermionic ${}^6{\rm Li}$, the choice of the Rydberg state $\ket{84{\rm S}}$ used in the main text is motivated by the maximization of the Rydberg state lifetime to overcome the decoherence, cf. Eq.~(\ref{eq:Lmax_comb}), as (at zero temperature) $\tau_0 \propto \left. n^* \right.^3$, where $n^*$ is the effective principal quantum number accounting for a quantum defect \cite{gallagher2005rydberg}. The associated Van der Waals coefficient $C_6=645\;{\rm GHz} \cdot \mu{\rm m}^6$ determines the Rydberg interaction energy. Choosing $\Omega=2\pi \times 10$ MHz and $\Omega/\Delta=1/10$ as experimentally sensible values, the lattice constant $r_0=2.5\;\mu{\rm m}$ is determined to satisfy the hierarchy $W(r_0) \gg W(2 r_0) \gg W(3 r_0)$. This then finally determines the actual energy scale of the Hamiltonian at the supersymmetric point $J=W(2 r_0) \approx 4 \; {\rm kHz}$. 

In the main text, we have focused mainly on the critical case $\lambda=1$. This was motivated by the relative experimental simplicity compared to the off-critical one described in Sec.~\ref{app:OffCritical}, but also by its prominence, since at criticality the propagating kinks reach their maximum velocity, the Fermi velocity $v_F$. In this case the chemical potential $\mu$ in Eq.~(\ref{eq:HQ_dimful}) contributes a uniform global offset and can be thus dropped everywhere but at the boundaries, where $\mu_1 = \mu_L = J$. 

With this, we can now enquire about the experimental feasibility of such scenario. Our construction relies on the tight-binding approximation for atoms tunneling in the optical lattice. For lattice potentials of the form $V(x) = V_0/2 \cos(2 k x)$ this occurs in the limit $V_0/E_r \gg 1$, where $E_r = (\hbar k)^2/(2m)$ is the recoil energy, $m$ the atom mass and $k=2\pi/\lambda$ the lattice wavevector. In the tight-binding regime, the tunneling rate is given by \cite{Arzamasovs_2017}
\beq
	J/E_r = \frac{4}{\sqrt{\pi}} \left(\frac{V_0}{E_r}\right)^\frac{3}{4} {\rm e}^{-2\sqrt{\frac{V_0}{E_r}}}
\eeq
and the band width $\Delta E = 4 J$. Furthermore, taking a harmonic approximation for the minima of the potential $V(x)$, one can identify the corresponding harmonic oscillator frequency separating the ground and the first excited state as $\omega = \sqrt{2 V_0 k^2/m}$. This indicates the relevant energy scale for the temperature of the atoms, namely $k_{\rm B} T \ll \omega$ in order to avoid thermal excitations of the higher bands of the lattice. Put together we thus require $V_0 \gg E_r$ and $\Delta E, k_{\rm B}T \ll \omega$. 

The above parameters fix the tunneling rate $J \approx 4\;{\rm kHz}$, which implies $V_0 \approx 5.5 \,E_r$ (for $\lambda = 2 r_0 = 5 \, \mu{\rm m}$ of the optical lattice laser light) and $\omega \approx 2\pi \times 6\;{\rm kHz}$ corresponding to $\hbar \omega/k_{\rm B} \approx 0.3 \; \mu{\rm K}$. To this end we note that temperatures of order $O(10\;{\rm nK})$ have been achieved when cooling fermions in optical lattices \cite{Taie_2012_NatPhys,Ozawa_2018_PRL,taie2020observation}. Clearly, the value $V_0/E_r \approx 5.5$ puts in question the appropriateness of the tight-binding approximation. One way to ensure its applicability is to increase $V_0/E_r$, reducing however the tunneling rate significantly. For instance for $V_0/E_r=40$ we get $J=6\;{\rm Hz}$. To remedy this, one could implement the Raman assisted hopping as suggested in Fig. \ref{fig:scheme_off} in the context of the off-critical implementation of the ${\rm M}_1$ model, leading to $J \propto \left.\Omega^R\right.^2/\Delta_R$.

~\\
\noindent
\emph{Quality of the approximations.}
In the proposed quantum simulation of the supersymmetric lattice Hamiltonian, we have made two crucial approximations, namely we have adiabatically eliminated the Rydberg states, cf. Sec. \ref{app:Elimination} and neglected the long range tails of the dressed interaction beyond next-to-nearest neighbours. The former approximation effectively neglects a finite population in the Rydberg state which results in photon scattering with associated energy scale $\gamma \propto \gamma_0 (\Omega/\Delta)^2$, cf. Sec.~\ref{app:Coherent}, while the energy scale of the latter is $W(3 r_0)$. These two scales then set a bound on the times where the approximations remain valid. Importantly, one also requires coherent evolution up to times $t J/l \sim O(1)$, where $J=W(2 r_0)$, to see the effects of the (s)kink propagation. 

Here we specifically focus on evaluating the effect of neglecting the tails of the interaction versus neglecting the scattering. We further parametrize $W(n r_0)=w_n \Omega^4/\Delta^3$ and ask what is the effect of both for a given time $tJ$. We get for the scattering and neglecting the tails respectively
\beq
	\frac{t \gamma}{t J} \propto \frac{1}{w_2}\frac{\Delta}{\Omega}\frac{\gamma_0}{\Omega}, \;\; \frac{t W(3 r_0)}{t J} = \frac{W(3 r_0)}{W(2 r_0)}.
	\label{eq:approx}
\eeq
The ratio $W(3 r_0)/W(2 r_0) \ll 1$ by construction. In comparison, $\Delta \gamma_0/(w_2 \Omega^2) \approx 290$ for the parameters used in the main text, i.e. $\Omega = 2\pi \times 10\,{\rm MHz}$, $\Omega/\Delta=1/10$, $C_6=645\,{\rm GHz}\cdot \mu{\rm m}^6$, $r_0=2.5 \,\mu{\rm m}$ and taking $\tau_0 = 1/\gamma_0 = 8.6\,{\rm ms}$ corresponding to zero temperature lifetime of $\ket{84{\rm}{\rm S}}$ state of ${}^6{\rm Li}$. It is thus clear that the off-resonant scattering is the dominant limiting factor for the coherent evolution. 
Furthermore, this trend will become even more pronounced in the double dressing scheme discussed in Sec.~\ref{app:Double_dressing}, which further reduces the ratio $W(3 r_0)/W(2 r_0)$ and increases the scattering due to the off-resonant coupling to the additional Rydberg state.
This is indeed compatible with the time dynamics generated by the Rydberg Hamiltonian with interactions truncated beyond next-to-nearest neighbour (gray line in Fig.~\ref{fig:3}c). In this case, the only discrepancy with respect to the dynamics generated by the supersymmetric Hamiltonian (blue dashed line in Fig.~\ref{fig:3}c) comes from the leakage to the Rydberg state. While this increases with increasing $\Omega/\Delta$, the overall optimization is dictated by the requirements in Eq.~(\ref{eq:approx}), namely it is favorable to increase $\Omega/\Delta$ while maximizing $\Omega$, cf. Sec.~\ref{app:Coherent}.

In summary, while further optimization is in principle possible, for realistic parameters the effect of omitting the tails remains negligible in comparison with the leakage of the atom population to the Rydberg state and the associated scattering. As mentioned in the main text, a promissing avenue to overcome this difficulty is to use cold molecules instead of Rydberg atoms \cite{Buchler_2007, Baranov_2012_ChemRev}. These are particularly interesting as they provide permanent or electric-field induced dipole moment, which is typically of order 1 Debye corresponding to the interaction energy of 10 kHz for $1\,\mu{\rm m}$ separation \cite{Carr_2009_NJP,Baranov_2012_ChemRev,Balakrishnan_2016_JChemPhys}, scale comparable to the present study. This would provide the blockade without the need of dressing to the optically excited state avoiding thus the decoherence due to off-resonant scattering.

~\\
\noindent
\emph{Symmetry considerations.} Let us recap the effects of the finite energy scales from the point of view of the (super)symmetry of the target Hamiltonian. As discussed in the previous paragraph, for the relevant parameter regime, the imperfect blockade $W(r_0)/W(2 r_0) < \infty$ dominates over the effect of the interaction tails $W(n r_0), n>2$, cf. Eq.~(\ref{eq:HQ_long_range}), and both represent a weak breaking of the supersymmetry.

In order to quantify the effect of the tails in the limit of large system sizes one could treat the effects of the $W(n r_0)$ terms, $n>2$, perturbatively (as suggested in a similar context in \cite{Lesanovsky_2012_PRA}) or using large-scale numerical simulations for real time evolution such as the DMRG \cite{Paeckel_2019_AnnPhys} or neural networks \cite{Carleo_2017_Science} based methods. Similarly, one could perform a systematic study of the imperfect blockade by replacing the projectors $P_i$ in the definition of the supercharges $Q$, cf. (\ref{eq:HQ_dimful}), 
by $P_i \rightarrow P_i + \eta n_i$, where $\eta = O(1/W(r_0))$ \cite{Lesanovsky_2012_PRA} and simultaneous deformation of $H_Q$ from the supersymmetric point with a potential $\sum_i W(r_0) n_i n_{i+1}$ to account for the nearest-neihbour interaction. However, such detailed analysis is beyond the scope of the present work.

Importantly, we note that $H_Q$, Eq.~(\ref{eq:HQ_dimful}), is recovered from the parent Rydberg Hamiltonian (\ref{eq:HRy}) in the limit of $W(r_0)/W(2 r_0), W(2 r_0)/W(n r_0) \rightarrow \infty, \; \forall n > 2$. This is reminiscent of a situation encountered in the early proposals \cite{Zohar_2012_PRL,Banerjee_2012_PRL} of the cold atom based quantum simulators of lattice gauge theories (LGT).
There, the local Hamiltonian symmetry corresponds to the Gauss law which was recovered only asymptotically in the limit of infinite interaction strength of the parent Hubbard-like model from which the effective Hamiltonian corresponding to the LGT was derived. 
The engineering of the Gauss law has then been improved in \cite{Kasamatsu_2013_PRL} by invoking the Higgs field, cf. also \cite{Haase_2021,Paulson_2020} for recent developments. 

In this context, it would be interesting to explore alternative avenues for engineering the supersymmetric lattice Hamiltonians in order to recover the exact supersymmetry, for instance through the mapping to spin Hamiltonians as discussed in the \emph{Outlook} of the main text.


\begin{thebibliography}{90}%
\makeatletter
\providecommand \@ifxundefined [1]{%
 \@ifx{#1\undefined}
}%
\providecommand \@ifnum [1]{%
 \ifnum #1\expandafter \@firstoftwo
 \else \expandafter \@secondoftwo
 \fi
}%
\providecommand \@ifx [1]{%
 \ifx #1\expandafter \@firstoftwo
 \else \expandafter \@secondoftwo
 \fi
}%
\providecommand \natexlab [1]{#1}%
\providecommand \enquote  [1]{``#1''}%
\providecommand \bibnamefont  [1]{#1}%
\providecommand \bibfnamefont [1]{#1}%
\providecommand \citenamefont [1]{#1}%
\providecommand \href@noop [0]{\@secondoftwo}%
\providecommand \href [0]{\begingroup \@sanitize@url \@href}%
\providecommand \@href[1]{\@@startlink{#1}\@@href}%
\providecommand \@@href[1]{\endgroup#1\@@endlink}%
\providecommand \@sanitize@url [0]{\catcode `\\12\catcode `\$12\catcode
  `\&12\catcode `\#12\catcode `\^12\catcode `\_12\catcode `\%12\relax}%
\providecommand \@@startlink[1]{}%
\providecommand \@@endlink[0]{}%
\providecommand \url  [0]{\begingroup\@sanitize@url \@url }%
\providecommand \@url [1]{\endgroup\@href {#1}{\urlprefix }}%
\providecommand \urlprefix  [0]{URL }%
\providecommand \Eprint [0]{\href }%
\providecommand \doibase [0]{http://dx.doi.org/}%
\providecommand \selectlanguage [0]{\@gobble}%
\providecommand \bibinfo  [0]{\@secondoftwo}%
\providecommand \bibfield  [0]{\@secondoftwo}%
\providecommand \translation [1]{[#1]}%
\providecommand \BibitemOpen [0]{}%
\providecommand \bibitemStop [0]{}%
\providecommand \bibitemNoStop [0]{.\EOS\space}%
\providecommand \EOS [0]{\spacefactor3000\relax}%
\providecommand \BibitemShut  [1]{\csname bibitem#1\endcsname}%
\let\auto@bib@innerbib\@empty
\bibitem [{\citenamefont {Suzuki}(1993)}]{Suzuki_1993_Book}%
  \BibitemOpen
  \bibfield  {author} {\bibinfo {author} {\bibfnamefont {M.}~\bibnamefont
  {Suzuki}},\ }\href@noop {} {\emph {\bibinfo {title} {Quantum Monte Carlo
  methods in condensed matter physics}}}\ (\bibinfo  {publisher} {World
  scientific},\ \bibinfo {year} {1993})\BibitemShut {NoStop}%
\bibitem [{\citenamefont {Gubernatis}\ \emph {et~al.}(2016)\citenamefont
  {Gubernatis}, \citenamefont {Kawashima},\ and\ \citenamefont
  {Werner}}]{Gubernatis_2016_Book}%
  \BibitemOpen
  \bibfield  {author} {\bibinfo {author} {\bibfnamefont {J.}~\bibnamefont
  {Gubernatis}}, \bibinfo {author} {\bibfnamefont {N.}~\bibnamefont
  {Kawashima}}, \ and\ \bibinfo {author} {\bibfnamefont {P.}~\bibnamefont
  {Werner}},\ }\href@noop {} {\emph {\bibinfo {title} {Quantum Monte Carlo
  Methods}}}\ (\bibinfo  {publisher} {Cambridge University Press},\ \bibinfo
  {year} {2016})\BibitemShut {NoStop}%
\bibitem [{\citenamefont {Becca}\ and\ \citenamefont
  {Sorella}(2017)}]{Becca_2017_Book}%
  \BibitemOpen
  \bibfield  {author} {\bibinfo {author} {\bibfnamefont {F.}~\bibnamefont
  {Becca}}\ and\ \bibinfo {author} {\bibfnamefont {S.}~\bibnamefont
  {Sorella}},\ }\href@noop {} {\emph {\bibinfo {title} {Quantum Monte Carlo
  approaches for correlated systems}}}\ (\bibinfo  {publisher} {Cambridge
  University Press},\ \bibinfo {year} {2017})\BibitemShut {NoStop}%
\bibitem [{\citenamefont {Or{\'u}s}(2014)}]{Orus_2014_AnnPhys}%
  \BibitemOpen
  \bibfield  {author} {\bibinfo {author} {\bibfnamefont {R.}~\bibnamefont
  {Or{\'u}s}},\ }\href@noop {} {\bibfield  {journal} {\bibinfo  {journal}
  {Annals of Physics}\ }\textbf {\bibinfo {volume} {349}},\ \bibinfo {pages}
  {117} (\bibinfo {year} {2014})}\BibitemShut {NoStop}%
\bibitem [{\citenamefont {Montangero}\ \emph {et~al.}(2018)\citenamefont
  {Montangero}, \citenamefont {Montangero},\ and\ \citenamefont
  {Evenson}}]{Montangero_2018_Book}%
  \BibitemOpen
  \bibfield  {author} {\bibinfo {author} {\bibfnamefont {S.}~\bibnamefont
  {Montangero}}, \bibinfo {author} {\bibfnamefont {S.}~\bibnamefont
  {Montangero}}, \ and\ \bibinfo {author} {\bibnamefont {Evenson}},\
  }\href@noop {} {\emph {\bibinfo {title} {Introduction to Tensor Network
  Methods}}}\ (\bibinfo  {publisher} {Springer},\ \bibinfo {year}
  {2018})\BibitemShut {NoStop}%
\bibitem [{\citenamefont {Zaanen}\ \emph {et~al.}(2015)\citenamefont {Zaanen},
  \citenamefont {Liu}, \citenamefont {Sun},\ and\ \citenamefont
  {Schalm}}]{Zaanen_2015_Book}%
  \BibitemOpen
  \bibfield  {author} {\bibinfo {author} {\bibfnamefont {J.}~\bibnamefont
  {Zaanen}}, \bibinfo {author} {\bibfnamefont {Y.}~\bibnamefont {Liu}},
  \bibinfo {author} {\bibfnamefont {Y.-W.}\ \bibnamefont {Sun}}, \ and\
  \bibinfo {author} {\bibfnamefont {K.}~\bibnamefont {Schalm}},\ }\href@noop {}
  {\emph {\bibinfo {title} {Holographic duality in condensed matter physics}}}\
  (\bibinfo  {publisher} {Cambridge University Press},\ \bibinfo {year}
  {2015})\BibitemShut {NoStop}%
\bibitem [{\citenamefont {Nicolai}(1976)}]{Nicolai_1976_JPA}%
  \BibitemOpen
  \bibfield  {author} {\bibinfo {author} {\bibfnamefont {H.}~\bibnamefont
  {Nicolai}},\ }\href@noop {} {\bibfield  {journal} {\bibinfo  {journal} {J.
  Phys. A}\ }\textbf {\bibinfo {volume} {9}},\ \bibinfo {pages} {1497}
  (\bibinfo {year} {1976})}\BibitemShut {NoStop}%
\bibitem [{\citenamefont {Fendley}\ \emph
  {et~al.}(2003{\natexlab{a}})\citenamefont {Fendley}, \citenamefont
  {Schoutens},\ and\ \citenamefont {de~Boer}}]{Fendley_2003_PRL}%
  \BibitemOpen
  \bibfield  {author} {\bibinfo {author} {\bibfnamefont {P.}~\bibnamefont
  {Fendley}}, \bibinfo {author} {\bibfnamefont {K.}~\bibnamefont {Schoutens}},
  \ and\ \bibinfo {author} {\bibfnamefont {J.}~\bibnamefont {de~Boer}},\ }\href
  {\doibase 10.1103/PhysRevLett.90.120402} {\bibfield  {journal} {\bibinfo
  {journal} {Phys. Rev. Lett.}\ }\textbf {\bibinfo {volume} {90}},\ \bibinfo
  {pages} {120402} (\bibinfo {year} {2003}{\natexlab{a}})}\BibitemShut
  {NoStop}%
\bibitem [{\citenamefont {Fendley}\ \emph
  {et~al.}(2003{\natexlab{b}})\citenamefont {Fendley}, \citenamefont
  {Nienhuis},\ and\ \citenamefont {Schoutens}}]{Fendley_2003_JPhysA}%
  \BibitemOpen
  \bibfield  {author} {\bibinfo {author} {\bibfnamefont {P.}~\bibnamefont
  {Fendley}}, \bibinfo {author} {\bibfnamefont {B.}~\bibnamefont {Nienhuis}}, \
  and\ \bibinfo {author} {\bibfnamefont {K.}~\bibnamefont {Schoutens}},\
  }\href@noop {} {\bibfield  {journal} {\bibinfo  {journal} {J. Phys. A}\
  }\textbf {\bibinfo {volume} {36}},\ \bibinfo {pages} {12399} (\bibinfo {year}
  {2003}{\natexlab{b}})}\BibitemShut {NoStop}%
\bibitem [{\citenamefont {Fu}\ \emph {et~al.}(2017)\citenamefont {Fu},
  \citenamefont {Gaiotto}, \citenamefont {Maldacena},\ and\ \citenamefont
  {Sachdev}}]{Fu_2017_PhysRevD}%
  \BibitemOpen
  \bibfield  {author} {\bibinfo {author} {\bibfnamefont {W.}~\bibnamefont
  {Fu}}, \bibinfo {author} {\bibfnamefont {D.}~\bibnamefont {Gaiotto}},
  \bibinfo {author} {\bibfnamefont {J.}~\bibnamefont {Maldacena}}, \ and\
  \bibinfo {author} {\bibfnamefont {S.}~\bibnamefont {Sachdev}},\ }\href@noop
  {} {\bibfield  {journal} {\bibinfo  {journal} {Phys. Rev. D}\ }\textbf
  {\bibinfo {volume} {95}},\ \bibinfo {pages} {026009} (\bibinfo {year}
  {2017})}\BibitemShut {NoStop}%
\bibitem [{\citenamefont {Sannomiya}\ \emph {et~al.}(2017)\citenamefont
  {Sannomiya}, \citenamefont {Katsura},\ and\ \citenamefont
  {Nakayama}}]{Sannomiya_2017_PhysRevD}%
  \BibitemOpen
  \bibfield  {author} {\bibinfo {author} {\bibfnamefont {N.}~\bibnamefont
  {Sannomiya}}, \bibinfo {author} {\bibfnamefont {H.}~\bibnamefont {Katsura}},
  \ and\ \bibinfo {author} {\bibfnamefont {Y.}~\bibnamefont {Nakayama}},\
  }\href@noop {} {\bibfield  {journal} {\bibinfo  {journal} {Phys. Rev. D}\
  }\textbf {\bibinfo {volume} {95}},\ \bibinfo {pages} {065001} (\bibinfo
  {year} {2017})}\BibitemShut {NoStop}%
\bibitem [{\citenamefont {O'Brien}\ and\ \citenamefont
  {Fendley}(2018)}]{OBrien_2017_PhysRevLett}%
  \BibitemOpen
  \bibfield  {author} {\bibinfo {author} {\bibfnamefont {E.}~\bibnamefont
  {O'Brien}}\ and\ \bibinfo {author} {\bibfnamefont {P.}~\bibnamefont
  {Fendley}},\ }\href@noop {} {\bibfield  {journal} {\bibinfo  {journal} {Phys.
  Rev. Lett.}\ }\textbf {\bibinfo {volume} {120}},\ \bibinfo {pages} {206403}
  (\bibinfo {year} {2018})}\BibitemShut {NoStop}%
\bibitem [{\citenamefont {Grover}\ \emph {et~al.}(2014)\citenamefont {Grover},
  \citenamefont {Sheng},\ and\ \citenamefont
  {Vishwanath}}]{Grover_2014_Science}%
  \BibitemOpen
  \bibfield  {author} {\bibinfo {author} {\bibfnamefont {T.}~\bibnamefont
  {Grover}}, \bibinfo {author} {\bibfnamefont {D.}~\bibnamefont {Sheng}}, \
  and\ \bibinfo {author} {\bibfnamefont {A.}~\bibnamefont {Vishwanath}},\
  }\href {\doibase 10.1126/science.1248253} {\bibfield  {journal} {\bibinfo
  {journal} {Science}\ }\textbf {\bibinfo {volume} {344}},\ \bibinfo {pages}
  {6181} (\bibinfo {year} {2014})}\BibitemShut {NoStop}%
\bibitem [{\citenamefont {Huijse}\ \emph {et~al.}(2014)\citenamefont {Huijse},
  \citenamefont {Bauer},\ and\ \citenamefont {Berg}}]{Huijse_2014_PRL}%
  \BibitemOpen
  \bibfield  {author} {\bibinfo {author} {\bibfnamefont {L.}~\bibnamefont
  {Huijse}}, \bibinfo {author} {\bibfnamefont {B.}~\bibnamefont {Bauer}}, \
  and\ \bibinfo {author} {\bibfnamefont {E.}~\bibnamefont {Berg}},\ }\href
  {\doibase 10.1103/PhysRevLett.114.090404} {\bibfield  {journal} {\bibinfo
  {journal} {Phys. Rev. Lett.}\ }\textbf {\bibinfo {volume} {114}},\ \bibinfo
  {pages} {9} (\bibinfo {year} {2014})}\BibitemShut {NoStop}%
\bibitem [{\citenamefont {Yu}\ and\ \citenamefont {Yang}(2010)}]{Yu_2010_PRL}%
  \BibitemOpen
  \bibfield  {author} {\bibinfo {author} {\bibfnamefont {Y.}~\bibnamefont
  {Yu}}\ and\ \bibinfo {author} {\bibfnamefont {K.}~\bibnamefont {Yang}},\
  }\href {\doibase 10.1103/PhysRevLett.105.150605} {\bibfield  {journal}
  {\bibinfo  {journal} {Phys. Rev. Lett.}\ }\textbf {\bibinfo {volume} {105}},\
  \bibinfo {pages} {150605} (\bibinfo {year} {2010})}\BibitemShut {NoStop}%
\bibitem [{\citenamefont {Witten}(1982)}]{Witten_1982_NuclPhysB}%
  \BibitemOpen
  \bibfield  {author} {\bibinfo {author} {\bibfnamefont {E.}~\bibnamefont
  {Witten}},\ }\href@noop {} {\bibfield  {journal} {\bibinfo  {journal}
  {Nuclear Physics B}\ }\textbf {\bibinfo {volume} {202}},\ \bibinfo {pages}
  {253} (\bibinfo {year} {1982})}\BibitemShut {NoStop}%
\bibitem [{\citenamefont {van Eerten}(2005)}]{vEerten_2005_JMathPhys}%
  \BibitemOpen
  \bibfield  {author} {\bibinfo {author} {\bibfnamefont {H.}~\bibnamefont {van
  Eerten}},\ }\href@noop {} {\bibfield  {journal} {\bibinfo  {journal} {J.
  Math. Phys.}\ }\textbf {\bibinfo {volume} {46}},\ \bibinfo {pages} {123302}
  (\bibinfo {year} {2005})}\BibitemShut {NoStop}%
\bibitem [{\citenamefont {Fendley}\ and\ \citenamefont
  {Schoutens}(2005)}]{Fendley_2005_PRL}%
  \BibitemOpen
  \bibfield  {author} {\bibinfo {author} {\bibfnamefont {P.}~\bibnamefont
  {Fendley}}\ and\ \bibinfo {author} {\bibfnamefont {K.}~\bibnamefont
  {Schoutens}},\ }\href {\doibase 10.1103/PhysRevLett.95.046403} {\bibfield
  {journal} {\bibinfo  {journal} {Phys. Rev. Lett.}\ }\textbf {\bibinfo
  {volume} {95}},\ \bibinfo {pages} {046403} (\bibinfo {year}
  {2005})}\BibitemShut {NoStop}%
\bibitem [{\citenamefont {Chepiga}\ \emph {et~al.}(2021)\citenamefont
  {Chepiga}, \citenamefont {Min{\'a}{\v{r}}},\ and\ \citenamefont
  {Schoutens}}]{Chepiga_2021_arXiv}%
  \BibitemOpen
  \bibfield  {author} {\bibinfo {author} {\bibfnamefont {N.}~\bibnamefont
  {Chepiga}}, \bibinfo {author} {\bibfnamefont {J.}~\bibnamefont
  {Min{\'a}{\v{r}}}}, \ and\ \bibinfo {author} {\bibfnamefont {K.}~\bibnamefont
  {Schoutens}},\ }\href@noop {} {\bibfield  {journal} {\bibinfo  {journal}
  {arXiv:2105.04359}\ } (\bibinfo {year} {2021})}\BibitemShut {NoStop}%
\bibitem [{\citenamefont {Browaeys}\ and\ \citenamefont
  {Lahaye}(2020)}]{Browaeys_2020_NatPhys}%
  \BibitemOpen
  \bibfield  {author} {\bibinfo {author} {\bibfnamefont {A.}~\bibnamefont
  {Browaeys}}\ and\ \bibinfo {author} {\bibfnamefont {T.}~\bibnamefont
  {Lahaye}},\ }\href@noop {} {\bibfield  {journal} {\bibinfo  {journal} {Nature
  Physics}\ }\textbf {\bibinfo {volume} {16}},\ \bibinfo {pages} {132}
  (\bibinfo {year} {2020})}\BibitemShut {NoStop}%
\bibitem [{\citenamefont {Barredo}\ \emph {et~al.}(2016)\citenamefont
  {Barredo}, \citenamefont {De~L{\'e}s{\'e}leuc}, \citenamefont {Lienhard},
  \citenamefont {Lahaye},\ and\ \citenamefont
  {Browaeys}}]{Barredo_2016_Science}%
  \BibitemOpen
  \bibfield  {author} {\bibinfo {author} {\bibfnamefont {D.}~\bibnamefont
  {Barredo}}, \bibinfo {author} {\bibfnamefont {S.}~\bibnamefont
  {De~L{\'e}s{\'e}leuc}}, \bibinfo {author} {\bibfnamefont {V.}~\bibnamefont
  {Lienhard}}, \bibinfo {author} {\bibfnamefont {T.}~\bibnamefont {Lahaye}}, \
  and\ \bibinfo {author} {\bibfnamefont {A.}~\bibnamefont {Browaeys}},\
  }\href@noop {} {\bibfield  {journal} {\bibinfo  {journal} {Science}\ }\textbf
  {\bibinfo {volume} {354}},\ \bibinfo {pages} {1021} (\bibinfo {year}
  {2016})}\BibitemShut {NoStop}%
\bibitem [{\citenamefont {Endres}\ \emph {et~al.}(2016)\citenamefont {Endres},
  \citenamefont {Bernien}, \citenamefont {Keesling}, \citenamefont {Levine},
  \citenamefont {Anschuetz}, \citenamefont {Krajenbrink}, \citenamefont
  {Senko}, \citenamefont {Vuletic}, \citenamefont {Greiner},\ and\
  \citenamefont {Lukin}}]{Endres_2016_Science}%
  \BibitemOpen
  \bibfield  {author} {\bibinfo {author} {\bibfnamefont {M.}~\bibnamefont
  {Endres}}, \bibinfo {author} {\bibfnamefont {H.}~\bibnamefont {Bernien}},
  \bibinfo {author} {\bibfnamefont {A.}~\bibnamefont {Keesling}}, \bibinfo
  {author} {\bibfnamefont {H.}~\bibnamefont {Levine}}, \bibinfo {author}
  {\bibfnamefont {E.~R.}\ \bibnamefont {Anschuetz}}, \bibinfo {author}
  {\bibfnamefont {A.}~\bibnamefont {Krajenbrink}}, \bibinfo {author}
  {\bibfnamefont {C.}~\bibnamefont {Senko}}, \bibinfo {author} {\bibfnamefont
  {V.}~\bibnamefont {Vuletic}}, \bibinfo {author} {\bibfnamefont
  {M.}~\bibnamefont {Greiner}}, \ and\ \bibinfo {author} {\bibfnamefont
  {M.~D.}\ \bibnamefont {Lukin}},\ }\href@noop {} {\bibfield  {journal}
  {\bibinfo  {journal} {Science}\ }\textbf {\bibinfo {volume} {354}},\ \bibinfo
  {pages} {1024} (\bibinfo {year} {2016})}\BibitemShut {NoStop}%
\bibitem [{\citenamefont {Barredo}\ \emph {et~al.}(2018)\citenamefont
  {Barredo}, \citenamefont {Lienhard}, \citenamefont {De~Leseleuc},
  \citenamefont {Lahaye},\ and\ \citenamefont
  {Browaeys}}]{Barredo_2018_Nature}%
  \BibitemOpen
  \bibfield  {author} {\bibinfo {author} {\bibfnamefont {D.}~\bibnamefont
  {Barredo}}, \bibinfo {author} {\bibfnamefont {V.}~\bibnamefont {Lienhard}},
  \bibinfo {author} {\bibfnamefont {S.}~\bibnamefont {De~Leseleuc}}, \bibinfo
  {author} {\bibfnamefont {T.}~\bibnamefont {Lahaye}}, \ and\ \bibinfo {author}
  {\bibfnamefont {A.}~\bibnamefont {Browaeys}},\ }\href@noop {} {\bibfield
  {journal} {\bibinfo  {journal} {Nature}\ }\textbf {\bibinfo {volume} {561}},\
  \bibinfo {pages} {79} (\bibinfo {year} {2018})}\BibitemShut {NoStop}%
\bibitem [{\citenamefont {Wang}\ \emph {et~al.}(2019)\citenamefont {Wang},
  \citenamefont {Shevate}, \citenamefont {Wintermantel}, \citenamefont
  {Morgado}, \citenamefont {Lochead},\ and\ \citenamefont
  {Whitlock}}]{Wang_2019}%
  \BibitemOpen
  \bibfield  {author} {\bibinfo {author} {\bibfnamefont {Y.}~\bibnamefont
  {Wang}}, \bibinfo {author} {\bibfnamefont {S.}~\bibnamefont {Shevate}},
  \bibinfo {author} {\bibfnamefont {T.}~\bibnamefont {Wintermantel}}, \bibinfo
  {author} {\bibfnamefont {M.}~\bibnamefont {Morgado}}, \bibinfo {author}
  {\bibfnamefont {G.}~\bibnamefont {Lochead}}, \ and\ \bibinfo {author}
  {\bibfnamefont {S.}~\bibnamefont {Whitlock}},\ }\href@noop {} {\bibfield
  {journal} {\bibinfo  {journal} {arXiv:1912.04200}\ } (\bibinfo {year}
  {2019})}\BibitemShut {NoStop}%
\bibitem [{\citenamefont {Schau{\ss}}\ \emph {et~al.}(2015)\citenamefont
  {Schau{\ss}}, \citenamefont {Zeiher}, \citenamefont {Fukuhara}, \citenamefont
  {Hild}, \citenamefont {Cheneau}, \citenamefont {Macr{\`\i}}, \citenamefont
  {Pohl}, \citenamefont {Bloch},\ and\ \citenamefont
  {Gro{\ss}}}]{Schauss_2015_Science}%
  \BibitemOpen
  \bibfield  {author} {\bibinfo {author} {\bibfnamefont {P.}~\bibnamefont
  {Schau{\ss}}}, \bibinfo {author} {\bibfnamefont {J.}~\bibnamefont {Zeiher}},
  \bibinfo {author} {\bibfnamefont {T.}~\bibnamefont {Fukuhara}}, \bibinfo
  {author} {\bibfnamefont {S.}~\bibnamefont {Hild}}, \bibinfo {author}
  {\bibfnamefont {M.}~\bibnamefont {Cheneau}}, \bibinfo {author} {\bibfnamefont
  {T.}~\bibnamefont {Macr{\`\i}}}, \bibinfo {author} {\bibfnamefont
  {T.}~\bibnamefont {Pohl}}, \bibinfo {author} {\bibfnamefont {I.}~\bibnamefont
  {Bloch}}, \ and\ \bibinfo {author} {\bibfnamefont {C.}~\bibnamefont
  {Gro{\ss}}},\ }\href@noop {} {\bibfield  {journal} {\bibinfo  {journal}
  {Science}\ }\textbf {\bibinfo {volume} {347}},\ \bibinfo {pages} {1455}
  (\bibinfo {year} {2015})}\BibitemShut {NoStop}%
\bibitem [{\citenamefont {Labuhn}\ \emph {et~al.}(2016)\citenamefont {Labuhn},
  \citenamefont {Barredo}, \citenamefont {Ravets}, \citenamefont
  {De~L{\'e}s{\'e}leuc}, \citenamefont {Macr{\`\i}}, \citenamefont {Lahaye},\
  and\ \citenamefont {Browaeys}}]{Labuhn_2016_Nature}%
  \BibitemOpen
  \bibfield  {author} {\bibinfo {author} {\bibfnamefont {H.}~\bibnamefont
  {Labuhn}}, \bibinfo {author} {\bibfnamefont {D.}~\bibnamefont {Barredo}},
  \bibinfo {author} {\bibfnamefont {S.}~\bibnamefont {Ravets}}, \bibinfo
  {author} {\bibfnamefont {S.}~\bibnamefont {De~L{\'e}s{\'e}leuc}}, \bibinfo
  {author} {\bibfnamefont {T.}~\bibnamefont {Macr{\`\i}}}, \bibinfo {author}
  {\bibfnamefont {T.}~\bibnamefont {Lahaye}}, \ and\ \bibinfo {author}
  {\bibfnamefont {A.}~\bibnamefont {Browaeys}},\ }\href@noop {} {\bibfield
  {journal} {\bibinfo  {journal} {Nature}\ }\textbf {\bibinfo {volume} {534}},\
  \bibinfo {pages} {667} (\bibinfo {year} {2016})}\BibitemShut {NoStop}%
\bibitem [{\citenamefont {Bernien}\ \emph {et~al.}(2017)\citenamefont
  {Bernien}, \citenamefont {Lukin}, \citenamefont {Pichler}, \citenamefont
  {Choi}, \citenamefont {Greiner}, \citenamefont {Vuleti{\'{c}}}, \citenamefont
  {Omran}, \citenamefont {Levine}, \citenamefont {Schwartz}, \citenamefont
  {Keesling}, \citenamefont {Endres},\ and\ \citenamefont
  {Zibrov}}]{Bernien_2017}%
  \BibitemOpen
  \bibfield  {author} {\bibinfo {author} {\bibfnamefont {H.}~\bibnamefont
  {Bernien}}, \bibinfo {author} {\bibfnamefont {M.~D.}\ \bibnamefont {Lukin}},
  \bibinfo {author} {\bibfnamefont {H.}~\bibnamefont {Pichler}}, \bibinfo
  {author} {\bibfnamefont {S.}~\bibnamefont {Choi}}, \bibinfo {author}
  {\bibfnamefont {M.}~\bibnamefont {Greiner}}, \bibinfo {author} {\bibfnamefont
  {V.}~\bibnamefont {Vuleti{\'{c}}}}, \bibinfo {author} {\bibfnamefont
  {A.}~\bibnamefont {Omran}}, \bibinfo {author} {\bibfnamefont
  {H.}~\bibnamefont {Levine}}, \bibinfo {author} {\bibfnamefont
  {S.}~\bibnamefont {Schwartz}}, \bibinfo {author} {\bibfnamefont
  {A.}~\bibnamefont {Keesling}}, \bibinfo {author} {\bibfnamefont
  {M.}~\bibnamefont {Endres}}, \ and\ \bibinfo {author} {\bibfnamefont {A.~S.}\
  \bibnamefont {Zibrov}},\ }\href {\doibase 10.1038/nature24622} {\bibfield
  {journal} {\bibinfo  {journal} {Nature}\ }\textbf {\bibinfo {volume} {551}},\
  \bibinfo {pages} {579} (\bibinfo {year} {2017})}\BibitemShut {NoStop}%
\bibitem [{\citenamefont {de~L{\'e}s{\'e}leuc}\ \emph
  {et~al.}(2019)\citenamefont {de~L{\'e}s{\'e}leuc}, \citenamefont {Lienhard},
  \citenamefont {Scholl}, \citenamefont {Barredo}, \citenamefont {Weber},
  \citenamefont {Lang}, \citenamefont {B{\"u}chler}, \citenamefont {Lahaye},\
  and\ \citenamefont {Browaeys}}]{deLeseleuc_2019_Science}%
  \BibitemOpen
  \bibfield  {author} {\bibinfo {author} {\bibfnamefont {S.}~\bibnamefont
  {de~L{\'e}s{\'e}leuc}}, \bibinfo {author} {\bibfnamefont {V.}~\bibnamefont
  {Lienhard}}, \bibinfo {author} {\bibfnamefont {P.}~\bibnamefont {Scholl}},
  \bibinfo {author} {\bibfnamefont {D.}~\bibnamefont {Barredo}}, \bibinfo
  {author} {\bibfnamefont {S.}~\bibnamefont {Weber}}, \bibinfo {author}
  {\bibfnamefont {N.}~\bibnamefont {Lang}}, \bibinfo {author} {\bibfnamefont
  {H.~P.}\ \bibnamefont {B{\"u}chler}}, \bibinfo {author} {\bibfnamefont
  {T.}~\bibnamefont {Lahaye}}, \ and\ \bibinfo {author} {\bibfnamefont
  {A.}~\bibnamefont {Browaeys}},\ }\href@noop {} {\bibfield  {journal}
  {\bibinfo  {journal} {Science}\ }\textbf {\bibinfo {volume} {365}},\ \bibinfo
  {pages} {775} (\bibinfo {year} {2019})}\BibitemShut {NoStop}%
\bibitem [{\citenamefont {Helmrich}\ \emph {et~al.}(2020)\citenamefont
  {Helmrich}, \citenamefont {Arias}, \citenamefont {Lochead}, \citenamefont
  {Wintermantel}, \citenamefont {Buchhold}, \citenamefont {Diehl},\ and\
  \citenamefont {Whitlock}}]{Helmrich_2020_Nature}%
  \BibitemOpen
  \bibfield  {author} {\bibinfo {author} {\bibfnamefont {S.}~\bibnamefont
  {Helmrich}}, \bibinfo {author} {\bibfnamefont {A.}~\bibnamefont {Arias}},
  \bibinfo {author} {\bibfnamefont {G.}~\bibnamefont {Lochead}}, \bibinfo
  {author} {\bibfnamefont {T.}~\bibnamefont {Wintermantel}}, \bibinfo {author}
  {\bibfnamefont {M.}~\bibnamefont {Buchhold}}, \bibinfo {author}
  {\bibfnamefont {S.}~\bibnamefont {Diehl}}, \ and\ \bibinfo {author}
  {\bibfnamefont {S.}~\bibnamefont {Whitlock}},\ }\href@noop {} {\bibfield
  {journal} {\bibinfo  {journal} {Nature}\ }\textbf {\bibinfo {volume} {577}},\
  \bibinfo {pages} {481} (\bibinfo {year} {2020})}\BibitemShut {NoStop}%
\bibitem [{\citenamefont {Balewski}\ \emph {et~al.}(2014)\citenamefont
  {Balewski}, \citenamefont {Krupp}, \citenamefont {Gaj}, \citenamefont
  {Hofferberth}, \citenamefont {L{\"o}w},\ and\ \citenamefont
  {Pfau}}]{Balewski_2014_NJP}%
  \BibitemOpen
  \bibfield  {author} {\bibinfo {author} {\bibfnamefont {J.~B.}\ \bibnamefont
  {Balewski}}, \bibinfo {author} {\bibfnamefont {A.~T.}\ \bibnamefont {Krupp}},
  \bibinfo {author} {\bibfnamefont {A.}~\bibnamefont {Gaj}}, \bibinfo {author}
  {\bibfnamefont {S.}~\bibnamefont {Hofferberth}}, \bibinfo {author}
  {\bibfnamefont {R.}~\bibnamefont {L{\"o}w}}, \ and\ \bibinfo {author}
  {\bibfnamefont {T.}~\bibnamefont {Pfau}},\ }\href@noop {} {\bibfield
  {journal} {\bibinfo  {journal} {New Journal of Physics}\ }\textbf {\bibinfo
  {volume} {16}},\ \bibinfo {pages} {063012} (\bibinfo {year}
  {2014})}\BibitemShut {NoStop}%
\bibitem [{\citenamefont {Jau}\ \emph {et~al.}(2016)\citenamefont {Jau},
  \citenamefont {Hankin}, \citenamefont {Keating}, \citenamefont {Deutsch},\
  and\ \citenamefont {Biedermann}}]{Jau_2016_NatPhys}%
  \BibitemOpen
  \bibfield  {author} {\bibinfo {author} {\bibfnamefont {Y.-Y.}\ \bibnamefont
  {Jau}}, \bibinfo {author} {\bibfnamefont {A.}~\bibnamefont {Hankin}},
  \bibinfo {author} {\bibfnamefont {T.}~\bibnamefont {Keating}}, \bibinfo
  {author} {\bibfnamefont {I.}~\bibnamefont {Deutsch}}, \ and\ \bibinfo
  {author} {\bibfnamefont {G.}~\bibnamefont {Biedermann}},\ }\href@noop {}
  {\bibfield  {journal} {\bibinfo  {journal} {Nature Physics}\ }\textbf
  {\bibinfo {volume} {12}},\ \bibinfo {pages} {71} (\bibinfo {year}
  {2016})}\BibitemShut {NoStop}%
\bibitem [{\citenamefont {Zeiher}\ \emph {et~al.}(2016)\citenamefont {Zeiher},
  \citenamefont {Van~Bijnen}, \citenamefont {Schau{\ss}}, \citenamefont {Hild},
  \citenamefont {Choi}, \citenamefont {Pohl}, \citenamefont {Bloch},\ and\
  \citenamefont {Gross}}]{Zeiher_2016_NatPhys}%
  \BibitemOpen
  \bibfield  {author} {\bibinfo {author} {\bibfnamefont {J.}~\bibnamefont
  {Zeiher}}, \bibinfo {author} {\bibfnamefont {R.}~\bibnamefont {Van~Bijnen}},
  \bibinfo {author} {\bibfnamefont {P.}~\bibnamefont {Schau{\ss}}}, \bibinfo
  {author} {\bibfnamefont {S.}~\bibnamefont {Hild}}, \bibinfo {author}
  {\bibfnamefont {J.-y.}\ \bibnamefont {Choi}}, \bibinfo {author}
  {\bibfnamefont {T.}~\bibnamefont {Pohl}}, \bibinfo {author} {\bibfnamefont
  {I.}~\bibnamefont {Bloch}}, \ and\ \bibinfo {author} {\bibfnamefont
  {C.}~\bibnamefont {Gross}},\ }\href@noop {} {\bibfield  {journal} {\bibinfo
  {journal} {Nature Physics}\ }\textbf {\bibinfo {volume} {12}},\ \bibinfo
  {pages} {1095} (\bibinfo {year} {2016})}\BibitemShut {NoStop}%
\bibitem [{\citenamefont {Arias}\ \emph {et~al.}(2019)\citenamefont {Arias},
  \citenamefont {Lochead}, \citenamefont {Wintermantel}, \citenamefont
  {Helmrich},\ and\ \citenamefont {Whitlock}}]{Arias_2019_PRL}%
  \BibitemOpen
  \bibfield  {author} {\bibinfo {author} {\bibfnamefont {A.}~\bibnamefont
  {Arias}}, \bibinfo {author} {\bibfnamefont {G.}~\bibnamefont {Lochead}},
  \bibinfo {author} {\bibfnamefont {T.~M.}\ \bibnamefont {Wintermantel}},
  \bibinfo {author} {\bibfnamefont {S.}~\bibnamefont {Helmrich}}, \ and\
  \bibinfo {author} {\bibfnamefont {S.}~\bibnamefont {Whitlock}},\ }\href
  {\doibase 10.1103/PhysRevLett.122.053601} {\bibfield  {journal} {\bibinfo
  {journal} {Phys. Rev. Lett.}\ }\textbf {\bibinfo {volume} {122}},\ \bibinfo
  {pages} {053601} (\bibinfo {year} {2019})}\BibitemShut {NoStop}%
\bibitem [{\citenamefont {Huijse}(2011)}]{Huijse_2011_JStatMech}%
  \BibitemOpen
  \bibfield  {author} {\bibinfo {author} {\bibfnamefont {L.}~\bibnamefont
  {Huijse}},\ }\href@noop {} {\bibfield  {journal} {\bibinfo  {journal} {J.
  Stat. Mech.}\ }\textbf {\bibinfo {volume} {2011}},\ \bibinfo {pages} {P04004}
  (\bibinfo {year} {2011})}\BibitemShut {NoStop}%
\bibitem [{\citenamefont {Milsted}\ \emph {et~al.}(2020)\citenamefont
  {Milsted}, \citenamefont {Liu}, \citenamefont {Preskill},\ and\ \citenamefont
  {Vidal}}]{Milsted_2020}%
  \BibitemOpen
  \bibfield  {author} {\bibinfo {author} {\bibfnamefont {A.}~\bibnamefont
  {Milsted}}, \bibinfo {author} {\bibfnamefont {J.}~\bibnamefont {Liu}},
  \bibinfo {author} {\bibfnamefont {J.}~\bibnamefont {Preskill}}, \ and\
  \bibinfo {author} {\bibfnamefont {G.}~\bibnamefont {Vidal}},\ }\href@noop {}
  {\bibfield  {journal} {\bibinfo  {journal} {arXiv:2012.07243}\ } (\bibinfo
  {year} {2020})}\BibitemShut {NoStop}%
\bibitem [{\citenamefont {Fendley}\ and\ \citenamefont
  {Hagendorf}(2011)}]{Fendley_2011_JStatMech}%
  \BibitemOpen
  \bibfield  {author} {\bibinfo {author} {\bibfnamefont {P.}~\bibnamefont
  {Fendley}}\ and\ \bibinfo {author} {\bibfnamefont {C.}~\bibnamefont
  {Hagendorf}},\ }\href@noop {} {\bibfield  {journal} {\bibinfo  {journal} {J.
  Stat. Mech.}\ }\textbf {\bibinfo {volume} {2011}},\ \bibinfo {pages} {P02014}
  (\bibinfo {year} {2011})}\BibitemShut {NoStop}%
\bibitem [{Sup()}]{Suppl}%
  \BibitemOpen
  \href@noop {} {}\bibinfo {note} {See Supplemental Material for (i) particle
  densities of the critical ground states, (ii) (s)kink profiles and
  observables, (iii) state preparation, (iv) saddle point approximation and (v)
  details of the experimental implementation.}\BibitemShut {Stop}%
\bibitem [{\citenamefont {Fokkema}\ and\ \citenamefont
  {Schoutens}(2017)}]{Fokkema_2017_SciPost}%
  \BibitemOpen
  \bibfield  {author} {\bibinfo {author} {\bibfnamefont {T.}~\bibnamefont
  {Fokkema}}\ and\ \bibinfo {author} {\bibfnamefont {K.}~\bibnamefont
  {Schoutens}},\ }\href@noop {} {\bibfield  {journal} {\bibinfo  {journal}
  {SciPost Phys.}\ }\textbf {\bibinfo {volume} {3}},\ \bibinfo {pages} {004}
  (\bibinfo {year} {2017})}\BibitemShut {NoStop}%
\bibitem [{\citenamefont {Min\'{a}\v{r}}\ and\ \citenamefont
  {Schoutens}()}]{InPrep}%
  \BibitemOpen
  \bibfield  {author} {\bibinfo {author} {\bibfnamefont {J.}~\bibnamefont
  {Min\'{a}\v{r}}}\ and\ \bibinfo {author} {\bibfnamefont {K.}~\bibnamefont
  {Schoutens}},\ }\href@noop {} {}\bibinfo {note} {In Preparation}\BibitemShut
  {NoStop}%
\bibitem [{\citenamefont {Fendley}\ and\ \citenamefont
  {Hagendorf}(2010)}]{Fendley_2010_JPhysA}%
  \BibitemOpen
  \bibfield  {author} {\bibinfo {author} {\bibfnamefont {P.}~\bibnamefont
  {Fendley}}\ and\ \bibinfo {author} {\bibfnamefont {C.}~\bibnamefont
  {Hagendorf}},\ }\href@noop {} {\bibfield  {journal} {\bibinfo  {journal} {J.
  Phys. A}\ }\textbf {\bibinfo {volume} {43}},\ \bibinfo {pages} {402004}
  (\bibinfo {year} {2010})}\BibitemShut {NoStop}%
\bibitem [{\citenamefont {Henkel}\ \emph {et~al.}(2010)\citenamefont {Henkel},
  \citenamefont {Nath},\ and\ \citenamefont {Pohl}}]{Henkel_2010_PRL}%
  \BibitemOpen
  \bibfield  {author} {\bibinfo {author} {\bibfnamefont {N.}~\bibnamefont
  {Henkel}}, \bibinfo {author} {\bibfnamefont {R.}~\bibnamefont {Nath}}, \ and\
  \bibinfo {author} {\bibfnamefont {T.}~\bibnamefont {Pohl}},\ }\href {\doibase
  10.1103/PhysRevLett.104.195302} {\bibfield  {journal} {\bibinfo  {journal}
  {Phys. Rev. Lett.}\ }\textbf {\bibinfo {volume} {104}},\ \bibinfo {pages}
  {195302} (\bibinfo {year} {2010})}\BibitemShut {NoStop}%
\bibitem [{\citenamefont {Pupillo}\ \emph {et~al.}(2010)\citenamefont
  {Pupillo}, \citenamefont {Micheli}, \citenamefont {Boninsegni}, \citenamefont
  {Lesanovsky},\ and\ \citenamefont {Zoller}}]{Pupillo_2010_PRL}%
  \BibitemOpen
  \bibfield  {author} {\bibinfo {author} {\bibfnamefont {G.}~\bibnamefont
  {Pupillo}}, \bibinfo {author} {\bibfnamefont {A.}~\bibnamefont {Micheli}},
  \bibinfo {author} {\bibfnamefont {M.}~\bibnamefont {Boninsegni}}, \bibinfo
  {author} {\bibfnamefont {I.}~\bibnamefont {Lesanovsky}}, \ and\ \bibinfo
  {author} {\bibfnamefont {P.}~\bibnamefont {Zoller}},\ }\href {\doibase
  10.1103/PhysRevLett.104.223002} {\bibfield  {journal} {\bibinfo  {journal}
  {Phys. Rev. Lett.}\ }\textbf {\bibinfo {volume} {104}},\ \bibinfo {pages}
  {223002} (\bibinfo {year} {2010})}\BibitemShut {NoStop}%
\bibitem [{\citenamefont {{\v{S}}ibali{\'c}}\ \emph {et~al.}(2017)\citenamefont
  {{\v{S}}ibali{\'c}}, \citenamefont {Pritchard}, \citenamefont {Adams},\ and\
  \citenamefont {Weatherill}}]{Sibalic_CompPhysComm_2017}%
  \BibitemOpen
  \bibfield  {author} {\bibinfo {author} {\bibfnamefont {N.}~\bibnamefont
  {{\v{S}}ibali{\'c}}}, \bibinfo {author} {\bibfnamefont {J.~D.}\ \bibnamefont
  {Pritchard}}, \bibinfo {author} {\bibfnamefont {C.~S.}\ \bibnamefont
  {Adams}}, \ and\ \bibinfo {author} {\bibfnamefont {K.~J.}\ \bibnamefont
  {Weatherill}},\ }\href@noop {} {\bibfield  {journal} {\bibinfo  {journal}
  {Computer Physics Communications}\ }\textbf {\bibinfo {volume} {220}},\
  \bibinfo {pages} {319} (\bibinfo {year} {2017})}\BibitemShut {NoStop}%
\bibitem [{\citenamefont {\v{S}ibali\'{c}}\ \emph {et~al.}()\citenamefont
  {\v{S}ibali\'{c}}, \citenamefont {Pritchard}, \citenamefont {Adams},\ and\
  \citenamefont {Weatherill}}]{ARC_package}%
  \BibitemOpen
  \bibfield  {author} {\bibinfo {author} {\bibfnamefont {N.}~\bibnamefont
  {\v{S}ibali\'{c}}}, \bibinfo {author} {\bibfnamefont {J.~D.}\ \bibnamefont
  {Pritchard}}, \bibinfo {author} {\bibfnamefont {C.~S.}\ \bibnamefont
  {Adams}}, \ and\ \bibinfo {author} {\bibfnamefont {K.~J.}\ \bibnamefont
  {Weatherill}},\ }\href@noop {} {\enquote {\bibinfo {title} {Arc package},}\
  }\bibinfo {howpublished}
  {\url{https://arc-alkali-rydberg-calculator.readthedocs.io/en/latest/}}\BibitemShut
  {NoStop}%
\bibitem [{\citenamefont {Arzamasovs}\ and\ \citenamefont
  {Liu}(2017)}]{Arzamasovs_2017}%
  \BibitemOpen
  \bibfield  {author} {\bibinfo {author} {\bibfnamefont {M.}~\bibnamefont
  {Arzamasovs}}\ and\ \bibinfo {author} {\bibfnamefont {B.}~\bibnamefont
  {Liu}},\ }\href@noop {} {\bibfield  {journal} {\bibinfo  {journal} {European
  Journal of Physics}\ }\textbf {\bibinfo {volume} {38}},\ \bibinfo {pages}
  {065405} (\bibinfo {year} {2017})}\BibitemShut {NoStop}%
\bibitem [{Note1()}]{Note1}%
  \BibitemOpen
  \bibinfo {note} {In order to be well in the deep lattice limit where the
  tight-binding approximation is applicable, one might further reduce $\Omega
  /\Delta $. This would in turn reduce $J$ and the achievable $L_{\protect \rm
  max}$. To overcome this limitation, one could use a Raman-assisted hopping as
  we discuss in \cite {Suppl}.}\BibitemShut {Stop}%
\bibitem [{\citenamefont {Weimer}\ \emph
  {et~al.}(2012{\natexlab{a}})\citenamefont {Weimer}, \citenamefont {Huijse},
  \citenamefont {Gorshkov}, \citenamefont {Pupillo}, \citenamefont {Zoller},
  \citenamefont {Lukin},\ and\ \citenamefont {Demler}}]{Weimer_APS_2012}%
  \BibitemOpen
  \bibfield  {author} {\bibinfo {author} {\bibfnamefont {H.}~\bibnamefont
  {Weimer}}, \bibinfo {author} {\bibfnamefont {L.}~\bibnamefont {Huijse}},
  \bibinfo {author} {\bibfnamefont {A.}~\bibnamefont {Gorshkov}}, \bibinfo
  {author} {\bibfnamefont {G.}~\bibnamefont {Pupillo}}, \bibinfo {author}
  {\bibfnamefont {P.}~\bibnamefont {Zoller}}, \bibinfo {author} {\bibfnamefont
  {M.}~\bibnamefont {Lukin}}, \ and\ \bibinfo {author} {\bibfnamefont
  {E.}~\bibnamefont {Demler}},\ }in\ \href
  {http://meetings.aps.org/link/BAPS.2012.DAMOP.T2.6} {\emph {\bibinfo
  {booktitle} {APS Division of Atomic, Molecular and Optical Physics Meeting
  Abstracts}}}\ (\bibinfo {year} {2012})\BibitemShut {NoStop}%
\bibitem [{\citenamefont {Weimer}\ \emph
  {et~al.}(2012{\natexlab{b}})\citenamefont {Weimer}, \citenamefont {Huijse},
  \citenamefont {Gorshkov}, \citenamefont {Pupillo}, \citenamefont {Zoller},
  \citenamefont {Lukin},\ and\ \citenamefont {Demler}}]{Weimer_DPG_2012}%
  \BibitemOpen
  \bibfield  {author} {\bibinfo {author} {\bibfnamefont {H.}~\bibnamefont
  {Weimer}}, \bibinfo {author} {\bibfnamefont {L.}~\bibnamefont {Huijse}},
  \bibinfo {author} {\bibfnamefont {A.}~\bibnamefont {Gorshkov}}, \bibinfo
  {author} {\bibfnamefont {G.}~\bibnamefont {Pupillo}}, \bibinfo {author}
  {\bibfnamefont {P.}~\bibnamefont {Zoller}}, \bibinfo {author} {\bibfnamefont
  {M.}~\bibnamefont {Lukin}}, \ and\ \bibinfo {author} {\bibfnamefont
  {E.}~\bibnamefont {Demler}},\ }in\ \href
  {https://www.dpg-verhandlungen.de/year/2012/conference/berlin/part/tt/session/19/contribution/10}
  {\emph {\bibinfo {booktitle} {76. annual conference of the DPG and DPG Spring
  meeting 2012}}}\ (\bibinfo {year} {2012})\BibitemShut {NoStop}%
\bibitem [{\citenamefont {Huijse}\ \emph {et~al.}(2008)\citenamefont {Huijse},
  \citenamefont {Halverson}, \citenamefont {Fendley},\ and\ \citenamefont
  {Schoutens}}]{Huijse_2008_PRL}%
  \BibitemOpen
  \bibfield  {author} {\bibinfo {author} {\bibfnamefont {L.}~\bibnamefont
  {Huijse}}, \bibinfo {author} {\bibfnamefont {J.}~\bibnamefont {Halverson}},
  \bibinfo {author} {\bibfnamefont {P.}~\bibnamefont {Fendley}}, \ and\
  \bibinfo {author} {\bibfnamefont {K.}~\bibnamefont {Schoutens}},\ }\href@noop
  {} {\bibfield  {journal} {\bibinfo  {journal} {Phys. Rev. Lett.}\ ,\ \bibinfo
  {pages} {146406}} (\bibinfo {year} {2008})}\BibitemShut {NoStop}%
\bibitem [{\citenamefont {Huijse}\ \emph {et~al.}(2012)\citenamefont {Huijse},
  \citenamefont {Mehta}, \citenamefont {Moran}, \citenamefont {Schoutens},\
  and\ \citenamefont {Vala}}]{Huijse_2012_NJP}%
  \BibitemOpen
  \bibfield  {author} {\bibinfo {author} {\bibfnamefont {L.}~\bibnamefont
  {Huijse}}, \bibinfo {author} {\bibfnamefont {D.}~\bibnamefont {Mehta}},
  \bibinfo {author} {\bibfnamefont {N.}~\bibnamefont {Moran}}, \bibinfo
  {author} {\bibfnamefont {K.}~\bibnamefont {Schoutens}}, \ and\ \bibinfo
  {author} {\bibfnamefont {J.}~\bibnamefont {Vala}},\ }\href@noop {} {\bibfield
   {journal} {\bibinfo  {journal} {New J. Phys.}\ }\textbf {\bibinfo {volume}
  {14}},\ \bibinfo {pages} {073002} (\bibinfo {year} {2012})}\BibitemShut
  {NoStop}%
\bibitem [{\citenamefont {Galanakis}\ \emph {et~al.}(2012)\citenamefont
  {Galanakis}, \citenamefont {Henley},\ and\ \citenamefont
  {Papanikolaou}}]{Galanakis_2012_PRB}%
  \BibitemOpen
  \bibfield  {author} {\bibinfo {author} {\bibfnamefont {D.}~\bibnamefont
  {Galanakis}}, \bibinfo {author} {\bibfnamefont {C.~L.}\ \bibnamefont
  {Henley}}, \ and\ \bibinfo {author} {\bibfnamefont {S.}~\bibnamefont
  {Papanikolaou}},\ }\href {\doibase 10.1103/PhysRevB.86.195105} {\bibfield
  {journal} {\bibinfo  {journal} {Phys. Rev. B}\ }\textbf {\bibinfo {volume}
  {86}},\ \bibinfo {pages} {195105} (\bibinfo {year} {2012})}\BibitemShut
  {NoStop}%
\bibitem [{\citenamefont {Surace}\ \emph {et~al.}(2020)\citenamefont {Surace},
  \citenamefont {Giudici},\ and\ \citenamefont {Dalmonte}}]{Surace_2020}%
  \BibitemOpen
  \bibfield  {author} {\bibinfo {author} {\bibfnamefont {F.~M.}\ \bibnamefont
  {Surace}}, \bibinfo {author} {\bibfnamefont {G.}~\bibnamefont {Giudici}}, \
  and\ \bibinfo {author} {\bibfnamefont {M.}~\bibnamefont {Dalmonte}},\
  }\href@noop {} {\bibfield  {journal} {\bibinfo  {journal} {arXiv:2003.11073}\
  } (\bibinfo {year} {2020})}\BibitemShut {NoStop}%
\bibitem [{\citenamefont {Gambetta}\ \emph {et~al.}(2020)\citenamefont
  {Gambetta}, \citenamefont {Li}, \citenamefont {Schmidt-Kaler},\ and\
  \citenamefont {Lesanovsky}}]{Gambetta_PRL_2020}%
  \BibitemOpen
  \bibfield  {author} {\bibinfo {author} {\bibfnamefont {F.~M.}\ \bibnamefont
  {Gambetta}}, \bibinfo {author} {\bibfnamefont {W.}~\bibnamefont {Li}},
  \bibinfo {author} {\bibfnamefont {F.}~\bibnamefont {Schmidt-Kaler}}, \ and\
  \bibinfo {author} {\bibfnamefont {I.}~\bibnamefont {Lesanovsky}},\ }\href
  {\doibase 10.1103/PhysRevLett.124.043402} {\bibfield  {journal} {\bibinfo
  {journal} {Phys. Rev. Lett.}\ }\textbf {\bibinfo {volume} {124}},\ \bibinfo
  {pages} {043402} (\bibinfo {year} {2020})}\BibitemShut {NoStop}%
\bibitem [{\citenamefont {B{\"u}chler}\ \emph {et~al.}(2007)\citenamefont
  {B{\"u}chler}, \citenamefont {Micheli},\ and\ \citenamefont
  {Zoller}}]{Buchler_2007}%
  \BibitemOpen
  \bibfield  {author} {\bibinfo {author} {\bibfnamefont {H.}~\bibnamefont
  {B{\"u}chler}}, \bibinfo {author} {\bibfnamefont {A.}~\bibnamefont
  {Micheli}}, \ and\ \bibinfo {author} {\bibfnamefont {P.}~\bibnamefont
  {Zoller}},\ }\href@noop {} {\bibfield  {journal} {\bibinfo  {journal} {Nature
  Physics}\ }\textbf {\bibinfo {volume} {3}},\ \bibinfo {pages} {726} (\bibinfo
  {year} {2007})}\BibitemShut {NoStop}%
\bibitem [{\citenamefont {Carr}\ \emph {et~al.}(2009)\citenamefont {Carr},
  \citenamefont {DeMille}, \citenamefont {Krems},\ and\ \citenamefont
  {Ye}}]{Carr_2009_NJP}%
  \BibitemOpen
  \bibfield  {author} {\bibinfo {author} {\bibfnamefont {L.~D.}\ \bibnamefont
  {Carr}}, \bibinfo {author} {\bibfnamefont {D.}~\bibnamefont {DeMille}},
  \bibinfo {author} {\bibfnamefont {R.~V.}\ \bibnamefont {Krems}}, \ and\
  \bibinfo {author} {\bibfnamefont {J.}~\bibnamefont {Ye}},\ }\href@noop {}
  {\bibfield  {journal} {\bibinfo  {journal} {New Journal of Physics}\ }\textbf
  {\bibinfo {volume} {11}},\ \bibinfo {pages} {055049} (\bibinfo {year}
  {2009})}\BibitemShut {NoStop}%
\bibitem [{\citenamefont {Baranov}\ \emph {et~al.}(2012)\citenamefont
  {Baranov}, \citenamefont {Dalmonte}, \citenamefont {Pupillo},\ and\
  \citenamefont {Zoller}}]{Baranov_2012_ChemRev}%
  \BibitemOpen
  \bibfield  {author} {\bibinfo {author} {\bibfnamefont {M.~A.}\ \bibnamefont
  {Baranov}}, \bibinfo {author} {\bibfnamefont {M.}~\bibnamefont {Dalmonte}},
  \bibinfo {author} {\bibfnamefont {G.}~\bibnamefont {Pupillo}}, \ and\
  \bibinfo {author} {\bibfnamefont {P.}~\bibnamefont {Zoller}},\ }\href@noop {}
  {\bibfield  {journal} {\bibinfo  {journal} {Chemical Reviews}\ }\textbf
  {\bibinfo {volume} {112}},\ \bibinfo {pages} {5012} (\bibinfo {year}
  {2012})}\BibitemShut {NoStop}%
\bibitem [{\citenamefont {Balakrishnan}(2016)}]{Balakrishnan_2016_JChemPhys}%
  \BibitemOpen
  \bibfield  {author} {\bibinfo {author} {\bibfnamefont {N.}~\bibnamefont
  {Balakrishnan}},\ }\href@noop {} {\bibfield  {journal} {\bibinfo  {journal}
  {The Journal of chemical physics}\ }\textbf {\bibinfo {volume} {145}},\
  \bibinfo {pages} {150901} (\bibinfo {year} {2016})}\BibitemShut {NoStop}%
\bibitem [{\citenamefont {Pedersen}\ \emph {et~al.}(2019)\citenamefont
  {Pedersen}, \citenamefont {Christensen},\ and\ \citenamefont
  {Zinner}}]{Pedersen_2019_PRR}%
  \BibitemOpen
  \bibfield  {author} {\bibinfo {author} {\bibfnamefont {S.~P.}\ \bibnamefont
  {Pedersen}}, \bibinfo {author} {\bibfnamefont {K.~S.}\ \bibnamefont
  {Christensen}}, \ and\ \bibinfo {author} {\bibfnamefont {N.~T.}\ \bibnamefont
  {Zinner}},\ }\href {\doibase 10.1103/PhysRevResearch.1.033123} {\bibfield
  {journal} {\bibinfo  {journal} {Phys. Rev. Research}\ }\textbf {\bibinfo
  {volume} {1}},\ \bibinfo {pages} {033123} (\bibinfo {year}
  {2019})}\BibitemShut {NoStop}%
\bibitem [{\citenamefont {Roy}\ \emph {et~al.}(2020)\citenamefont {Roy},
  \citenamefont {Hazra}, \citenamefont {Kundu}, \citenamefont {Chand},
  \citenamefont {Patankar},\ and\ \citenamefont {Vijay}}]{Roy_2020_PRApp}%
  \BibitemOpen
  \bibfield  {author} {\bibinfo {author} {\bibfnamefont {T.}~\bibnamefont
  {Roy}}, \bibinfo {author} {\bibfnamefont {S.}~\bibnamefont {Hazra}}, \bibinfo
  {author} {\bibfnamefont {S.}~\bibnamefont {Kundu}}, \bibinfo {author}
  {\bibfnamefont {M.}~\bibnamefont {Chand}}, \bibinfo {author} {\bibfnamefont
  {M.~P.}\ \bibnamefont {Patankar}}, \ and\ \bibinfo {author} {\bibfnamefont
  {R.}~\bibnamefont {Vijay}},\ }\href {\doibase
  10.1103/PhysRevApplied.14.014072} {\bibfield  {journal} {\bibinfo  {journal}
  {Phys. Rev. Applied}\ }\textbf {\bibinfo {volume} {14}},\ \bibinfo {pages}
  {014072} (\bibinfo {year} {2020})}\BibitemShut {NoStop}%
\bibitem [{\citenamefont {Press}\ \emph {et~al.}(2007)\citenamefont {Press},
  \citenamefont {Teukolsky}, \citenamefont {Vetterling},\ and\ \citenamefont
  {Flannery}}]{Press_2007}%
  \BibitemOpen
  \bibfield  {author} {\bibinfo {author} {\bibfnamefont {W.~H.}\ \bibnamefont
  {Press}}, \bibinfo {author} {\bibfnamefont {S.~A.}\ \bibnamefont
  {Teukolsky}}, \bibinfo {author} {\bibfnamefont {W.~T.}\ \bibnamefont
  {Vetterling}}, \ and\ \bibinfo {author} {\bibfnamefont {B.~P.}\ \bibnamefont
  {Flannery}},\ }\href@noop {} {\emph {\bibinfo {title} {Numerical recipes 3rd
  edition: The art of scientific computing}}}\ (\bibinfo  {publisher}
  {Cambridge university press},\ \bibinfo {year} {2007})\BibitemShut {NoStop}%
\bibitem [{\citenamefont {Messiah}(1999)}]{Messiah_1999}%
  \BibitemOpen
  \bibinfo {editor} {\bibfnamefont {A.}~\bibnamefont {Messiah}},\ ed.,\
  \href@noop {} {\emph {\bibinfo {title} {Quantum Mechanics}}}\ (\bibinfo
  {publisher} {Dover Publications},\ \bibinfo {address} {New York},\ \bibinfo
  {year} {1999})\BibitemShut {NoStop}%
\bibitem [{\citenamefont {Teufel}(2003)}]{Teufel_2003_Book}%
  \BibitemOpen
  \bibfield  {author} {\bibinfo {author} {\bibfnamefont {S.}~\bibnamefont
  {Teufel}},\ }\href@noop {} {\emph {\bibinfo {title} {Adiabatic perturbation
  theory in quantum dynamics}}}\ (\bibinfo  {publisher} {Springer},\ \bibinfo
  {year} {2003})\BibitemShut {NoStop}%
\bibitem [{\citenamefont {Albash}\ and\ \citenamefont
  {Lidar}(2018)}]{Albash_2018_RMP}%
  \BibitemOpen
  \bibfield  {author} {\bibinfo {author} {\bibfnamefont {T.}~\bibnamefont
  {Albash}}\ and\ \bibinfo {author} {\bibfnamefont {D.~A.}\ \bibnamefont
  {Lidar}},\ }\href {\doibase 10.1103/RevModPhys.90.015002} {\bibfield
  {journal} {\bibinfo  {journal} {Rev. Mod. Phys.}\ }\textbf {\bibinfo {volume}
  {90}},\ \bibinfo {pages} {015002} (\bibinfo {year} {2018})}\BibitemShut
  {NoStop}%
\bibitem [{\citenamefont {Avron}\ and\ \citenamefont
  {Elgart}(1998)}]{Avron_1998_PRA}%
  \BibitemOpen
  \bibfield  {author} {\bibinfo {author} {\bibfnamefont {J.~E.}\ \bibnamefont
  {Avron}}\ and\ \bibinfo {author} {\bibfnamefont {A.}~\bibnamefont {Elgart}},\
  }\href {\doibase 10.1103/PhysRevA.58.4300} {\bibfield  {journal} {\bibinfo
  {journal} {Phys. Rev. A}\ }\textbf {\bibinfo {volume} {58}},\ \bibinfo
  {pages} {4300} (\bibinfo {year} {1998})}\BibitemShut {NoStop}%
\bibitem [{\citenamefont {Rigolin}\ and\ \citenamefont
  {Ortiz}(2012)}]{Rigolin_2012_PRA}%
  \BibitemOpen
  \bibfield  {author} {\bibinfo {author} {\bibfnamefont {G.}~\bibnamefont
  {Rigolin}}\ and\ \bibinfo {author} {\bibfnamefont {G.}~\bibnamefont
  {Ortiz}},\ }\href {\doibase 10.1103/PhysRevA.85.062111} {\bibfield  {journal}
  {\bibinfo  {journal} {Phys. Rev. A}\ }\textbf {\bibinfo {volume} {85}},\
  \bibinfo {pages} {062111} (\bibinfo {year} {2012})}\BibitemShut {NoStop}%
\bibitem [{Note2()}]{Note2}%
  \BibitemOpen
  \bibinfo {note} {This is in contrast to the gap $E_{\protect \rm gap}$ used
  in the main text in scenario \protect \emph {(ii)}, which was the energy of
  the lowest excited state, i.e. its distance from the zero energy
  corresponding to the true ground state of the supersymmetric Hamiltonian on a
  chain with periodic boundaries. In the thermodynamic limit the lowest excited
  states become however equidistant as a consequence of the conformal symmetry
  and $E_{\protect \rm gap}$ coincides with $\Delta _{sg}$.}\BibitemShut
  {Stop}%
\bibitem [{Note3()}]{Note3}%
  \BibitemOpen
  \bibinfo {note} {We numerically implement the time evolution using
  Crank-Nicholson discrete time-step evolution given by $\mathinner {|{\psi
  (t+\Delta t)}\protect \rangle } = (1+ i H(t) \Delta t/2)^{-1} (1 - i H(t)
  \Delta t/2) \mathinner {|{\psi (t)}\protect \rangle }$ \cite
  {Press_2007}.}\BibitemShut {Stop}%
\bibitem [{\citenamefont {Agarwal}\ \emph {et~al.}(2018)\citenamefont
  {Agarwal}, \citenamefont {Bhatt},\ and\ \citenamefont
  {Sondhi}}]{Agarwal_2018_PRL}%
  \BibitemOpen
  \bibfield  {author} {\bibinfo {author} {\bibfnamefont {K.}~\bibnamefont
  {Agarwal}}, \bibinfo {author} {\bibfnamefont {R.~N.}\ \bibnamefont {Bhatt}},
  \ and\ \bibinfo {author} {\bibfnamefont {S.~L.}\ \bibnamefont {Sondhi}},\
  }\href {\doibase 10.1103/PhysRevLett.120.210604} {\bibfield  {journal}
  {\bibinfo  {journal} {Phys. Rev. Lett.}\ }\textbf {\bibinfo {volume} {120}},\
  \bibinfo {pages} {210604} (\bibinfo {year} {2018})}\BibitemShut {NoStop}%
\bibitem [{\citenamefont {Macr\`{\i}}\ and\ \citenamefont
  {Pohl}(2014)}]{Macri_2014_PRA}%
  \BibitemOpen
  \bibfield  {author} {\bibinfo {author} {\bibfnamefont {T.}~\bibnamefont
  {Macr\`{\i}}}\ and\ \bibinfo {author} {\bibfnamefont {T.}~\bibnamefont
  {Pohl}},\ }\href {\doibase 10.1103/PhysRevA.89.011402} {\bibfield  {journal}
  {\bibinfo  {journal} {Phys. Rev. A}\ }\textbf {\bibinfo {volume} {89}},\
  \bibinfo {pages} {011402} (\bibinfo {year} {2014})}\BibitemShut {NoStop}%
\bibitem [{Note4()}]{Note4}%
  \BibitemOpen
  \bibinfo {note} {Here it becomes apparent why we have introduced the
  site-dependent factor $(-1)^i$ in the definition of the supercharge $Q =
  \DOTSB \sum@ \slimits@ _i (-1)^i \lambda _i c^\protect \dag _i P_{\mathinner
  {\protect \langle {i}\protect \rangle }}$ - this is required, for $\lambda
  _i$ real, for the kinetic term to be \protect \emph {negative}. Another
  consequence of this choice is the appearance of triplets in the ground state
  $\mathinner {|{\protect \rm I}\protect \rangle }$. Conversely, the omission
  of the $(-1)^i$ factor leads to positive kinetic term and singlets instead of
  triplets in $\mathinner {|{{\protect \rm I}}\protect \rangle }$}\BibitemShut
  {NoStop}%
\bibitem [{\citenamefont {Jaksch}\ and\ \citenamefont
  {Zoller}(2003)}]{Jaksch_2003_NJP}%
  \BibitemOpen
  \bibfield  {author} {\bibinfo {author} {\bibfnamefont {D.}~\bibnamefont
  {Jaksch}}\ and\ \bibinfo {author} {\bibfnamefont {P.}~\bibnamefont
  {Zoller}},\ }\href@noop {} {\bibfield  {journal} {\bibinfo  {journal} {New
  Journal of Physics}\ }\textbf {\bibinfo {volume} {5}},\ \bibinfo {pages} {56}
  (\bibinfo {year} {2003})}\BibitemShut {NoStop}%
\bibitem [{\citenamefont {Aidelsburger}\ \emph {et~al.}(2011)\citenamefont
  {Aidelsburger}, \citenamefont {Atala}, \citenamefont {Nascimb\`ene},
  \citenamefont {Trotzky}, \citenamefont {Chen},\ and\ \citenamefont
  {Bloch}}]{Aidelsburger_2011_PRL}%
  \BibitemOpen
  \bibfield  {author} {\bibinfo {author} {\bibfnamefont {M.}~\bibnamefont
  {Aidelsburger}}, \bibinfo {author} {\bibfnamefont {M.}~\bibnamefont {Atala}},
  \bibinfo {author} {\bibfnamefont {S.}~\bibnamefont {Nascimb\`ene}}, \bibinfo
  {author} {\bibfnamefont {S.}~\bibnamefont {Trotzky}}, \bibinfo {author}
  {\bibfnamefont {Y.-A.}\ \bibnamefont {Chen}}, \ and\ \bibinfo {author}
  {\bibfnamefont {I.}~\bibnamefont {Bloch}},\ }\href {\doibase
  10.1103/PhysRevLett.107.255301} {\bibfield  {journal} {\bibinfo  {journal}
  {Phys. Rev. Lett.}\ }\textbf {\bibinfo {volume} {107}},\ \bibinfo {pages}
  {255301} (\bibinfo {year} {2011})}\BibitemShut {NoStop}%
\bibitem [{\citenamefont {Miyake}\ \emph {et~al.}(2013)\citenamefont {Miyake},
  \citenamefont {Siviloglou}, \citenamefont {Kennedy}, \citenamefont {Burton},\
  and\ \citenamefont {Ketterle}}]{Miyake_2013_PRL}%
  \BibitemOpen
  \bibfield  {author} {\bibinfo {author} {\bibfnamefont {H.}~\bibnamefont
  {Miyake}}, \bibinfo {author} {\bibfnamefont {G.~A.}\ \bibnamefont
  {Siviloglou}}, \bibinfo {author} {\bibfnamefont {C.~J.}\ \bibnamefont
  {Kennedy}}, \bibinfo {author} {\bibfnamefont {W.~C.}\ \bibnamefont {Burton}},
  \ and\ \bibinfo {author} {\bibfnamefont {W.}~\bibnamefont {Ketterle}},\
  }\href {\doibase 10.1103/PhysRevLett.111.185302} {\bibfield  {journal}
  {\bibinfo  {journal} {Phys. Rev. Lett.}\ }\textbf {\bibinfo {volume} {111}},\
  \bibinfo {pages} {185302} (\bibinfo {year} {2013})}\BibitemShut {NoStop}%
\bibitem [{\citenamefont {Lan}\ \emph {et~al.}(2015)\citenamefont {Lan},
  \citenamefont {Min\'a\ifmmode~\check{r}\else \v{r}\fi{}}, \citenamefont
  {Levi}, \citenamefont {Li},\ and\ \citenamefont {Lesanovsky}}]{Lan_2015_PRL}%
  \BibitemOpen
  \bibfield  {author} {\bibinfo {author} {\bibfnamefont {Z.}~\bibnamefont
  {Lan}}, \bibinfo {author} {\bibfnamefont {J.~c.~v.}\ \bibnamefont
  {Min\'a\ifmmode~\check{r}\else \v{r}\fi{}}}, \bibinfo {author} {\bibfnamefont
  {E.}~\bibnamefont {Levi}}, \bibinfo {author} {\bibfnamefont {W.}~\bibnamefont
  {Li}}, \ and\ \bibinfo {author} {\bibfnamefont {I.}~\bibnamefont
  {Lesanovsky}},\ }\href {\doibase 10.1103/PhysRevLett.115.203001} {\bibfield
  {journal} {\bibinfo  {journal} {Phys. Rev. Lett.}\ }\textbf {\bibinfo
  {volume} {115}},\ \bibinfo {pages} {203001} (\bibinfo {year}
  {2015})}\BibitemShut {NoStop}%
\bibitem [{\citenamefont {W{\"u}ster}\ \emph {et~al.}(2011)\citenamefont
  {W{\"u}ster}, \citenamefont {Ates}, \citenamefont {Eisfeld},\ and\
  \citenamefont {Rost}}]{Wuster_2011_NJP}%
  \BibitemOpen
  \bibfield  {author} {\bibinfo {author} {\bibfnamefont {S.}~\bibnamefont
  {W{\"u}ster}}, \bibinfo {author} {\bibfnamefont {C.}~\bibnamefont {Ates}},
  \bibinfo {author} {\bibfnamefont {A.}~\bibnamefont {Eisfeld}}, \ and\
  \bibinfo {author} {\bibfnamefont {J.}~\bibnamefont {Rost}},\ }\href@noop {}
  {\bibfield  {journal} {\bibinfo  {journal} {New J. Phys.}\ }\textbf {\bibinfo
  {volume} {13}},\ \bibinfo {pages} {073044} (\bibinfo {year}
  {2011})}\BibitemShut {NoStop}%
\bibitem [{\citenamefont {Goldschmidt}\ \emph {et~al.}(2016)\citenamefont
  {Goldschmidt}, \citenamefont {Boulier}, \citenamefont {Brown}, \citenamefont
  {Koller}, \citenamefont {Young}, \citenamefont {Gorshkov}, \citenamefont
  {Rolston},\ and\ \citenamefont {Porto}}]{Goldschmidt_2016_PRL}%
  \BibitemOpen
  \bibfield  {author} {\bibinfo {author} {\bibfnamefont {E.~A.}\ \bibnamefont
  {Goldschmidt}}, \bibinfo {author} {\bibfnamefont {T.}~\bibnamefont
  {Boulier}}, \bibinfo {author} {\bibfnamefont {R.~C.}\ \bibnamefont {Brown}},
  \bibinfo {author} {\bibfnamefont {S.~B.}\ \bibnamefont {Koller}}, \bibinfo
  {author} {\bibfnamefont {J.~T.}\ \bibnamefont {Young}}, \bibinfo {author}
  {\bibfnamefont {A.~V.}\ \bibnamefont {Gorshkov}}, \bibinfo {author}
  {\bibfnamefont {S.~L.}\ \bibnamefont {Rolston}}, \ and\ \bibinfo {author}
  {\bibfnamefont {J.~V.}\ \bibnamefont {Porto}},\ }\href {\doibase
  10.1103/PhysRevLett.116.113001} {\bibfield  {journal} {\bibinfo  {journal}
  {Phys. Rev. Lett.}\ }\textbf {\bibinfo {volume} {116}},\ \bibinfo {pages}
  {113001} (\bibinfo {year} {2016})}\BibitemShut {NoStop}%
\bibitem [{\citenamefont {Boulier}\ \emph {et~al.}(2017)\citenamefont
  {Boulier}, \citenamefont {Magnan}, \citenamefont {Bracamontes}, \citenamefont
  {Maslek}, \citenamefont {Goldschmidt}, \citenamefont {Young}, \citenamefont
  {Gorshkov}, \citenamefont {Rolston},\ and\ \citenamefont
  {Porto}}]{Boulier_2017_PRA}%
  \BibitemOpen
  \bibfield  {author} {\bibinfo {author} {\bibfnamefont {T.}~\bibnamefont
  {Boulier}}, \bibinfo {author} {\bibfnamefont {E.}~\bibnamefont {Magnan}},
  \bibinfo {author} {\bibfnamefont {C.}~\bibnamefont {Bracamontes}}, \bibinfo
  {author} {\bibfnamefont {J.}~\bibnamefont {Maslek}}, \bibinfo {author}
  {\bibfnamefont {E.~A.}\ \bibnamefont {Goldschmidt}}, \bibinfo {author}
  {\bibfnamefont {J.~T.}\ \bibnamefont {Young}}, \bibinfo {author}
  {\bibfnamefont {A.~V.}\ \bibnamefont {Gorshkov}}, \bibinfo {author}
  {\bibfnamefont {S.~L.}\ \bibnamefont {Rolston}}, \ and\ \bibinfo {author}
  {\bibfnamefont {J.~V.}\ \bibnamefont {Porto}},\ }\href {\doibase
  10.1103/PhysRevA.96.053409} {\bibfield  {journal} {\bibinfo  {journal} {Phys.
  Rev. A}\ }\textbf {\bibinfo {volume} {96}},\ \bibinfo {pages} {053409}
  (\bibinfo {year} {2017})}\BibitemShut {NoStop}%
\bibitem [{\citenamefont {Young}\ \emph {et~al.}(2018)\citenamefont {Young},
  \citenamefont {Boulier}, \citenamefont {Magnan}, \citenamefont {Goldschmidt},
  \citenamefont {Wilson}, \citenamefont {Rolston}, \citenamefont {Porto},\ and\
  \citenamefont {Gorshkov}}]{Young_2018_PRA}%
  \BibitemOpen
  \bibfield  {author} {\bibinfo {author} {\bibfnamefont {J.~T.}\ \bibnamefont
  {Young}}, \bibinfo {author} {\bibfnamefont {T.}~\bibnamefont {Boulier}},
  \bibinfo {author} {\bibfnamefont {E.}~\bibnamefont {Magnan}}, \bibinfo
  {author} {\bibfnamefont {E.~A.}\ \bibnamefont {Goldschmidt}}, \bibinfo
  {author} {\bibfnamefont {R.~M.}\ \bibnamefont {Wilson}}, \bibinfo {author}
  {\bibfnamefont {S.~L.}\ \bibnamefont {Rolston}}, \bibinfo {author}
  {\bibfnamefont {J.~V.}\ \bibnamefont {Porto}}, \ and\ \bibinfo {author}
  {\bibfnamefont {A.~V.}\ \bibnamefont {Gorshkov}},\ }\href {\doibase
  10.1103/PhysRevA.97.023424} {\bibfield  {journal} {\bibinfo  {journal} {Phys.
  Rev. A}\ }\textbf {\bibinfo {volume} {97}},\ \bibinfo {pages} {023424}
  (\bibinfo {year} {2018})}\BibitemShut {NoStop}%
\bibitem [{\citenamefont {Gallagher}(2005)}]{gallagher2005rydberg}%
  \BibitemOpen
  \bibfield  {author} {\bibinfo {author} {\bibfnamefont {T.~F.}\ \bibnamefont
  {Gallagher}},\ }\href@noop {} {\emph {\bibinfo {title} {Rydberg atoms}}},\
  \bibinfo {number} {3}\ (\bibinfo  {publisher} {Cambridge University Press},\
  \bibinfo {year} {2005})\BibitemShut {NoStop}%
\bibitem [{\citenamefont {Taie}\ \emph {et~al.}(2012)\citenamefont {Taie},
  \citenamefont {Yamazaki}, \citenamefont {Sugawa},\ and\ \citenamefont
  {Takahashi}}]{Taie_2012_NatPhys}%
  \BibitemOpen
  \bibfield  {author} {\bibinfo {author} {\bibfnamefont {S.}~\bibnamefont
  {Taie}}, \bibinfo {author} {\bibfnamefont {R.}~\bibnamefont {Yamazaki}},
  \bibinfo {author} {\bibfnamefont {S.}~\bibnamefont {Sugawa}}, \ and\ \bibinfo
  {author} {\bibfnamefont {Y.}~\bibnamefont {Takahashi}},\ }\href@noop {}
  {\bibfield  {journal} {\bibinfo  {journal} {Nature Physics}\ }\textbf
  {\bibinfo {volume} {8}},\ \bibinfo {pages} {825} (\bibinfo {year}
  {2012})}\BibitemShut {NoStop}%
\bibitem [{\citenamefont {Ozawa}\ \emph {et~al.}(2018)\citenamefont {Ozawa},
  \citenamefont {Taie}, \citenamefont {Takasu},\ and\ \citenamefont
  {Takahashi}}]{Ozawa_2018_PRL}%
  \BibitemOpen
  \bibfield  {author} {\bibinfo {author} {\bibfnamefont {H.}~\bibnamefont
  {Ozawa}}, \bibinfo {author} {\bibfnamefont {S.}~\bibnamefont {Taie}},
  \bibinfo {author} {\bibfnamefont {Y.}~\bibnamefont {Takasu}}, \ and\ \bibinfo
  {author} {\bibfnamefont {Y.}~\bibnamefont {Takahashi}},\ }\href {\doibase
  10.1103/PhysRevLett.121.225303} {\bibfield  {journal} {\bibinfo  {journal}
  {Phys. Rev. Lett.}\ }\textbf {\bibinfo {volume} {121}},\ \bibinfo {pages}
  {225303} (\bibinfo {year} {2018})}\BibitemShut {NoStop}%
\bibitem [{\citenamefont {Taie}\ \emph {et~al.}(2020)\citenamefont {Taie},
  \citenamefont {Ibarra-Garc{\'\i}a-Padilla}, \citenamefont {Nishizawa},
  \citenamefont {Takasu}, \citenamefont {Kuno}, \citenamefont {Wei},
  \citenamefont {Scalettar}, \citenamefont {Hazzard},\ and\ \citenamefont
  {Takahashi}}]{taie2020observation}%
  \BibitemOpen
  \bibfield  {author} {\bibinfo {author} {\bibfnamefont {S.}~\bibnamefont
  {Taie}}, \bibinfo {author} {\bibfnamefont {E.}~\bibnamefont
  {Ibarra-Garc{\'\i}a-Padilla}}, \bibinfo {author} {\bibfnamefont
  {N.}~\bibnamefont {Nishizawa}}, \bibinfo {author} {\bibfnamefont
  {Y.}~\bibnamefont {Takasu}}, \bibinfo {author} {\bibfnamefont
  {Y.}~\bibnamefont {Kuno}}, \bibinfo {author} {\bibfnamefont {H.-T.}\
  \bibnamefont {Wei}}, \bibinfo {author} {\bibfnamefont {R.~T.}\ \bibnamefont
  {Scalettar}}, \bibinfo {author} {\bibfnamefont {K.~R.}\ \bibnamefont
  {Hazzard}}, \ and\ \bibinfo {author} {\bibfnamefont {Y.}~\bibnamefont
  {Takahashi}},\ }\href@noop {} {\bibfield  {journal} {\bibinfo  {journal}
  {arXiv:2010.07730}\ } (\bibinfo {year} {2020})}\BibitemShut {NoStop}%
\bibitem [{\citenamefont {Lesanovsky}\ and\ \citenamefont
  {Katsura}(2012)}]{Lesanovsky_2012_PRA}%
  \BibitemOpen
  \bibfield  {author} {\bibinfo {author} {\bibfnamefont {I.}~\bibnamefont
  {Lesanovsky}}\ and\ \bibinfo {author} {\bibfnamefont {H.}~\bibnamefont
  {Katsura}},\ }\href {\doibase 10.1103/PhysRevA.86.041601} {\bibfield
  {journal} {\bibinfo  {journal} {Phys. Rev. A}\ }\textbf {\bibinfo {volume}
  {86}},\ \bibinfo {pages} {041601} (\bibinfo {year} {2012})}\BibitemShut
  {NoStop}%
\bibitem [{\citenamefont {Paeckel}\ \emph {et~al.}(2019)\citenamefont
  {Paeckel}, \citenamefont {K{\"o}hler}, \citenamefont {Swoboda}, \citenamefont
  {Manmana}, \citenamefont {Schollw{\"o}ck},\ and\ \citenamefont
  {Hubig}}]{Paeckel_2019_AnnPhys}%
  \BibitemOpen
  \bibfield  {author} {\bibinfo {author} {\bibfnamefont {S.}~\bibnamefont
  {Paeckel}}, \bibinfo {author} {\bibfnamefont {T.}~\bibnamefont {K{\"o}hler}},
  \bibinfo {author} {\bibfnamefont {A.}~\bibnamefont {Swoboda}}, \bibinfo
  {author} {\bibfnamefont {S.~R.}\ \bibnamefont {Manmana}}, \bibinfo {author}
  {\bibfnamefont {U.}~\bibnamefont {Schollw{\"o}ck}}, \ and\ \bibinfo {author}
  {\bibfnamefont {C.}~\bibnamefont {Hubig}},\ }\href@noop {} {\bibfield
  {journal} {\bibinfo  {journal} {Annals of Physics}\ }\textbf {\bibinfo
  {volume} {411}},\ \bibinfo {pages} {167998} (\bibinfo {year}
  {2019})}\BibitemShut {NoStop}%
\bibitem [{\citenamefont {Carleo}\ and\ \citenamefont
  {Troyer}(2017)}]{Carleo_2017_Science}%
  \BibitemOpen
  \bibfield  {author} {\bibinfo {author} {\bibfnamefont {G.}~\bibnamefont
  {Carleo}}\ and\ \bibinfo {author} {\bibfnamefont {M.}~\bibnamefont
  {Troyer}},\ }\href@noop {} {\bibfield  {journal} {\bibinfo  {journal}
  {Science}\ }\textbf {\bibinfo {volume} {355}},\ \bibinfo {pages} {602}
  (\bibinfo {year} {2017})}\BibitemShut {NoStop}%
\bibitem [{\citenamefont {Zohar}\ \emph {et~al.}(2012)\citenamefont {Zohar},
  \citenamefont {Cirac},\ and\ \citenamefont {Reznik}}]{Zohar_2012_PRL}%
  \BibitemOpen
  \bibfield  {author} {\bibinfo {author} {\bibfnamefont {E.}~\bibnamefont
  {Zohar}}, \bibinfo {author} {\bibfnamefont {J.~I.}\ \bibnamefont {Cirac}}, \
  and\ \bibinfo {author} {\bibfnamefont {B.}~\bibnamefont {Reznik}},\ }\href
  {\doibase 10.1103/PhysRevLett.109.125302} {\bibfield  {journal} {\bibinfo
  {journal} {Phys. Rev. Lett.}\ }\textbf {\bibinfo {volume} {109}},\ \bibinfo
  {pages} {125302} (\bibinfo {year} {2012})}\BibitemShut {NoStop}%
\bibitem [{\citenamefont {Banerjee}\ \emph {et~al.}(2012)\citenamefont
  {Banerjee}, \citenamefont {Dalmonte}, \citenamefont {M\"uller}, \citenamefont
  {Rico}, \citenamefont {Stebler}, \citenamefont {Wiese},\ and\ \citenamefont
  {Zoller}}]{Banerjee_2012_PRL}%
  \BibitemOpen
  \bibfield  {author} {\bibinfo {author} {\bibfnamefont {D.}~\bibnamefont
  {Banerjee}}, \bibinfo {author} {\bibfnamefont {M.}~\bibnamefont {Dalmonte}},
  \bibinfo {author} {\bibfnamefont {M.}~\bibnamefont {M\"uller}}, \bibinfo
  {author} {\bibfnamefont {E.}~\bibnamefont {Rico}}, \bibinfo {author}
  {\bibfnamefont {P.}~\bibnamefont {Stebler}}, \bibinfo {author} {\bibfnamefont
  {U.-J.}\ \bibnamefont {Wiese}}, \ and\ \bibinfo {author} {\bibfnamefont
  {P.}~\bibnamefont {Zoller}},\ }\href {\doibase
  10.1103/PhysRevLett.109.175302} {\bibfield  {journal} {\bibinfo  {journal}
  {Phys. Rev. Lett.}\ }\textbf {\bibinfo {volume} {109}},\ \bibinfo {pages}
  {175302} (\bibinfo {year} {2012})}\BibitemShut {NoStop}%
\bibitem [{\citenamefont {Kasamatsu}\ \emph {et~al.}(2013)\citenamefont
  {Kasamatsu}, \citenamefont {Ichinose},\ and\ \citenamefont
  {Matsui}}]{Kasamatsu_2013_PRL}%
  \BibitemOpen
  \bibfield  {author} {\bibinfo {author} {\bibfnamefont {K.}~\bibnamefont
  {Kasamatsu}}, \bibinfo {author} {\bibfnamefont {I.}~\bibnamefont {Ichinose}},
  \ and\ \bibinfo {author} {\bibfnamefont {T.}~\bibnamefont {Matsui}},\ }\href
  {\doibase 10.1103/PhysRevLett.111.115303} {\bibfield  {journal} {\bibinfo
  {journal} {Phys. Rev. Lett.}\ }\textbf {\bibinfo {volume} {111}},\ \bibinfo
  {pages} {115303} (\bibinfo {year} {2013})}\BibitemShut {NoStop}%
\bibitem [{\citenamefont {Haase}\ \emph {et~al.}(2021)\citenamefont {Haase},
  \citenamefont {Dellantonio}, \citenamefont {Celi}, \citenamefont {Paulson},
  \citenamefont {Kan}, \citenamefont {Jansen},\ and\ \citenamefont
  {Muschik}}]{Haase_2021}%
  \BibitemOpen
  \bibfield  {author} {\bibinfo {author} {\bibfnamefont {J.~F.}\ \bibnamefont
  {Haase}}, \bibinfo {author} {\bibfnamefont {L.}~\bibnamefont {Dellantonio}},
  \bibinfo {author} {\bibfnamefont {A.}~\bibnamefont {Celi}}, \bibinfo {author}
  {\bibfnamefont {D.}~\bibnamefont {Paulson}}, \bibinfo {author} {\bibfnamefont
  {A.}~\bibnamefont {Kan}}, \bibinfo {author} {\bibfnamefont {K.}~\bibnamefont
  {Jansen}}, \ and\ \bibinfo {author} {\bibfnamefont {C.~A.}\ \bibnamefont
  {Muschik}},\ }\href@noop {} {\bibfield  {journal} {\bibinfo  {journal}
  {Quantum}\ }\textbf {\bibinfo {volume} {5}},\ \bibinfo {pages} {393}
  (\bibinfo {year} {2021})}\BibitemShut {NoStop}%
\bibitem [{\citenamefont {Paulson}\ \emph {et~al.}(2020)\citenamefont
  {Paulson}, \citenamefont {Dellantonio}, \citenamefont {Haase}, \citenamefont
  {Celi}, \citenamefont {Kan}, \citenamefont {Jena}, \citenamefont {Kokail},
  \citenamefont {van Bijnen}, \citenamefont {Jansen}, \citenamefont {Zoller}
  \emph {et~al.}}]{Paulson_2020}%
  \BibitemOpen
  \bibfield  {author} {\bibinfo {author} {\bibfnamefont {D.}~\bibnamefont
  {Paulson}}, \bibinfo {author} {\bibfnamefont {L.}~\bibnamefont
  {Dellantonio}}, \bibinfo {author} {\bibfnamefont {J.~F.}\ \bibnamefont
  {Haase}}, \bibinfo {author} {\bibfnamefont {A.}~\bibnamefont {Celi}},
  \bibinfo {author} {\bibfnamefont {A.}~\bibnamefont {Kan}}, \bibinfo {author}
  {\bibfnamefont {A.}~\bibnamefont {Jena}}, \bibinfo {author} {\bibfnamefont
  {C.}~\bibnamefont {Kokail}}, \bibinfo {author} {\bibfnamefont
  {R.}~\bibnamefont {van Bijnen}}, \bibinfo {author} {\bibfnamefont
  {K.}~\bibnamefont {Jansen}}, \bibinfo {author} {\bibfnamefont
  {P.}~\bibnamefont {Zoller}},  \emph {et~al.},\ }\href@noop {} {\bibfield
  {journal} {\bibinfo  {journal} {arXiv:2008.09252}\ } (\bibinfo {year}
  {2020})}\BibitemShut {NoStop}%
\end{thebibliography}
\end{document}